\documentclass[%
 reprint,
 twocolumn,
 aps,
 prb,
 amsmath,amssymb,
 floatfix,
]{revtex4-2}

\usepackage[utf8]{inputenc}
\usepackage{graphicx}
\usepackage{dcolumn}
\usepackage{bm}
\usepackage{amsmath}
\usepackage{mathtools}
\usepackage{booktabs}
\usepackage{array}
\usepackage{tabularx}
\usepackage{xcolor}
\makeatletter
\def\switch@array{\let\@array@sw\@array@sw@LaTeX}
\makeatother
\usepackage[colorlinks=true,linkcolor=blue,citecolor=blue,urlcolor=blue,
            bookmarks=true,bookmarksnumbered=true]{hyperref}

\newcommand{\rs}{\ensuremath{r_s}}
\newcommand{\kf}{\ensuremath{k_{\text{F}}}}
\newcommand{\ef}{\ensuremath{E_{\text{F}}}}
\newcommand{\nz}{\ensuremath{n_0}}
\newcommand{\rlrt}{\ensuremath{r_{\text{LRT}}}}
\newcommand{\vh}{\ensuremath{V_{\text{H}}}}

\newcommand{\dnr}{\ensuremath{\Delta n(r)}}
\newcommand{\dnq}{\ensuremath{\Delta n(q)}}
\newcommand{\dn}[1]{\ensuremath{\Delta n(#1)}}
\newcommand{\rmin}{\ensuremath{r_{\text{min}}}}
\newcommand{\rmax}{\ensuremath{r_{\text{max}}}}

\begin{document}

\title{Nonlinear Static Screening of Positive Charges in an Electron Gas: Contact Hartree Potential}

\author{Mohammad Sherafati}
\email{Contact author: mohammadsherafati@alpharing.com}
\author{Gilles Rodway-Gant}
\author{Allan Xi Chen}
\affiliation{Alpha Ring International Limited, 1631 W. 135th St., Gardena, California 90249, USA}

\date{November 11, 2026}

\begin{abstract}
Electron screening of positive charges in metals is most strongly nonlinear in the static near-field regime. We revisit the static screening of a proton embedded in a homogeneous electron gas, focusing on the induced electron density and the density-derived contact Hartree potential $V_{\rm H}(0)$. Although evaluated at the impurity site, $V_{\rm H}(0)$ is not a strictly local observable but an $r$-weighted moment of the full induced density, in a formulation applicable to both linear-response and nonlinear density-functional descriptions. We compare Thomas--Fermi, random-phase-approximation, and local-field-corrected linear response with nonlinear density-functional-theory reference calculations within the local-density approximation and the Perdew--Burke--Ernzerhof generalized-gradient approximation. The Estreicher--Meier LDA parametrization closely reproduces the contact Hartree potential from direct LDA and historical self-consistent results and separates its core and Friedel contributions, although its radial fit obeys only an approximate induced-density cusp. For the two ground-state local-field-factor parametrizations examined here, Kaplan--Kukkonen and Corradini--Del Sole--Onida--Palummo, the differences in the contact density and $V_{\rm H}(0)$ are negligible at the scale of the present comparisons; this does not imply a general insensitivity to local-field corrections. We then use a phase-plus-amplitude variable-phase formulation to reconstruct the full quantal density generated by constrained screened model potentials. Friedel-constrained Yukawa, hydrogenic, and Hulth\'en potentials satisfy the same total-screening condition but generate different contact responses, showing that the Friedel sum rule does not uniquely determine $V_{\rm H}(0)$. In the Apagyi--Nagy construction, the exact cusp is imposed on an auxiliary density ansatz used in the Poisson construction, whereas reconstructed quantum densities obey it automatically through the Coulomb singularity; neither condition fixes the complete screening cloud or its Hartree moment. The constrained model-potential results approach the nonlinear density-functional-theory references toward high density but exhibit progressive deviation with increasing $\rs$ as the screening cloud becomes more spatially extended. In contrast, density-informed two-parameter Popovi\'c--Stott potentials, propagated through the same quantal variable-phase reconstruction of the density and Hartree potential without refitting, reproduce the corresponding historical Hartree-only and exchange--correlation-modified contact values to within $0.05\%$ relative accuracy. Taken together, these results show that the constraints used to construct a reduced potential fix only selected properties of the screening problem, not the full density response generated by the resulting potential. The Friedel sum rule constrains a global scattering quantity and the cusp constrains local short-range behavior; potentials satisfying these constraints can nevertheless generate different radial screening densities and different $V_{\rm H}(0)$. The Popovi\'c--Stott construction is distinct in additionally incorporating information from the generated screening density into the determination of the potential. The resulting one-center benchmark provides a baseline for genuine two-center screening calculations in metallic environments.
\end{abstract}

\maketitle

\section{Introduction}
\label{sec:introduction}

Electron screening of a charged impurity such as a proton in a metal is one of the oldest and most central problems in condensed-matter physics
\cite{1952FriedelXIVDistributionElectrons,1954LindhardPROPERTIESGASCHARGED}.
The electronic structure of hydrogen in metals provides a canonical example, motivated both by fundamental impurity physics and by applications including hydrogen storage, sensing, and catalysis
\cite{2005FukaiAtomisticStatesHydrogen,2006Pundt,2011JenaMaterialsHydrogenStorage}.
In the present work we isolate the static one-center problem of a positive charge embedded in an electron gas, with particular emphasis on the proton. Its screening cloud contains a strongly perturbed near-field region together with the long-range Friedel-oscillatory response characteristic of a degenerate electron system
\cite{1952FriedelXIVDistributionElectrons,2025FriedelOsc_Review}.

The natural first theoretical description of a charged impurity in an electron gas is linear response theory (LRT), in which the impurity is treated as an external perturbation and the induced density is determined by the density--density response function of the homogeneous electron gas
\cite{1994BowenStaticDielectricResponse,2005GiulianiQuantumTheoryElectron}.
The random-phase approximation (RPA) provides the basic linear-response description, while exchange--correlation effects beyond RPA can be incorporated through an exchange--correlation kernel or, equivalently in the present static density channel, a local-field correction (LFC) $G_{+}(q)$.
Static LFCs have been developed in analytical and quantum-Monte-Carlo-consistent forms, providing increasingly accurate descriptions of exchange--correlation effects in the metallic-density regime
\cite{1998CorradiniAnalyticalExpressionsLocalfield,2019ChenCombinedVariationalDiagrammatic,2021Kukkonen-Chen,2023KaplanQMCconsistentStaticSpin}.

Here the host metal is represented by the uniform electron gas within the jellium approximation
\cite{2005GiulianiQuantumTheoryElectron,2016LoosUniformElectronGas}.
The bare Coulomb field of a positive impurity in jellium produces a strong rearrangement of the electron density in its immediate vicinity, where the perturbation is no longer weak.
Linear response theory \cite{2005GiulianiQuantumTheoryElectron} is therefore not expected to describe contact quantities such as the on-top induced density and the contact Hartree potential quantitatively.

Self-consistent density-functional theory (DFT) provides a direct nonlinear treatment of this static screening problem
\cite{1964HohenbergInhomogeneousElectronGas,1965KohnSelfConsistentEquationsIncluding}.
Popovi\'c and Stott \cite{1974PopovicNonlinearSelfConsistentTheory} gave an early self-consistent treatment of a hydrogen impurity in Al and Mg, explicitly comparing the induced density with linear-response results.
Almbladh, von Barth, Popovi\'c, and Stott subsequently demonstrated the substantial failure of LRT for a proton in jellium over the metallic-density regime: their nonlinear calculation produced a much larger near-proton induced density and Hartree potential, a shifted Friedel-oscillation phase, and, at lower metallic densities, a shallow bound or incipient bound state
\cite{1976AlmbladhScreeningProtonElectron}.

Beyond these early DFT studies, nonlinear screening of charged impurities has been investigated through several complementary approaches, including approximate self-consistent DFT using a two-parameter trial electron--proton potential
\cite{1976PopovicTheoryHeatSolution},
the first gradient correction beyond LDA to the exchange--correlation potential
\cite{1978JenaElectronicStructureHydrogen},
and an analytical parametrization of the self-consistent DFT-LDA induced density across the metallic-density range
\cite{1983_Estreicher-Meier}.
Other treatments include collective-boson and hypernetted-chain methods
\cite{1979ArponenElectronLiquidCollective,1985GondzikScreeningPositiveParticles},
Green's-function approaches
\cite{1979NorskovElectronStructureSingle},
and quantum Monte Carlo
\cite{1988SugiyamaQuantumMonteCarlo}.
Together, these studies provide broader theoretical context for the charged-impurity problem; the present work focuses specifically on static proton screening in jellium and on the induced density and density-derived contact Hartree potential.
This body of work is complemented by historical reviews on the electronic structure of hydrogen and point defects in metals
\cite{1981Jena_Review,1983Jena_Book,1984GuptaElectronicStructureHydrogen,1987Demangeat}.

The central single-impurity electrostatic quantity considered in this work is the contact induced Hartree potential,
\begin{align}
V_{\rm H}(0)
=
4\pi\int_0^\infty r\,\Delta n(r)\,dr
=
\frac{2}{\pi}\int_0^\infty \Delta n(q)\,dq,
\label{eq:VH_contact_intro}
\end{align}
where $\Delta n(r)=n(r)-n_0$ is the induced electron density around the positive impurity.
With the sign convention used below, $V_{\rm H}(0)>0$ denotes the magnitude of the attractive electronic potential at the impurity site after the bare Coulomb singularity has been separated, and the corresponding impurity--screening-cloud interaction energy is
$U_{\rm H}(0)=ZV_{\rm H}(0)$.

Although evaluated at the impurity position, $V_{\rm H}(0)$ is not a strictly local observable.
The contact density $\Delta n(0)$ measures the electron pile-up exactly at the impurity, whereas $V_{\rm H}(0)$ is a Coulomb-weighted radial moment of the entire induced screening cloud.
It therefore provides a natural electrostatic measure of the screening experienced by the positive impurity and of the associated one-center screening-energy scale.
Because the strongest nonlinear rearrangement occurs in the near-field region, $V_{\rm H}(0)$ is particularly sensitive to the failure of weak-perturbation descriptions there, while still retaining contributions from the longer-range screening density.

Complementary to self-consistent DFT, screened model potentials provide a reduced description of the same electron--impurity scattering problem. In Kohn--Sham DFT, the effective potential is determined self-consistently with the electron density and its radial shape is therefore an output of the calculation. In a model-potential treatment, by contrast, the functional form of the screened interaction is prescribed in advance, while one or a small number of screening parameters are determined by imposing physical scattering constraints such as the Friedel sum rule (FSR) \cite{1952FriedelXIVDistributionElectrons}. The resulting description sacrifices full density--potential self-consistency in exchange for a compact representation whose scattering properties can be analyzed directly.

Rogers and co-workers, Grandjouan and Deutsch, and Meier used one-parameter screened Coulomb-type potentials, including Yukawa- and Hulth\'en-type forms, in closely related studies of the static screening of positive charges in metals and electron gases, including calculations of bound states, scattering phase shifts, induced densities, and related impurity observables
\cite{1970RogersBoundEigenstatesStatic,1971RogersPhaseShiftsStatic,1975GrandjouanPhaseShiftsStatic,1975MeierElectronDensitiesCharged}.
A common feature of these approaches is the reduction of the screening problem to electron--impurity scattering. Such scattering problems can be treated efficiently with the variable-phase approach (VPA)
\cite{1933MorseEffectExchangeScatteringa,1967Calogero_VariablePhaseApproach,1967Babikov_phase_shift},
in which the scattering phase shift is obtained from a first-order nonlinear phase equation. The parameters of a screened model potential can then be constrained by the FSR.

Apagyi and Nagy subsequently introduced a two-parameter hydrogenic-like screened potential by supplementing the FSR with the exact electron--proton cusp condition \cite{1988ApagyiNagyTwoParameter,1957Kato_cusp}. Crucially, they imposed this condition on an auxiliary total-density ansatz from which the model potential is obtained through Poisson's equation. This construction constraint is distinct from the cusp obeyed by the quantum density generated from that Coulomb-singular potential: for regular bound and continuum eigenstates the latter follows from the near-origin Schr\"odinger equation, without being separately imposed in the quantum reconstruction. The cusp fixes a local density slope relative to its contact amplitude, not the amplitude itself (see Appendix~\ref{app:qm_vpa_cusp}).

The same scattering framework also extends naturally beyond the strictly zero-temperature limit. At finite temperature, the spectral capacity of a screened potential to support bound states must be distinguished from the thermal occupation of those states, a distinction that enters finite-temperature screening formulations such as that of Perrot
\cite{1982PerrotTemperatureDependent}.
More recently, Arista generalized the static FSR to arbitrary electron degeneracy for a projectile impurity of charge $\pm Ze$, using parametrized Yukawa, hydrogenic, and Hulth\'en interactions
\cite{2025AristaNonlinearScreeningIons}.
The present work specializes to the exact $T=0$ static limit but uses the same phase-shift language to examine what Friedel-constrained screened potentials do, and do not, determine about the induced density and the contact Hartree response.

Because such analytic potentials have a prescribed short-range form, their screening parameters or finite near-origin terms can suggest simple contact-screening measures, but these quantities need not coincide with the contact Hartree potential generated by the actual quantum-mechanical screening density.

The role of the VPA in the present work should be distinguished from these earlier applications. The phase equation and the use of the FSR to constrain screened-potential parameters are established methods. Here we additionally solve the companion variable-amplitude equation to reconstruct the properly normalized scattering wavefunctions, combine the occupied continuum states with any discrete bound-state contributions, and thereby construct the full quantal induced density $\Delta n(r)$. This permits direct evaluation of the density-derived $V_{\rm H}(0)$ from the generated screening cloud, rather than identification of a contact quantity with the regular short-range coefficient of the analytic model potential. The calculation therefore provides a direct test of what information the Friedel constraint does, and does not, determine about the local contact density and the nonlocal contact Hartree response. Earlier screened-potential studies provide important benchmarks for phase shifts, FSR-constrained parameters, and bound-state spectra, but did not evaluate the full quantal $\Delta n(r)$ and its density-derived contact Hartree moment for the model potentials considered here.

Electron-screening enhancement of low-energy nuclear fusion in deuterated metals is an active problem in beam--target nuclear reactions
\cite{2025ChenElectrochemicalLoadingEnhances}.
In this context, the screening electrons reduce the electrostatic repulsion experienced by approaching positively charged nuclei and thereby lower the effective Coulomb barrier.
The contact Hartree potential $V_{\rm H}(0)$ of an isolated proton provides a natural one-center measure of the strength of this electronic barrier lowering, because it is the electrostatic potential produced by the full induced screening cloud at the position of the positive charge.

A simple additive picture based on two isolated screening clouds would suggest a barrier reduction of order twice the one-center contact Hartree potential.
Such an estimate is necessarily heuristic: as two protons approach one another, the electron gas undergoes a genuinely nonlinear two-center rearrangement that cannot in general be represented by superposing isolated one-center screening clouds.
We therefore treat $V_{\rm H}(0)$ here as a one-center electrostatic diagnostic of the nonlinear screening cloud and as a baseline for future work; it cannot determine the full electron-mediated proton--proton interaction without a genuine two-center calculation.

The proton-in-jellium problem, the nonlinear character of its short-range screening, and the use of screened model potentials constrained by scattering information are well established; the present work does not claim novelty for those ingredients. Instead, we ask how much information different levels of screening theory and different constraints on reduced potentials actually provide about the contact density and the density-derived contact Hartree potential. We first reproduce the well-established breakdown of Thomas--Fermi, RPA, and local-field-corrected screening for these contact quantities against nonlinear DFT-LDA and DFT-PBE reference calculations and historical self-consistent results. A specific feature of this comparison is the use of the recent Kaplan--Kukkonen (KK) LFC parametrization
\cite{2023KaplanQMCconsistentStaticSpin}
to quantify how much the predicted $V_{\rm H}(0)$ differs from that obtained with the earlier Corradini--Del Sole--Onida--Palummo (CDOP) parametrization
\cite{1998CorradiniAnalyticalExpressionsLocalfield}.

We then use the quantal VPA reconstruction described above to test what different constraints on reduced screened potentials actually determine. Friedel-constrained Yukawa, hydrogenic, and Hulth\'en potentials satisfy the same global total-screening condition but need not generate the same $\Delta n(0)$ or $V_{\rm H}(0)$. Imposing the exact cusp on the auxiliary Poisson-derived density in the two-parameter Apagyi--Nagy construction adds information to the construction of the potential, but does not remove this underdetermination of its independently reconstructed quantum density. For the FSR-only Yukawa, hydrogenic, and Hulth\'en potentials, Appendix~\ref{app:qm_vpa_cusp} provides an independent numerical check of the exact quantum-density cusp at $r_s=2$; no additional cusp constraint was used to determine their screening parameters. As a complementary test, we revisit the density-informed two-parameter potential of Popovi\'c--Stott
\cite{1974PopovicNonlinearSelfConsistentTheory, 1976PopovicTheoryHeatSolution},
for which the density generated by the trial potential enters the approximate potential--density self-consistency condition. Their published parameters, propagated through the present quantal VPA density reconstruction without refitting, closely reproduce the corresponding historical Hartree contacts for both the Hartree-only and exchange--correlation-modified constructions. Taken together, these comparisons separate the roles of global charge screening, local short-range constraints, and nonlocal information carried by the full radial density in determining $V_{\rm H}(0)$.

The remainder of the paper is organized as follows. Sec.~\ref{sec:theory} develops the theoretical framework for the jellium screening problem, including the contact Hartree potential, linear-response and DFT descriptions, the Friedel and cusp constraints, and the VPA reconstruction of bound and scattering states for screened model potentials. Sec.~\ref{sec:results} presents the numerical results, first comparing linear-response predictions with nonlinear DFT references and then testing one- and two-parameter screened potentials to determine how global, local, and density-informed constraints affect $\Delta n(0)$ and $V_{\rm H}(0)$. Sec.~\ref{sec:discussion} discusses the role of bound states in the quantal density reconstruction and examines the utility and limitations of constrained model-potential methods, including how their performance changes with electron-gas density and how those limitations should be distinguished from the VPA used to solve the corresponding scattering problem. Sec.~\ref{sec:conclusion} summarizes the main conclusions.

\section{Theory}
\label{sec:theory}

Our starting point is the so-called jellium model of interacting electrons in elemental metals \cite{1976AshcroftSolidStatePhysics, 2005GiulianiQuantumTheoryElectron, 2016LoosUniformElectronGas}. Throughout this work, we will be using Hartree atomic units (a.u.) for which $e = a_0 = m_{\text{e}} = \hbar = 4\pi \epsilon_0 = 1$ with Bohr radius $a_0$ being the atomic unit of length. One caveat is that in this unit system, the electrostatic potential and potential energy are given by the same expressions; we distinguish them clearly wherever needed. 

In the absence of any charge impurity, we assume that $N$ electrons confined to a three-dimensional (3D) cube of volume $V$ form a uniform electron gas such that as $V$ approaches infinity the gas density $\nz = N/V$ is held constant. To preserve charge neutrality, a homogeneous positive background of density $\nz$ is added to the electron system such that the equilibrium electron density is spatially constant. The electron gas density is characterized by the Wigner--Seitz radius of an electron, $\rs$ given by 
\begin{align}
    \frac{V}{N} = \frac{1}{\nz} = \frac{4 \pi \rs^3}{3} ; \quad  \rs = \left(\frac{3}{4\pi \nz}\right)^{1/3}
    \label{eq:rs}
\end{align}
We further assume a paramagnetic electron gas with equal numbers of spin-up and spin-down electrons at zero temperature obeying Fermi--Dirac statistics. The kinetic energy of free electrons is characterized by the Fermi energy $E_{\text{F}} = k_{\text{F}}^2 /2$ (a.u.) where Fermi momentum $k_{\text{F}}$ is related to gas density via 
\begin{align}
    \nz = \frac{3}{4 \pi \rs^3} = \frac{\kf^3}{3 \pi^2} ; \quad \kf = (9\pi/4)^{1/3} / \rs. 
    \label{eq:kF}
\end{align}
In this work, we focus mostly on the metallic densities with $\rs \in [2,6]$. For aluminum (Al) with valence electron density of $n_0 = 1.81\times 10^{23}\,\text{cm}^{-3} = 0.0268$ (a.u.), Eqs.~\eqref{eq:rs} and \eqref{eq:kF} yield $\rs = 2.07$ and $\kf = 0.925$, respectively. 

Due to an exact cancellation of the mutual electron--electron, electron--ion, and ion--ion electrostatic Coulomb interactions, the reduced ground-state energy per electron is expressed by the sum of only kinetic energy $\epsilon_{\text{kin}}(\rs) = (3/5)E_{\text{F}} = 1.105/\rs^2$ and quantal contributions from electron--electron interactions beyond electrostatic repulsion, known as exchange--correlation energies, namely $\epsilon_{\text{XC}}(\rs) = \epsilon_{\text{X}}(\rs) + \epsilon_{\text{C}} (\rs)$. The exchange energy arises entirely from the antisymmetry of the many-body electron wavefunction under the exchange of two electrons. Within the jellium model, this energy is obtained to be \cite{2005GiulianiQuantumTheoryElectron}   
\begin{align}
   \epsilon_{\text{X}}(\rs) \ \text{[a.u.]} = -\frac{3}{4 \pi} \kf =-\frac{0.4582}{\rs}.
\label{eq:Ex}
\end{align}
For the gas density regime with $\rs \in[1, 10]$, one of the best known analytical representations of the electron gas correlation energy was proposed by Perdew and Wang \cite{1992PerdewAccurateSimpleAnalytic}. Their parametrized correlation energy was found by fitting to the Green's-function Monte Carlo data of Ceperley and Alder densely sampled for $\rs = 1, 2, 5, 10$ \cite{1980CeperleyGroundStateElectron}. The Perdew--Wang (PW) reduced correlation energy $\epsilon_{\text{C}}$ is given by   
\begin{align}
   & \epsilon_C(\rs)  \ \text{[a.u.]}
= -2A_{\text{PW}}(1 + B_{\text{PW}} \rs) \times \nonumber \\
 &\ln\left[
1 + \frac{1/(2 A_{\text{PW}})}{
C_{\text{PW}} \rs^{1/2}
+ D_{\text{PW}} \rs
+ E_{\text{PW}} \rs^{3/2}
+ F_{\text{PW}} \rs^{2}
}
\right]
\label{eq:Ecorr_PW}
\end{align}
with parameters tabulated in Table~\ref{tab:PW_parameters}.
\begin{table}[htbp!]
\centering
\caption{Parameters of the Perdew--Wang (PW) correlation energy.}
\label{tab:PW_parameters}
\begin{tabular}{cc}
\toprule
Parameter & Value \\
\midrule
$A_{\text{PW}}$ & 0.031091 \\
$B_{\text{PW}}$ & 0.21370 \\
$C_{\text{PW}}$ & 7.5957 \\
$D_{\text{PW}}$ & 3.5876 \\
$E_{\text{PW}}$ & 1.6382 \\
$F_{\text{PW}}$ & 0.49294 \\
\bottomrule
\end{tabular}
\end{table}
Owing to its $\rs^{-2}$ dependence, $\epsilon_{\text{kin}}$ dominates the XC energy in the high-density regime ($\rs \ll 1$). This renders perturbative methods such as linear response theory rather accurate in this regime. However, in the metallic regime ($2 \le \rs \le 6$), the kinetic and exchange--correlation energies become comparable, namely $\epsilon_{\text{kin}} \sim \vert\epsilon_{\text{XC}}\vert$.

An electron gas in the presence of a single positive charge impurity such as proton forms an inhomogeneous electron gas. The ground-state properties of this system, the alterations in the otherwise uniform electron charge density by the impurity and the impurity potential, and how electrons screen the charge impurity to preserve the charge neutrality are among the most widely studied problems in condensed-matter physics pioneered by Friedel \cite{1952FriedelXIVDistributionElectrons} and Hohenberg, Kohn, and Sham \cite{1964HohenbergInhomogeneousElectronGas, 1965KohnSelfConsistentEquationsIncluding}. 

The central quantity of interest in this paper is the contact Hartree potential, namely the magnitude of the electrostatic potential generated by the induced electron cloud at the position of the positive charge impurity. The two well-established theoretical approaches used here to obtain this quantity are linear response theory and density-functional theory. We first outline a unified formulation of the Hartree potential associated with the screening electron cloud that is equally applicable to linear and nonlinear descriptions.

Suppose we denote the induced electron charge density by $\rho_{\text{ind}}(\bm r) = -e \ \Delta n(\bm r) = -e  \left[\ n(\bm r)-n_0\right] = -e  \ n_{\text{ind}}(\bm r)$, and the Coulomb function by 
$v(\bm r) =1/r$.
Therefore, the magnitude of the electrostatic Hartree potential generated by the screening electron cloud is given by 
\begin{align}
\vh(\bm r)
&= \int d^3r'\, v(\bm r-\bm r') \Delta n(\bm r') \nonumber \\
&= (v*\Delta n)(\bm r) \nonumber \\
&= \int \frac{d^3q}{(2\pi)^3}
e^{i\bm q\cdot\bm r} v_q \dnq \nonumber \\
&= \left(\frac{2}{\pi r}\right)
\int_0^{\infty} \frac{\dnq\sin(qr)}{q}\,dq
\label{eq:VH_q}
\end{align}
where the second equality results from the convolution theorem, which immediately yields $\vh(q) = v_q \dnq$ in momentum space for the Fourier transformation of $\vh(\bm r)$ given by the third equality with $v_q=4\pi/q^2$, where $v_q$ and $\Delta n(q)$ are the Fourier transforms of $v(\bm r)$ and $\Delta n(\bm r)$, respectively. To obtain the last expression, we have used the identity $\int f(q) e^{\pm i \bm q \cdot \bm r} d^3q =(4\pi/r)\int_0^{\infty} f(q) q \sin(qr) dq$. Now, the strength of the screening is characterized by the contact Hartree potential, namely $\vh(0) = \lim_{r \to 0} \vh(r)$. Applying $\lim_{r \to 0}\sin(qr)/(qr) = 1$ to Eq.~\eqref{eq:VH_q} immediately yields
\begin{align}
\vh(0) = \frac{2}{\pi}\int_{0}^{\infty}\dnq\,dq
\label{eq:VH0_q}    
\end{align}
For the sake of completeness, here we present an equivalent expression for the real-space Hartree potential, $\vh(r)$ when $\Delta n(r)$ is available. Due to the spherical symmetry of the Coulomb potential of a stationary positive charge impurity, one can apply the expansion of $1/|\bm r - \bm r'|$ in terms of the Legendre polynomials in the first equation in Eq.~\eqref{eq:VH_q}. After angular integration is performed, only the radial dependence remains which is presented as  
\begin{align}
\vh(r)
= \frac{1}{r} \int_0^r 4\pi r'^2 \,\Delta n(r') \,dr'
  + \int_r^{\infty} 4\pi r' \,\Delta n(r') \,dr'.
\label{eq:VH_r_sphsym}
\end{align}
To obtain the contact Hartree potential, once again we take $\vh(0) = \lim_{r \to 0} \vh(r)$, which makes the first term vanish due to successive applications of L'H\^opital's rule and the Leibniz formula for the derivative of an integral, which yields
\begin{align}
\vh(0) = 4\pi\int_{0}^{\infty} r\,\dnr\,dr
\label{eq:VH0_r}
\end{align}
We emphasize that Eqs.~\eqref{eq:VH_q}, \eqref{eq:VH0_q},
\eqref{eq:VH_r_sphsym}, and \eqref{eq:VH0_r} are exact electrostatic identities for a given induced electron number density. With our sign convention, $V_{\rm H}(0)$ denotes the positive magnitude of the Hartree potential generated by the screening electron cloud at the impurity
center. The momentum-space expressions, Eqs.~\eqref{eq:VH_q} and \eqref{eq:VH0_q}, are useful when $\Delta n(q)$ is available, whereas the real-space expressions, Eqs.~\eqref{eq:VH_r_sphsym} and \eqref{eq:VH0_r}, are useful when $\Delta n(r)$ is available. We now apply this formulation to linear-response and nonlinear density-functional
descriptions of the contact Hartree potential.

\subsection{Static linear-response evaluation of the induced contact Hartree potential
}
\label{sec:LRT_VH0}

As a perturbative method valid only for sufficiently weak perturbations, the LRT estimates the ground-state properties of an inhomogeneous electron gas in the presence of a positive charge $+Ze$ in terms of the response functions of the unperturbed system \cite{1999PinesElectronsPlasmons}. Taking only the linear term in the perturbation expansions sets a crucial limit to the validity of the LRT results for contact quantities including $\Delta n(0)$ and $\vh(0)$. As an order-of-magnitude estimate, the radial scale ($r_{\mathrm{LRT}}$) at which the bare impurity Coulomb energy becomes comparable to the Fermi energy is given by  
\begin{align}
 Ze^2/\rlrt = \ef \Longrightarrow \rlrt =0.543 Z \rs^2  
 \label{eq:rLRT}
\end{align} 
where we have used Eq.~\eqref{eq:kF}. Equation \eqref{eq:rLRT} provides a heuristic estimate of the radial region where nonlinear screening effects may become significant, rather than a sharp boundary for the validity of LRT. The estimated scale decreases with increasing electron density (decreasing $\rs$). For a proton in a palladium electron gas with $\rs=1.33$, Equation \eqref{eq:rLRT} gives $r_{\mathrm{LRT}}\simeq 0.96$, suggesting that nonlinear corrections may become important at distances of this order or smaller. Any nonlinear rearrangement of the screening cloud, in particular, its nonlinear dependence on $Z$ and the strong charge pile-up immediately around an attractive center lies outside the approximation imposed by the LRT, and it is the main reason that LRT is known to fail at the impurity itself despite the fact that it remains accurate at large distances.   

The most widely used formulation of the LRT for the screened potential of a positive test charge of magnitude $+Ze$ is based on the premise of a linear relationship in momentum space between the induced charge number density $\dnq$ and the isotropic external potential $V_{\text{ext}}(q)$ \cite{2005GiulianiQuantumTheoryElectron, 2005SimionFriedelOscillationsFermi}. This linearity is expressed in terms of the density-density response function of interacting electron gas denoted by $\chi(q) < 0$, viz. $\Delta n (q) =  - \chi(q) V_{\text{ext}}(q)$, where
\begin{align}
    \chi(q) = \frac{\chi_0(q)}{1-v_q\left[1-G_+(q)\right]\chi_0(q)}
    \label{eq:chi_q}
\end{align}
In Eq.~\eqref{eq:chi_q}, again $v_q = 4\pi/q^2$, $\chi_0$ is the density-density response function of the noninteracting electron gas also known as the Lindhard function \cite{1954LindhardPROPERTIESGASCHARGED}, and it is given by
\begin{align}
\chi_0(q)&= -\frac{\kf}{\pi^2} F_{\text{Lindhard}} (x) \nonumber\\
&=-\frac{\kf}{\pi^2}\left[\frac{1}{2}+\frac{x^2-4}{8x}\ln \left\vert\frac{x-2}{x+2}\right\vert\right] 
\label{eq:chi0_q}
\end{align}
where $x(q) = q/\kf$. The function $G_+(q)$ is the static density LFC, which carries the information on exchange--correlation effects in the electron gas. Appendix~\ref{sec:Gplusq} is devoted to the parametrizations of the LFC proposed by Corradini--Del Sole--Onida--Palummo (CDOP) \cite{1998CorradiniAnalyticalExpressionsLocalfield} consistent with quantum Monte Carlo data from Moroni and co-workers \cite{1995MoroniStaticResponseLocal} and a modern parametrization of Kaplan--Kukkonen (KK)  \cite{2023KaplanQMCconsistentStaticSpin}, which is consistent with the old data by Moroni \emph{et al.} and the most recent data \cite{2021Kukkonen-Chen, 2019ChenCombinedVariationalDiagrammatic}.  

The response of the system can also be formulated in terms of the so-called  test-charge-test-charge (TC) dielectric function, $\varepsilon_{\text{TC}}(q)$, which is defined to measure the strength of the screened potential $V_{\text{SC}}(q) = V_{\text{ext}}(q) + V_{\text{ind}}(q) = V_{\text{ext}}/\varepsilon_{\text{TC}}(q)$ in terms of the external potential. The TC dielectric function is given by \cite{2005GiulianiQuantumTheoryElectron} 
\begin{align}
\frac{1}{\varepsilon_{\text{TC}}(q)} = 1+v_q\chi(q) = \frac{1+v_qG_+(q)\chi_0(q)}{1-v_q\chi_0(q)\left[1-G_+(q)\right]}. 
\label{eq:eps_LFC}
\end{align}
Finally, the real-space induced electron density is obtained by taking the inverse Fourier transformation of $\dnq$, viz.     
\begin{align}
\Delta n(\bm r) &= n(\bm r) - \nz \nonumber\\
&= \int \frac{d^{3}q}{(2\pi)^{3}} e^{i\bm q\cdot\bm r}\,\dnq \nonumber\\
&= \int \frac{d^{3}q}{(2\pi)^{3}} e^{i\bm q\cdot\bm r}\left[1- \frac{1}{\varepsilon_{\text{TC}}(q)}\right] \frac{V_{\text{ext}}(q)}{v_q} \nonumber\\
&= \frac{Z}{2 \pi^2 r}\int^{\infty}_0 \left[1-\frac{1}{\varepsilon_{\text{TC}}(q)}\right] q\sin(qr) \, dq
\label{eq:dn_LRT}
\end{align}
where we have used the Fourier transformation of the Coulomb potential $V_{\text{ext}}(q) = 4\pi Ze/q^2$ and once again the identity used to obtain Eq.~\eqref{eq:VH_q}. The magnitude of the pile-up of the screening charge at the position of the test charge or the contact electron density is given from Eq.~\eqref{eq:dn_LRT} by 
\begin{align}
\Delta n(0) = n(0)-n_0 = \frac{Z}{2\pi^2}\int_0^{\infty} \left[1-\frac{1}{\varepsilon_{\text{TC}}(q)}\right]q^2 dq
\label{eq:dn0_LRT}
\end{align}
From this point on, we will drop the test-charge subscript and denote the dielectric function of the interacting electron gas given in Eq.~\eqref{eq:eps_LFC} as $\varepsilon_{\text{LFC}}(q)$. Within the random-phase approximation (RPA), ($G_+(q)=0$), so exchange–correlation local-field corrections are neglected while the self-consistent Hartree response to the Coulomb interaction is retained. The corresponding dielectric function is given by 
\begin{align}
\varepsilon_{\text{RPA}}(q)= 1-v_q\chi_0(q) = 1+\frac{k^2_{\text{TF}}}{q^2} F_{\text{Lindhard}}(x),  
\label{eq:eps_RPA}
\end{align}
where Eq.~\eqref{eq:chi0_q} has been used and $k^2_{\text{TF}} = (4/\pi) \kf$ is known as the Thomas--Fermi momentum. The simplest approximation for the response of a noninteracting electron gas is characterized by the long-wavelength limit of $\varepsilon_{\text{RPA}}(q)$, namely $x=q/\kf \ll 1 $ known as the Thomas--Fermi (TF) approximation. In this regime, $F_{\text{Lindhard}}(x)\to 1$ and Eq.~\eqref{eq:eps_RPA} gives the TF dielectric function, viz.
\begin{align}
\varepsilon_{\text{TF}}(q)= 1 - v_q \lim_{x \to 0 } \chi_0(q) = 1+ \frac{k^2_{\text{TF}}}{q^2}
\label{eq:eps_TF}
\end{align}
Substituting the linear-response form of $\dnq = Z\left[1-1/\varepsilon(q)\right]$ into Eq.~\eqref{eq:VH0_q} yields a master formula for all LRT regimes, namely
\begin{align}
\vh^{\text{LRT}}(0) =
\frac{2Z}{\pi}
\int_{0}^{\infty}
\left[\,1-\frac{1}{\varepsilon(q)}\,\right]\,dq.
\label{eq:VH0_LRT}
\end{align}
The three linear-response approximations used in this work differ only in
the choice of $\varepsilon(q)$. An analytical expression for
$\vh^{\text{LRT}}(0)$ can be found within the TF approximation, in which
the induced density is
$\Delta n_{\text{TF}}(q)=Zk_{\text{TF}}^{2}/(q^{2}+k_{\text{TF}}^{2})$.
Equation~\eqref{eq:VH0_LRT} then gives
\begin{align}
V_{\text{H, TF}}^{\text{LRT}}(0)
&=
\frac{2Z}{\pi}
\int_{0}^{\infty}
\frac{k_{\text{TF}}^{2}}{q^{2}+k_{\text{TF}}^{2}}\,dq
\nonumber\\
&=
Z\,k_{\text{TF}}
=
\frac{1.563Z}{\sqrt{\rs}}.
\label{eq:VH0_TF}
\end{align}
Equation~\eqref{eq:VH0_TF} shows that the contact Hartree potential within
the TF approximation is proportional to the TF inverse screening length,
$k_{\text{TF}}$. For a proton impurity in Al ($\rs=2.07$),
Eq.~\eqref{eq:VH0_TF} gives
$V_{\text{H, TF}}^{\text{LRT, Al}}(0)\simeq 1.086$~Ha in Hartree atomic
units. As a useful scale-setting reference, the electron cloud of an
isolated hydrogen atom gives
$V_{\text{H}}^{\text{free-H}}(0)=1~{\rm Ha}$ for $Z=1$.
This follows directly from the $1s$ density,
$n_{1s}(r)=|\Psi_{1s}(r)|^2=\frac{\alpha^3}{8\pi}e^{-\alpha r}$ with
$\alpha=2$, inserted into Eq.~\eqref{eq:VH0_r}, which gives
$\vh(0)=\alpha/2=1~{\rm Ha}$. This comparison is not meant
to identify a metal-screened proton with a free hydrogen atom; it only
provides a transparent reference scale for the contact Hartree potential.
The fact that the TF estimate for Al is already slightly larger than the
free-H value is therefore physically reasonable, while the nonlinear DFT
results below show that the actual proton-in-jellium contact Hartree
potential is larger still.

\subsection{Nonlinear evaluation of the induced contact Hartree potential: DFT}
\label{sec:DFT_VH0}

A more complete description of the static one-center problem comes from self-consistent electronic-structure methods such as DFT. In a DFT pipeline, the electron density $n(\bm r)$ is obtained self-consistently by minimizing an approximate ground-state energy functional subject to charge neutrality. In the Kohn--Sham (KS) formulation of DFT, the interacting electron gas with true $n(\bm r)$ is replaced by a noninteracting system whose density is found self-consistently to be $n(\bm r)$. As the ground-state energy must be a unique functional of $n(r)$ required by Hohenberg--Kohn \cite{1964HohenbergInhomogeneousElectronGas} theorems, the noninteracting KS system is governed by a series of single-particle equations known as KS equations for the KS orbitals $\psi_{i}(\bm r)$ satisfying
\begin{align}
\left[-\tfrac{1}{2}\nabla^{2} + V_{\text{eff}}(r)\right]\psi_{i}(\bm r)
 = \varepsilon_{i}\,\psi_{i}(\bm r),
 \label{eq:KSEq}
\end{align}
and generate the electron density $n(\bm r) = \sum_i |\psi_{i}(\bm r)|^2$. The effective potential, or KS potential, is constructed as a functional of $n(\bm r)$ and is given by \cite{1976AlmbladhScreeningProtonElectron}
\begin{align}
 V_{\text{eff}}\left[n(r)\right] & = -\frac{Z}{r} + \vh\left[n(r)\right] + V_{\text{XC}}\left[n(r)\right] \nonumber \\
 & = -\frac{Z}{r} + \int\frac{\Delta n (r') d^3r'}{|\bm r-\bm r'|} \nonumber \\
 & \quad +\mu_{\text{XC}}\left[n(r)\right]-\mu_{\text{XC}}(\nz) 
 \label{eq:VKS}
\end{align}
where the second equation is valid within the local-density approximation. Within the LDA, a local electron density is related to a variable Wigner--Seitz parameter $\rs$ via Eq.~\eqref{eq:kF}, namely $n(r) = 3/(4\pi \rs^3)$ whereas $\nz = 3/\left[4\pi (r_s^0)^3\right]$ with $r^0_s$ being constant in the absence of the impurity, and $\mu_{\text{XC}} = d\left[n \epsilon_{\text{XC}}\right]/dn = \epsilon_{\text{XC}} - (\rs/3)d\epsilon_{\text{XC}}/dr_s$ in terms of the local XC energy per particle $\epsilon_{\text{XC}}$ evaluated at the local $\rs$, for example the sum of Eqs.~\eqref{eq:Ex} and \eqref{eq:Ecorr_PW}. In the self-consistent impurity calculations reported in this work, $\epsilon_{\text{XC}}$ is taken in the Hedin--Lundqvist parametrization~\cite{1971HedinExplicitLocalExchangecorrelationa}, the functional employed by Almbladh \emph{et al.}~\cite{1976AlmbladhScreeningProtonElectron} before the Perdew--Wang fit became available; using the same functional enables a like-for-like comparison with their results (see Appendix~\ref{sec:almbladh_method}). The subtraction of $\mu_{\text{XC}}(n_0)$ merely sets the reference of the XC potential to be at its value for the homogeneous system. Along with minimizing the variational ground-state energy, $n(\bm r)$ must satisfy three crucial conditions: 1) charge neutrality, 2) the cusp condition, and 3) the Friedel oscillatory tail.  

In the DFT calculations beyond the LDA, exchange--correlation effects are treated within the Perdew--Burke--Ernzerhof (PBE) generalized-gradient approximation \cite{1996PerdewGeneralizedGradientApproximation}, which improves upon the local-density approximation by including density-gradient corrections while retaining a semilocal form for $E_{\mathrm{xc}}[n]$. The corresponding spherically symmetric PBE exchange–correlation potential is evaluated from the full density-gradient functional derivative, with its radial expression and numerical implementation specified in Appendix~\ref{sec:sc_cycle}.

\subsubsection{Quantal induced electron density $\Delta n^{\rm QM--DFT}(r)$}
\label{sec:dft_quantal_density}

Within the spherically symmetric Kohn--Sham formulation, the induced electron density relative to the uniform background is constructed explicitly from the occupied continuum
scattering states together with any discrete bound states supported by the
self-consistent Kohn--Sham potential. We denote this density by
$\Delta n^{\rm QM--DFT}(r)$ in order to distinguish it from the
Poisson-based model densities introduced in Sec.~\ref{sec:model_potentials}.

For each angular momentum $\ell$ and momentum $k\in[0,k_{\rm F}]$, let
$u_{\ell}(r;k)$ denote the properly normalized reduced radial Kohn--Sham
scattering wavefunction, and let $u_{\ell}^{(0)}(r;k)$ denote the
corresponding free-electron solution. The continuum contribution to the
induced density is then
\begin{align}
\Delta n_{\rm sc}^{\rm QM--DFT}(r)
={}&
\frac{1}{\pi^{2}r^{2}}
\sum_{\ell=0}^{\infty}(2\ell+1)
\int_{0}^{k_{\rm F}} dk\,
\nonumber\\
&\times
\left[
|u_{\ell}(r;k)|^{2}
-
|u_{\ell}^{(0)}(r;k)|^{2}
\right].
\label{eq:dn_sc_qm_dft}
\end{align}
Thus, the continuum density is obtained from the occupied Fermi sea rather
than from the scattering state at $k_{\rm F}$ alone.

If the self-consistent Kohn--Sham potential supports bound states, they enter
as separate discrete spectral contributions. For a normalized reduced radial
bound-state wavefunction $u_{\nu\ell}(r)$,
\begin{align}
\int_{0}^{\infty}|u_{\nu\ell}(r)|^{2}\,dr=1,
\end{align}
the bound-state contribution is
\begin{align}
\Delta n_{\rm b}^{\rm QM--DFT}(r)
=
\sum_{\nu\ell}
N_{\nu\ell}
\frac{|u_{\nu\ell}(r)|^{2}}{4\pi r^{2}},
\label{eq:dn_bound_qm_dft}
\end{align}
where $N_{\nu\ell}$ denotes the total occupation associated with the
$(\nu,\ell)$ level, including spin and magnetic degeneracy. The full quantal induced density is therefore
\begin{align}
\Delta n^{\rm QM--DFT}(r)
=
\Delta n_{\rm sc}^{\rm QM--DFT}(r)
+
\Delta n_{\rm b}^{\rm QM--DFT}(r).
\label{eq:dn_total_qm_dft}
\end{align}
This spectral decomposition avoids double counting: true bound states are included discretely, whereas the continuum contribution is obtained from the occupied scattering states. Its integrated charge is constrained by charge neutrality and, equivalently, by the Friedel sum rule discussed in Sec.~\ref{sec:FSR}. Numerical details concerning the radial Kohn--Sham solution, continuum normalization, treatment of shallow bound states, and
convergence are given in Appendix~\ref{sec:almbladh_method}.

\subsubsection{Charge neutrality and Friedel sum rule (FSR)}
\label{sec:FSR}

For an impurity of positive charge $Z$, complete static screening requires
the total induced electronic charge to satisfy
\begin{align}
Q
\equiv
4\pi\int_{0}^{\infty}
r^{2}\Delta n^{\rm QM--DFT}(r)\,dr
=
Z .
\label{eq:charge_neutrality}
\end{align}
The same screening condition may be expressed in scattering language through
the Friedel sum rule (FSR). Before specializing to the zero-temperature case
considered in this work, it is useful to distinguish two bound-state
quantities that coincide at $T=0$ but are physically different at finite
temperature.

Let $n_{\ell}^{\rm b}$ denote the number of negative-energy radial bound
levels supported by the potential in angular-momentum channel $\ell$,
\begin{align}
n_{\ell}^{\rm b}
=
\sum_{\nu:E_{\nu l}<0}1 .
\label{eq:nlb_def}
\end{align}
Each such radial level has the one-electron degeneracy $2(2\ell+1)$, including
spin and magnetic degeneracy. The total bound-state capacity of the
potential is therefore
\begin{align}
N_{\rm cap}
=
2\sum^{\infty}_{\ell=0}(2\ell+1)n_{\ell}^{\rm b}.
\label{eq:Ncap_def}
\end{align}
The quantity $N_{\rm cap}$ is an integer property of the spectrum supported
by the potential and does not specify how strongly those levels are
thermally occupied.

The occupation of an individual bound radial level $(\nu,l)$ is instead
determined by the chemical potential $\mu$ and temperature $T$. With
\begin{align}
f_{\rm FD}(\epsilon;T,\mu)
=
\frac{1}
{1+\exp[\beta(\epsilon-\mu)]},
\qquad
\beta=(k_{\rm B}T)^{-1},
\label{eq:FD_def}
\end{align}
we may write
\begin{align}
N_{\nu \ell}(T,\mu)
=
2(2\ell+1)
f_{\rm FD}(\epsilon_{\nu \ell};T,\mu),
\label{eq:Nnul_occupation}
\end{align}
so that the total occupied bound charge is
\begin{align}
Z_{\rm b}(T,\mu)
=
\sum_{\nu \ell}N_{\nu l}(T,\mu)
=
2\sum_{\nu \ell}(2\ell+1)
f_{\rm FD}(\epsilon_{\nu \ell};T,\mu).
\label{eq:Zb_def}
\end{align}
Thus, $N_{\rm cap}$ measures how many one-electron bound states the potential
can accommodate, whereas $Z_{\rm b}(T,\mu)$ measures the charge actually occupying
those states.

The connection between $N_{\rm cap}$ and scattering theory follows from
Levinson's theorem,
\begin{align}
\delta_{\ell}(0)
=
n_{\ell}^{\rm b}\pi ,
\label{eq:Levinson}
\end{align}
which gives
\begin{align}
N_{\rm cap}
=
\frac{2}{\pi}
\sum^{\infty}_{\ell=0}(2\ell+1)\delta_{\ell}(0).
\label{eq:Ncap_phase}
\end{align}
Defining
\begin{align}
s(k)
\equiv
\frac{2}{\pi}
\sum^{\infty}_{\ell=0}(2\ell+1)\delta_{\ell}(k),
\label{eq:s_of_k}
\end{align}
one therefore has
\begin{align}
N_{\rm cap}=s(0).
\label{eq:Ncap_s0}
\end{align}

At finite temperature, the generalized FSR may be written as
\begin{align}
Z
=
Z_{\rm b}(T,\mu)
+
\int_0^\infty
f_{\rm FD}(\epsilon_k;T,\mu)
\frac{d s(k)}{dk}\,dk ,
\qquad
\epsilon_k=\frac{k^2}{2}.
\label{eq:FSR_finiteT_derivative}
\end{align}
This form makes explicit the separate contributions of the thermally
occupied bound spectrum and the displaced continuum density of states
\cite{1982PerrotTemperatureDependent}. Integrating by parts and using
$s(0)=N_{\rm cap}$ gives the equivalent derivative-free form
\begin{align}
Z
&=
Z_{\rm b}(T,\mu)
-
N_{\rm cap}f_{\rm FD}(0;T,\mu)
\nonumber\\
&\quad+
\beta\int_0^\infty
k\,f_{\rm FD}(\epsilon_k;T,\mu)
\left[
1-f_{\rm FD}(\epsilon_k;T,\mu)
\right]
s(k)\,dk .
\label{eq:FSR_finiteT}
\end{align}
The first two terms,
\begin{align}
B_{\rm bound}(T,\mu)
\equiv
Z_{\rm b}(T,\mu)
-
N_{\rm cap}f_{\rm FD}(0;T,\mu),
\label{eq:Bbound}
\end{align}
show explicitly why the bound-state capacity and occupied bound charge must
not be conflated at finite temperature.

The present work concerns the exact $T=0$ limit of a degenerate electron gas,
for which $\mu=E_F>0$. Every negative-energy bound state is then fully
occupied,
\begin{align}
f_{\rm FD}(\epsilon_{\nu \ell};T=0,\mu=E_F)=1,
\qquad
\epsilon_{\nu \ell}<0,
\label{eq:zeroT_FD_bound}
\end{align}
and therefore
\begin{align}
N_{\nu l}(T=0,\mu=E_F)
=
2(2\ell+1),
\qquad
\epsilon_{\nu \ell}<0.
\label{eq:zeroT_Nnul}
\end{align}
Consequently,
\begin{align}
Z_{\rm b}(T=0,\mu=E_F)
&=
N_{\rm cap},
\nonumber\\
f_{\rm FD}(0;T=0,\mu=E_F)&=1,
\label{eq:zeroT_bound_equalities}
\end{align}
and the explicit bound-sector term in
Eq.~\eqref{eq:FSR_finiteT} cancels exactly,
\begin{align}
B_{\rm bound}(T=0,\mu=E_F)
={}&
Z_{\rm b}(T=0,\mu=E_F)
\nonumber\\
&-
N_{\rm cap}f_{\rm FD}(0;T=0,\mu=E_F)
\nonumber\\
={}&
0.
\label{eq:zeroT_Bbound}
\end{align}
At the same time, the thermal kernel approaches the Fermi-surface delta
function,
\begin{align}
\lim_{T\rightarrow0\atop \mu\rightarrow E_F}
\beta k\,
f_{\rm FD}(\epsilon_k;T,\mu)
\left[
1-f_{\rm FD}(\epsilon_k;T,\mu)
\right]
=
\delta(k-k_{\rm F}),
\label{eq:zeroT_kernel}
\end{align}
so that Eq.~\eqref{eq:FSR_finiteT} reduces to the familiar
zero-temperature FSR,
\begin{align}
Z
=
s(k_{\rm F})
=
\frac{2}{\pi}
\sum^{\infty}_{\ell=0}(2\ell+1)\delta_{\ell}(k_{\rm F}).
\label{eq:FSR}
\end{align}

Although $Z_{\rm b}(T=0,\mu=E_F)$ and $N_{\rm cap}$ are numerically identical in
the present zero-temperature problem, their physical meanings remain
distinct: $N_{\rm cap}$ is fixed by the bound-state spectrum of the
potential, whereas $Z_{\rm b}(T,\mu)$ is fixed by the thermodynamic occupation of
that spectrum. This distinction becomes essential at finite temperature,
where bound levels may be only partially occupied.

For explicit bound--continuum bookkeeping in the present $T=0$ limit,
Eqs.~\eqref{eq:Ncap_phase} and \eqref{eq:FSR} give the displaced continuum
charge
\begin{align}
Z_{\rm sc}(T=0,\mu=E_F)
&=
Z-Z_{\rm b}(T=0,\mu=E_F)
\nonumber\\
&=
\frac{2}{\pi}
\sum^{\infty}_{\ell=0}(2\ell+1)
\left[
\delta_{\ell}(k_{\rm F})-\delta_{\ell}(0)
\right].
\label{eq:Zsc}
\end{align}
Thus,
\begin{align}
Z
&=
Z_{\rm b}(T=0,\mu=E_F)
+
Z_{\rm sc}(T=0,\mu=E_F)
\nonumber\\
&=
N_{\rm cap}
+
Z_{\rm sc}(T=0,\mu=E_F),
\label{eq:charge_partition}
\end{align}
consistent with the real-space decomposition of
$\Delta n^{\rm QM--DFT}(r)$ into bound and scattering contributions in
Sec.~\ref{sec:dft_quantal_density}. In particular, when a doubly occupied
$1s$ state is present for a proton impurity,
\begin{align}
&n_0^{\rm b}=1,
\qquad
N_{\rm cap}
=
Z_{\rm b}(T=0,\mu=E_F)
=
2,
\nonumber\\
&Z_{\rm sc}(T=0,\mu=E_F)
=
-1,
\end{align}
so that the two bound electrons together with a continuum depletion of one
electron give the required total screening charge $Z=1$.

In the self-consistent DFT calculations, exact screening requires both the integrated induced charge $Q$ from Eq.~\eqref{eq:charge_neutrality} and the Friedel sum $s(k_{\rm F})$ from
Eq.~\eqref{eq:FSR} to equal the impurity charge $Z = 1$ in the infinite-domain limit. As discussed in Appendix~\ref{sec:almbladh_method}, numerical evaluation of the real-space charge integral is sensitive to finite-domain truncation and the treatment of the long-range induced-density tail, particularly near a shallow bound-state threshold. We therefore treat the finite-domain charge integral as a supplementary diagnostic, while assessing spectral consistency through the Friedel sum rule and Levinson's theorem and evaluating the convergence of the contact Hartree potential separately.

\subsubsection{Kato cusp condition}
\label{sec:cusp}

The cusp condition in the impurity problem follows from Kato's theorem for quantum many-body systems with Coulomb interactions \cite{1957Kato_cusp}. It relates the one-sided radial derivative of the total electron density $n(r)$ at the impurity to its contact value \cite{1957Kato_cusp,1982_Carlsson_cusp}:
\begin{align}
\left.\frac{dn(r)}{dr}\right|_{r=0^+}=-2Z\,n(0).
\label{eq:cusp}
\end{align}
Since $n(r)=n_0+\Delta n(r)$ and the uniform background $n_0$ has zero radial derivative, the exact induced-density relation is $\Delta n'(0^+)=-2Z[n_0+\Delta n(0)]$. Full-wave-vector linear-response theory (LRT), however, also predicts a nonzero one-sided contact slope, $n'_{\rm LRT}(0^+)=-2Zn_0$. This reproduces the cusp to first order in the impurity charge $Z$, but not the full nonlinear relation involving the enhanced contact density $n(0)$. The apparent zero slope obtained by differentiating the Fourier kernel in Eq.~\eqref{eq:dn_LRT} at $r=0$ results from an unjustified interchange of the $r\to0^+$ and $q\to\infty$ limits. The large-wave-vector derivation, an exactly solvable free-hydrogen illustration, and a numerical RPA consistency check are presented in Appendix~\ref{sec:lrt_cusp}.

The Estreicher--Meier (EM) parametrization, by contrast, satisfies $\Delta n'_{\rm EM}(0^+)=-2\Delta n_{\rm EM}(0)$ for $Z=1$, omitting the background-density term $-2n_0$ (Sec.~\ref{sec:EM_fit} and Appendix~\ref{sec:EM_cusp}). Thus LRT and the EM fit retain different terms of the exact cusp relation, without either reproducing that relation when evaluated at its own predicted total contact density. For the Coulomb-singular screened potentials used below, regular bound and continuum eigenfunctions instead inherit the full total-density cusp independently of how the potential parameters are fitted. Appendix~\ref{app:qm_vpa_cusp} explains the wavefunction origin and verifies the derivative numerically for the three FSR-only one-parameter potentials at $r_s=2$. This automatic property of the quantum density must be distinguished from the cusp explicitly imposed on the auxiliary Poisson-derived density in the Apagyi--Nagy potential construction.

It is important to emphasize that the exact cusp condition constrains the local shape of the density at the origin, but does not determine its contact amplitude. Specifically, it fixes the logarithmic slope $n'(0^+)/n(0)=-2Z$ once $n(0)$ is specified, but does not determine $n(0)$ itself. Consequently, satisfaction of the exact cusp condition does not by itself determine either the contact induced density $\Delta n(0)$ or the density-derived contact Hartree potential $V_{\rm H}(0)$, which depends on the full radial form of $\Delta n(r)$ rather than only on its short-range behavior. This distinction becomes particularly relevant below when the two-parameter Apagyi--Nagy model is constrained simultaneously by the Friedel sum rule and the exact cusp condition.

\subsubsection{Asymptotic induced-density tail: Friedel oscillations}
\label{sec:FO}

The asymptotic form of the induced electron density is constrained by Friedel oscillations. For a positive impurity in a degenerate three-dimensional electron gas, the sharp Fermi surface leads to a nonanalytic response at $q=2k_{\rm F}$, so that the screening cloud cannot decay as a purely monotonic function. Instead, at $T=0$, the large-distance tail must exhibit damped oscillations of the form
\begin{align*}
\dnr \sim \frac{A}{r^3}
\cos\!\left(2k_{\rm F} r+\phi\right),
\qquad r\to\infty.    
\end{align*}
Therefore, any physically consistent representation of $\dnr$ must satisfy not only the near-origin constraints and overall charge-neutrality condition, but also the correct Friedel-oscillatory asymptotic behavior.

\subsubsection{Estreicher--Meier fit to DFT-LDA data}
\label{sec:EM_fit}

Estreicher and Meier \cite{1983_Estreicher-Meier} calculated electronic densities
around light interstitial impurities in jellium using DFT-LDA. For an impurity of $Z=1$, they fitted the resulting induced densities to a compact analytic expression designed to reproduce charge screening and the Friedel-oscillatory tail over the metallic-density interval $\rs\in[2,6]$. Its near-origin core satisfies an approximate induced-density cusp, namely $\Delta n'(0) \approx -2Z\Delta n(0)$, rather than the exact cusp of the total density in Eq.~\eqref{eq:cusp}, as demonstrated in Appendix~\ref{sec:EM_cusp}. The parametrized $\dnr$ nevertheless gave good agreement with the self-consistent densities available at the time.

Estreicher--Meier proposed an ansatz for $\dnr$ that contains two exponentially-decaying core terms resembling hydrogen $1s$ electron density with a contact electron density that is expected to approach that of a free hydrogen ($1/\pi$) in the limit of highly dilute gas ($\rs \gg 1$) and a third Friedel-generating term. 
\begin{align}
\dnr
 = &\frac{1}{\pi}\,e^{-2r}
 + \!\left[\dn{0} - \frac{1}{\pi}\right] e^{-2r(1+r)} \nonumber \\
 &+ f(2 \kf r),
\label{eq:EM-dnr}
\end{align}
where the contact density is given by 
\begin{align}
\dn{0}=\frac{1}{\pi}+\exp\left(-0.72-1.28\ln \rs -0.385\ln^2\rs\right), 
\label{eq:EM-dn0}
\end{align}
and $f(2\kf r)$ produces the required Friedel oscillations based on Riccati--Bessel functions, and it is found with 40 parameters such that it also vanishes at the origin along with its slope, namely $f(0) = f'(0)=0$. The reader is referred to Appendix~\ref{sec:EM-FO} for the form of $f(2 \kf r)$ and its fit parameters. Since the Friedel term contributes no constant or linear term at the origin, the exponential core yields $\Delta n_{\rm EM}'(0)=-2\Delta n_{\rm EM}(0)$ rather than $-2[n_0+\Delta n_{\rm EM}(0)]$ for $Z=1$. Appendix~\ref{sec:EM_cusp} quantifies this local slope deficit and explains why it must not be interpreted as an error estimate for $V_{\rm H}(0)$. 

With the self-consistent $\dnr$ available from
Eqs.~\eqref{eq:EM-dnr} and \eqref{eq:EM-dn0}, we can calculate the DFT-LDA
contact Hartree potential directly from Eq.~\eqref{eq:VH0_r}. In particular,
the integration of $r\Delta n(r)$ for the first two exponentially decaying
terms in Eq.~\eqref{eq:EM-dnr} can be done analytically. We denote this
contribution the core part and the integral of the third term the
Friedel part. The final expression for the DFT-LDA contact Hartree
potential based on the Estreicher--Meier fit is therefore
\begin{align}
\vh^{\text{LDA}}(0)
&=
\vh^{\text{LDA-core}}(0)
+
\vh^{\text{LDA-Friedel}}(0).
\label{eq:VH0_LDA}
\end{align}
where we have defined 
\begin{align}
\vh^{\text{LDA-core}}(0) = 1 + \frac{\pi}{2}\left[\dn{0} - \frac{1}{\pi}\right] \times \nonumber \\
\left[2 - \sqrt{2 e \pi}\,\operatorname{erfc}\left(1/\sqrt{2}\right)\right]
\label{eq:VH0_LDA_core} 
\end{align}
\begin{align}
\vh^{\text{LDA-Friedel}}(0) = 4\pi \int_0^\infty r\, f(2\kf r)\, dr
\label{eq:VH0_LDA_friedel}    
\end{align}
Exact ground-state DFT would, in principle, determine the ground-state density for the specified external potential. In practice, the Kohn--Sham calculation requires an approximate exchange--correlation functional, so the resulting self-consistent density and the contact Hartree potential $\vh(0)$ acquire a functional dependence. In this work, the Estreicher--Meier parametrization is treated specifically as an LDA-based representation of the induced density, and therefore as a compact surrogate for DFT-LDA screening, not for DFT in general. The comparison with our direct PBE calculations in Section~\ref{sec:res_VH0} is used to estimate the sensitivity of $\vh(0)$ to a semilocal gradient correction.

\subsection{Nonlinear evaluation of the induced contact Hartree potential: Model Potentials and Variable Phase Approach (VPA)}
\label{sec:VPA_VH0}

The KS-DFT formulation of the static-screening problem may also be viewed as
the elastic scattering of electrons from the self-consistent central
effective potential defined by Eq.~\eqref{eq:KSEq}. The scattering
information is encoded in the partial-wave phase shifts, which in a direct
radial solution are normally extracted by matching the numerical
wavefunctions to their asymptotic form. The variable-phase approach (VPA)
provides an alternative formulation in which the phase shifts are obtained
directly from a first-order differential equation, without an explicit
asymptotic matching procedure.

The variable-phase approach to quantum scattering
\cite{1967Calogero_VariablePhaseApproach,1967Babikov_phase_shift} is based on
the Riccati reduction of the radial Schr\"odinger equation to a first-order
nonlinear equation for the phase function. The phase-function formulation
traces back to the work of Morse and Allis
\cite{1933MorseEffectExchangeScatteringa}. In the present work, however, the
VPA is used for more than the determination of asymptotic phase shifts.
Together with the corresponding variable-amplitude equation, it provides the
properly normalized scattering wavefunctions required to construct the full
quantal induced density and, from that density, the corresponding contact
Hartree potential.

For the static screening of a charge $Z$ in an electron gas, we consider
central screened model potentials
$V_{\rm model}(Z,r;\boldsymbol{\lambda})$ that retain the Coulombic
short-range behavior and decay sufficiently rapidly at large distance. The
parameter set $\boldsymbol{\lambda}$ denotes either a single screening
parameter or, for a two-parameter model, the corresponding parameter pair.
The models considered below include the Yukawa, hydrogenic, Hulth\'en, and
two-parameter Apagyi--Nagy forms.

For a given electron gas characterized by density parameter $\rs$ and Fermi
momentum $\kf$, the VPA yields the partial-wave phase shifts
$\delta_{\ell}(k;\boldsymbol{\lambda})$. The model parameters are then constrained
by the zero-temperature Friedel sum rule derived in
Sec.~\ref{sec:FSR}. For the one-parameter models, the FSR determines the
screening parameter through a root-finding procedure. For the
two-parameter Apagyi--Nagy model, the FSR is supplemented by imposing the exact cusp
condition of Sec.~\ref{sec:cusp} on an auxiliary total-density ansatz used to construct the potential via Poisson's equation. By contrast, the quantum densities generated from all of these Coulomb-singular model potentials obey the cusp through their regular short-range eigenfunctions, even when the cusp played no part in the parameter fit (Appendix~\ref{app:qm_vpa_cusp}).

It is important that these constraints not be interpreted as determining the
contact properties of the screening cloud uniquely. The FSR fixes a global
screening-charge constraint, whereas the cusp condition fixes a local
short-range derivative relation. Neither the FSR alone nor, as examined
below, the combined FSR-plus-cusp constraints uniquely determine the
quantal contact density or the density-derived contact Hartree potential.
For this reason, after determining the constrained model potential, we use
the phase and amplitude equations to reconstruct its scattering states,
include any discrete bound states explicitly, construct
$\Delta n^{\rm QM--VPA}(r)$, and finally evaluate
$V_{\rm H}^{\rm QM--VPA}(0)$ from the resulting density.

\subsubsection{Model potentials: Poisson-based induced electron density and
contact Hartree potential $V_{\rm H}^{\rm Poisson}(0)$}
\label{sec:model_potentials}

We first introduce the screened model potentials used below and establish
the induced densities associated with them through Poisson's equation.
This distinction is important because a given analytic model potential can
be associated with a density in two conceptually different ways. In the
present subsection, $\Delta n_{\nu}^{\rm Poisson}(r)$ denotes the density
obtained by interpreting the model potential electrostatically through
Poisson's equation. In Sec.~\ref{sec:vpa_density}, by contrast, the same
potential will be treated as a quantum scattering potential from which a
quantal density $\Delta n_{\nu}^{\rm QM--VPA}(r)$ is constructed explicitly
from its bound and continuum eigenstates.

For the one-parameter models considered here, the attractive
electron--impurity potentials are
\begin{align}
V_{\text{Yuk}}(Z,r;\alpha)
&=
-\frac{Z}{r}e^{-\alpha r},
\label{eq:yuk_pot}
\\
V_{\text{Hyd}}(Z,r;\alpha)
&=
-\frac{Z}{r}
\left(1+\frac{\alpha r}{2}\right)e^{-\alpha r},
\label{eq:hyd_pot}
\\
V_{\text{Hult}}(Z,r;\alpha)
&=
-\frac{Z\alpha}{e^{\alpha r}-1}.
\label{eq:hult_pot}
\end{align}
These expressions represent electron potential energies in Hartree atomic
units. They may all be written as a bare Coulomb potential multiplied by a
dimensionless screening function,
\begin{align}
V_\nu(Z,r;\alpha)
=
-\frac{Z}{r}\phi_\nu(\alpha r),
\qquad
\nu\in\{\text{Yuk},\text{Hyd},\text{Hult}\},
\label{eq:mod_pot_gen}
\end{align}
where
\begin{align}
\phi_{\text{Yuk}}(x)
&=
e^{-x},
\label{eq:yuk_Scr_fct}
\\
\phi_{\text{Hyd}}(x)
&=
\left(1+\frac{x}{2}\right)e^{-x},
\label{eq:hyd_scr_fct}
\\
\phi_{\text{Hult}}(x)
&=
\frac{x}{e^x-1}.
\label{eq:hult_scr_fct}
\end{align}
All three satisfy $\phi_\nu(0)=1$ and $\phi_\nu(\infty)=0$, so that the
bare Coulomb singularity $-Z/r$ is recovered at the origin and the
interaction is screened at large distance.

The electrostatic content of these model potentials can be made explicit
by writing
\begin{align}
V_\nu(Z,r)
=
-\frac{Z}{r}
+
V_{{\rm H},\nu}^{\rm Poisson}(r),
\label{eq:model_poisson_decomposition}
\end{align}
where $V_{{\rm H},\nu}^{\rm Poisson}(r)$ is the Hartree potential generated
by the corresponding Poisson-based induced electron density
$\Delta n_{\nu}^{\rm Poisson}(r)$. In atomic units,
%
\begin{align}
\nabla^2
\left[
-\frac{Z\phi_\nu(r)}{r}
\right]
&=
4\pi Z\delta^{(3)}(\mathbf r)
-
\frac{Z}{r}\phi_\nu''(r).
\label{eq:model_poisson_distributional}
\end{align}
For $r>0$, using $V_\nu=-Z\phi_\nu(r)/r$ and
$\nabla^2[\phi(r)/r]=\phi''(r)/r$, one obtains
\begin{align}
\Delta n_{\nu}^{\rm Poisson}(r)
=
\frac{Z}{4\pi r}
\frac{d^2\phi_\nu(r)}{dr^2}.
\label{eq:poisson_density_general}
\end{align}
Here and below, when $\phi_\nu$ is written as a function of $r$, its
screening parameters are understood implicitly.

The normalization of the Poisson-based density follows directly from
Eq.~\eqref{eq:poisson_density_general}. Provided that
$r\phi_\nu'(r)\rightarrow0$ at both integration boundaries,
\begin{align}
4\pi\int_0^\infty
r^2\Delta n_{\nu}^{\rm Poisson}(r)\,dr
&=
Z\int_0^\infty r\,\phi_\nu''(r)\,dr
\nonumber\\
&=
Z,
\label{eq:poisson_density_normalization}
\end{align}
where $\phi_\nu(0)=1$ and $\phi_\nu(\infty)=0$ have been used. Thus each
model defines a completely screening Poisson density.

Of particular relevance here is its first radial moment. From
Eq.~\eqref{eq:poisson_density_general},
\begin{align}
V_{{\rm H},\nu}^{\rm Poisson}(0)
&=
4\pi\int_0^\infty
r\,\Delta n_{\nu}^{\rm Poisson}(r)\,dr
\nonumber\\
&=
Z\int_0^\infty\phi_\nu''(r)\,dr
\nonumber\\
&=
-Z\phi_\nu'(0),
\label{eq:VH0_poisson_general}
\end{align}
where $\phi_\nu'(\infty)=0$ has been used. Equivalently, the same quantity
appears as the finite regular term in the short-range expansion
\begin{align}
V_\nu(Z,r)
=
-\frac{Z}{r}
+
V_{{\rm H},\nu}^{\rm Poisson}(0)
+
\mathcal{O}(r).
\label{eq:model_short_range_poisson}
\end{align}
Equation~\eqref{eq:VH0_poisson_general} therefore gives a precise
electrostatic interpretation to the regular short-range coefficient:
it is the contact Hartree potential of the Poisson-based density
associated with the analytic model potential. It should not, in general,
be identified with the contact Hartree potential obtained from the
quantal density generated by that same potential.

\medskip
\noindent\textbf{\underline{Yukawa potential}.}

For the Yukawa screening function
$\phi_{\rm Yuk}(r)=e^{-\alpha r}$,
Eq.~\eqref{eq:poisson_density_general} gives
\begin{align}
\Delta n_{\rm Yuk}^{\rm Poisson}(r)
=
\frac{Z\alpha^2}{4\pi r}e^{-\alpha r},
\label{eq:dn_poisson_yuk}
\end{align}
and hence
\begin{align}
V_{{\rm H},{\rm Yuk}}^{\rm Poisson}(0)
=
Z\alpha.
\label{eq:VH0_poisson_yuk}
\end{align}
Although $\Delta n_{\rm Yuk}^{\rm Poisson}(r)$ diverges as $1/r$ at the
origin, both its total screening charge and its first radial moment remain
finite.

For the Thomas–Fermi dielectric function, the corresponding momentum-space induced density is $\Delta n_{\rm TF}(q)=\frac{Zk_{\rm TF}^2}{q^2+k_{\rm TF}^2},$ which is identical to the Poisson-based Yukawa construction Eq.~\eqref{eq:dn_poisson_yuk} under $\alpha=k_{\rm TF}$. Since $\Delta n_{\rm TF}(q)\sim Zk_{\rm TF}^2/q^2$ for $q\to\infty$, the contact-density integral contains $q^2\Delta n_{\rm TF}(q)$ and diverges, whereas the contact-Hartree integral $\int dq\,\Delta n_{\rm TF}(q)$ converges. Equivalently,
\begin{align*}
\Delta n_{\rm TF}(r)
&=
\frac{Zk_{\rm TF}^2}{4\pi r}e^{-k_{\rm TF}r},
\qquad
V_{\rm H,TF}^{\rm Poisson}(0)
=
Zk_{\rm TF}.
\end{align*}
This singular Poisson density is distinct from the finite quantal density obtained when the same screened interaction is treated as a scattering potential in Sec.~\ref{sec:vpa_density}.

\medskip
\noindent\textbf{\underline{Hydrogenic potential}.}

For the hydrogenic screening function
\begin{align}
\phi_{\rm Hyd}(r)
=
\left(1+\frac{\alpha r}{2}\right)e^{-\alpha r},
\end{align}
one obtains
\begin{align}
\frac{d^2\phi_{\rm Hyd}}{dr^2}
=
\frac{\alpha^3r}{2}e^{-\alpha r},
\end{align}
and therefore
\begin{align}
\Delta n_{\rm Hyd}^{\rm Poisson}(r)
=
\frac{Z\alpha^3}{8\pi}e^{-\alpha r}.
\label{eq:dn_poisson_hyd}
\end{align}
Unlike the Yukawa case, this density remains finite at the origin,
\begin{align}
\Delta n_{\rm Hyd}^{\rm Poisson}(0)
=
\frac{Z\alpha^3}{8\pi},
\label{eq:dn0_poisson_hyd}
\end{align}
and its first radial moment gives
\begin{align}
V_{{\rm H},{\rm Hyd}}^{\rm Poisson}(0)
=
\frac{Z\alpha}{2}.
\label{eq:VH0_poisson_hyd}
\end{align}

\medskip
\noindent\textbf{\underline{Hulth\'en potential}.}

For the Hulth\'en form it is convenient to introduce $x=\alpha r$,
so that
\begin{align}
\phi_{\rm Hult}(x)
=
\frac{x}{e^x-1}.
\end{align}
Its second derivative gives
\begin{align}
\Delta n_{\rm Hult}^{\rm Poisson}(r)
=
\frac{Z\alpha^3}{4\pi}
\frac{
e^x\left[(x-2)e^x+x+2\right]
}{
x(e^x-1)^3
},
\qquad
x=\alpha r.
\label{eq:dn_poisson_hult}
\end{align}
The short-range expansion
$\phi_{\rm Hult}(x)=1-x/2+x^2/12+\mathcal{O}(x^4)$ gives
\begin{align}
\Delta n_{\rm Hult}^{\rm Poisson}(r)
\underset{r\rightarrow0}{\sim}
\frac{Z\alpha^2}{24\pi r},
\label{eq:dn_poisson_hult_short}
\end{align}
so that the Hulth\'en Poisson density is also singular at contact.
Its first radial moment is nevertheless finite and gives
\begin{align}
V_{{\rm H},{\rm Hult}}^{\rm Poisson}(0)
=
\frac{Z\alpha}{2}.
\label{eq:VH0_poisson_hult}
\end{align}
\begin{figure}[htbp!]
    \centering
    \includegraphics[width=\columnwidth]{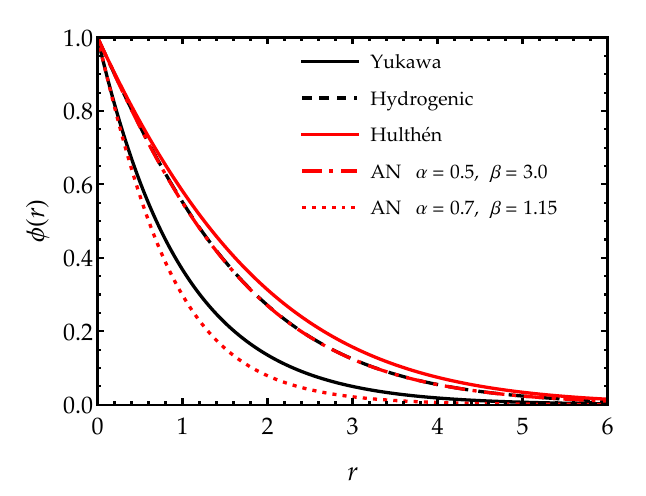}
    \caption{
Screening functions $\phi(r)$ for the Yukawa, hydrogenic, Hulth\'en, and
Apagyi--Nagy model potentials. For the one-parameter Yukawa, hydrogenic,
and Hulth\'en forms, the screening parameter is set to $\alpha=1\,$.
Two representative Apagyi--Nagy screening functions are shown for
$(\alpha,\beta)=(0.5,3.0)\,$ and
$(0.7,1.15)\,$ at $r_s=2$.
All models satisfy $\phi(0)=1$ and therefore retain the same Coulombic
near-origin behavior, $V(r)\sim -Z/r$, while $\phi(r)\rightarrow0$ at
large $r$. The Apagyi--Nagy form with $\alpha=0.5\,$ and
$\beta=3.0\,$ lies close to the hydrogenic screening function,
consistent with the hydrogenic limiting behavior of the two-parameter
model, whereas the $(0.7,1.15)\,$ case exhibits a more distinct
two-scale radial dependence.
}
\label{fig:mod_pot_phi}
\end{figure}

\medskip
\noindent\textbf{\underline{Apagyi--Nagy two-parameter potential}.}

We additionally consider the two-parameter model introduced by Apagyi and
Nagy for the static screening of a proton in an electron gas
\cite{1988ApagyiNagyTwoParameter}. In contrast to the preceding potentials,
this model was constructed by first postulating an induced-density ansatz
and subsequently obtaining the corresponding effective potential from
Poisson's equation. For $Z=1$, their induced density may be written as
\begin{align}
\Delta n_{\rm AN}^{\rm Poisson}(r)
&=
\mathcal{W}(\alpha,\beta;r_s)
\frac{\alpha^3}{\pi}e^{-2\alpha r}
\nonumber\\
&\quad+
\left[
1-\mathcal{W}(\alpha,\beta;r_s)
\right]
\frac{\beta^3}{\pi}e^{-2\beta r},
\label{eq:dn_poisson_AN}
\end{align}
where $\alpha\neq\beta$ and $\alpha,\beta>0$. Each exponential component is
individually normalized to one electron, so that the ansatz automatically
satisfies
\begin{align}
4\pi\int_0^\infty
r^2\Delta n_{\rm AN}^{\rm Poisson}(r)\,dr
=
1.
\label{eq:AN_normalization}
\end{align}

The mixing coefficient $\mathcal{W}$ is fixed by imposing the exact Kato
cusp condition on the total electron density
$n(r)=n_0+\Delta n_{\rm AN}^{\rm Poisson}(r)$. Following Apagyi and Nagy,
we define
\begin{align}
a
&=
\alpha^3(\alpha-1),
&
b
&=
\beta^3(\beta-1),
\end{align}
which gives
\begin{align}
\mathcal{W}(\alpha,\beta;r_s)
=
\frac{\pi n_0-b}{a-b}.
\label{eq:AN_weight}
\end{align}
Thus $\mathcal{W}$ is not an additional independent parameter: after
imposition of the cusp condition, the model is characterized by the two
range parameters $\alpha$ and $\beta$. The Friedel sum rule subsequently
constrains the allowed pairs $(\alpha,\beta)$ at each $r_s$. Importantly, the cusp-imposed density here is $n_0+\Delta n_{\rm AN}^{\rm Poisson}(r)$, not the independent QM--VPA density reconstructed from the resulting potential. The latter has its own exact Coulomb cusp but need not reproduce the Poisson-ansatz contact density or radial profile.

Solving Poisson's equation for
Eq.~\eqref{eq:dn_poisson_AN} gives the electron potential energy
\begin{align}
V_{\rm AN}(r;\alpha,\beta)
&=
-\mathcal{W}
e^{-2\alpha r}
\left(
\alpha+\frac{1}{r}
\right)
\nonumber\\
&\quad-
(1-\mathcal{W})
e^{-2\beta r}
\left(
\beta+\frac{1}{r}
\right).
\label{eq:AN_potential}
\end{align}
It can therefore be cast in the same bare-Coulomb-times-screening-function
form as the one-parameter potentials,
\begin{align}
V_{\rm AN}(r;\alpha,\beta)
=
-\frac{1}{r}
\phi_{\rm AN}(r;\alpha,\beta,r_s),
\label{eq:AN_potential_phi}
\end{align}
with
\begin{align}
\phi_{\rm AN}(r;\alpha,\beta,r_s)
&=
\mathcal{W}(\alpha,\beta;r_s)
(1+\alpha r)e^{-2\alpha r}
\nonumber\\
&\quad+
\left[
1-\mathcal{W}(\alpha,\beta;r_s)
\right]
(1+\beta r)e^{-2\beta r}.
\label{eq:AN_screening_function}
\end{align}
Since $\phi_{\rm AN}(0)=1$, the Apagyi--Nagy potential has the required
Coulombic short-range singularity.

The model is symmetric under interchange of its two range parameters.
Indeed, Eq.~\eqref{eq:AN_weight} gives
\begin{align}
\mathcal{W}(\beta,\alpha;r_s)
=
1-\mathcal{W}(\alpha,\beta;r_s),
\label{eq:AN_weight_symmetry}
\end{align}
and consequently
\begin{align}
\phi_{\rm AN}(r;\beta,\alpha,r_s)
&=
\phi_{\rm AN}(r;\alpha,\beta,r_s),
\nonumber\\
V_{\rm AN}(r;\beta,\alpha)
&=
V_{\rm AN}(r;\alpha,\beta).
\label{eq:AN_symmetry}
\end{align}
The two parameters can therefore be ordered, for example as
$\alpha<\beta$, without loss of generality, as done by Apagyi and Nagy.

The one-parameter hydrogenic potential is recovered as a limiting member of
the same family. For finite $\alpha$ and $\beta\rightarrow\infty$,
Eq.~\eqref{eq:AN_weight} gives
$\mathcal{W}\rightarrow1$, and hence
\begin{align}
\phi_{\rm AN}(r;\alpha,\beta,r_s)
\underset{\beta\rightarrow\infty}{\longrightarrow}
(1+\alpha r)e^{-2\alpha r}.
\label{eq:AN_beta_inf_limit}
\end{align}
Comparison with Eq.~\eqref{eq:hyd_scr_fct} shows that this is precisely the
hydrogenic screening function with
\begin{align}
\alpha_{\rm Hyd}=2\alpha.
\label{eq:AN_hyd_parameter_mapping}
\end{align}
By symmetry, for finite $\beta$ and $\alpha\rightarrow\infty$,
\begin{align}
\phi_{\rm AN}(r;\alpha,\beta,r_s)
\underset{\alpha\rightarrow\infty}{\longrightarrow}
(1+\beta r)e^{-2\beta r},
\end{align}
corresponding to the hydrogenic form with
$\alpha_{\rm Hyd}=2\beta$. Thus the hydrogenic model occurs at either
large-parameter boundary of the symmetric Apagyi--Nagy two-parameter
family.

The Poisson-based contact Hartree potential associated with the
Apagyi--Nagy ansatz follows either from
Eq.~\eqref{eq:VH0_poisson_general} or directly from its first radial moment,
\begin{align}
V_{{\rm H},{\rm AN}}^{\rm Poisson}(0)
&=
4\pi\int_0^\infty
r\,\Delta n_{\rm AN}^{\rm Poisson}(r)\,dr
\nonumber\\
&=
\mathcal{W}\alpha
+
(1-\mathcal{W})\beta.
\label{eq:VH0_poisson_AN}
\end{align}
The same result follows from $-\phi_{\rm AN}'(0)$, as required by
Eq.~\eqref{eq:VH0_poisson_general} for $Z=1$.

The Poisson-based densities and their first radial moments are summarized
in Table~\ref{tab:poisson_model_densities}.
\begin{table*}[tp]
\centering
\caption{Poisson-based induced densities associated with the screened model
potentials and their corresponding contact Hartree potentials. For the
Apagyi--Nagy model, $Z=1$ and
$\mathcal{W}=\mathcal{W}(\alpha,\beta;r_s)$.}
\label{tab:poisson_model_densities}
\begin{tabular}{lll}
\hline
Model
&
$\Delta n_{\nu}^{\rm Poisson}(r)$
&
$V_{{\rm H},\nu}^{\rm Poisson}(0)$
\\
\hline

Yukawa
&
$\displaystyle
\frac{Z\alpha^2}{4\pi r}e^{-\alpha r}$
&
$\displaystyle Z\alpha$
\\[2.0ex]

Hydrogenic
&
$\displaystyle
\frac{Z\alpha^3}{8\pi}e^{-\alpha r}$
&
$\displaystyle \frac{Z\alpha}{2}$
\\[2.0ex]

Hulth\'en
&
$\displaystyle
\frac{Z\alpha^3}{4\pi}
\frac{
e^x[(x-2)e^x+x+2]
}{
x(e^x-1)^3
},
\quad x=\alpha r$
&
$\displaystyle \frac{Z\alpha}{2}$
\\[3.0ex]

Apagyi--Nagy
&
$\displaystyle
\mathcal{W}\frac{\alpha^3}{\pi}e^{-2\alpha r}
+
(1-\mathcal{W})\frac{\beta^3}{\pi}e^{-2\beta r}$
&
$\displaystyle
\mathcal{W}\alpha+(1-\mathcal{W})\beta$
\\
\hline
\end{tabular}
\end{table*}

This Poisson construction provides a reference against which the quantum
calculation developed below can be compared. For every model potential
considered here, one may define the electrostatic sequence
\begin{align}
V_\nu(r)
\quad
\xleftrightarrow{\;\rm Poisson\;}
\quad
\Delta n_{\nu}^{\rm Poisson}(r)
\quad
\longrightarrow
\quad
V_{{\rm H},\nu}^{\rm Poisson}(0).
\label{eq:poisson_model_chain}
\end{align}
In the VPA treatment below, the same analytic potential is instead used as
a quantum scattering potential,
\begin{align}
V_\nu(r)
\quad
&\xrightarrow{\;\rm QM/VPA\;}
\quad
\Delta n_{\nu}^{\rm QM--VPA}(r)
\nonumber\\
\quad
&\longrightarrow
\quad
V_{{\rm H},\nu}^{\rm QM--VPA}(0).
\label{eq:quantal_model_chain}
\end{align}
The two constructions need not produce the same density. In particular,
satisfaction of the global Friedel constraint, or the simultaneous
satisfaction of the Friedel constraint and the cusp imposed on the auxiliary
Poisson density in the Apagyi--Nagy construction, does not establish equality between
$\Delta n_{\nu}^{\rm Poisson}(r)$ and
$\Delta n_{\nu}^{\rm QM--VPA}(r)$. Their comparison therefore provides a
direct test of how much information about the contact density and the
nonlocal contact Hartree potential is actually fixed by the imposed
screening constraints.

\subsubsection{Variable-phase equation}
\label{sec:vpa_phase}

The VPA account outlined in this section closely follows that presented in
Grandjouan and Deutsch \cite{1975GrandjouanPhaseShiftsStatic}. We start from
the Sturm--Liouville form of the radial Schr\"odinger equation given by
\begin{align}
\frac{d^2\psi_{\ell}}{dr^2}
+
\left[
\epsilon
-
U(r)
-
\frac{\ell(\ell+1)}{r^2}
\right]\psi_{\ell}(r)
&=0,
\label{eq:gd_schrodinger_sturm_liouville}
\end{align}
where the radial coordinate is measured in Bohr radii and
\begin{align}
\epsilon
&=
\frac{2\mu}{\hbar^2}E
=
k^2,
&
U(r)
&=
\frac{2\mu}{\hbar^2}V(r).
\label{eq:gd_energy_potential_definitions}
\end{align}
Here $\mu$ is the reduced mass of the scattering pair. For electron scattering
from a fixed positive center such as a proton in Hartree atomic units,
$\mu \approx 1$ is used.

The variable-phase construction requires that the potential be sufficiently
regular at the origin in the sense
\begin{align}
V(r)
&\sim
V_0 r^{-m},
\qquad
r\to0,
\qquad
m<2.
\label{eq:origin_regular_condition}
\end{align}
Physically, $m<2$ requires the short-range singularity of the scattering
potential to be weaker than the critical inverse-square singularity,
represented by the centrifugal term $\ell(\ell+1)/r^2$ for $l>0$. This condition
guarantees that the origin remains a regular scattering boundary: the reduced
radial solution starts with the same regular form as a free particle,
$\psi_l(r)\propto r^{l+1}$, so the variable phase may be initialized by
$\delta_{\ell}(0)=0$. The physical phase shift is then generated by integrating the
phase equation outward through the region where the potential acts. All
screened Coulomb model potentials considered in Sec.~\ref{sec:model_potentials}
satisfy Eq.~\eqref{eq:origin_regular_condition} with $m=1$, since
$V(r)\sim -Z/r$ as $r\to0$.

Following Grandjouan and Deutsch~\cite{1975GrandjouanPhaseShiftsStatic}, the
phase equation for the \emph{running phase} $\delta_{\ell}(k;r)$ for a fixed
partial wave $l$ and wave number $k$, accumulated from the origin up to the
radius $r$, is expressed as follows:
\begin{align}
\delta'_{\ell}(k;r)
={}&
-\frac{U(r)}{k}
\Bigl[
\cos\delta_{\ell}(k;r)\,\hat{j}_{\ell}(kr)
\nonumber\\
&-
\sin\delta_{\ell}(k;r)\,\hat{n}_{\ell}(kr)
\Bigr]^2,
\label{eq:phase_Eq}
\end{align}
where $\delta'_{\ell}(k;r)=d\delta_{\ell}(k;r)/dr$.
Grandjouan and Deutsch, following Calogero~\cite{1967Calogero_VariablePhaseApproach},
derived this equation using auxiliary scattering functions $S_{\ell}(k;r)$
and $C_{\ell}(k;r)$, defining the running tangent phase
$t_{\ell}(k;r)=S_{\ell}(k;r)/C_{\ell}(k;r)$, with the physical phase
shift obtained asymptotically from
$\tan\delta_{\ell}(k)=S_{\ell}(k;\infty)/C_{\ell}(k;\infty)$.
Writing $t_{\ell}(k;r)=\tan\delta_{\ell}(k;r)$ and using
$t'_{\ell}=(1+\tan^2\delta_{\ell})\,\delta'_{\ell}$
converts their tangent-phase equation into Eq.~\eqref{eq:phase_Eq}.
The solution of the phase equation
[Eq.~\eqref{eq:phase_Eq}] is monotonic if $U(r)$ does not change sign; it
decreases for repulsive potentials and increases for attractive ones. As
$U(r)\to0$, the phase function saturates. The physical phase shift is
the saturated value of the phase function,
\begin{align}
\delta_{\ell}(k)
&=
\lim_{r\rightarrow\infty}\delta_{\ell}(k;r).
\label{eq:phase_shift_saturation_definition}
\end{align}
It is this physical phase shift that enters the FSR in
Eq.~\eqref{eq:FSR}. The initial condition for the phase function at the
origin is
\begin{align}
\delta_{\ell}(k;0)
&=0.
\label{eq:phase_origin_condition}
\end{align}
This fixes the regular no-accumulated-phase boundary condition at the origin.

Numerically, Eq.~\eqref{eq:phase_Eq} is not integrated from exactly $r=0$ to
$r=\infty$. Instead, it is propagated on a finite radial interval
\begin{align*}
r_{\min}
\le r \le
r_{\max},
\end{align*}
where $r_{\min}$ is the inner starting radius and $r_{\max}$ is the outer
asymptotic integration radius. These radii are numerical endpoints, not physical cutoffs of
the screened potential. The inner radius $r_{\min}$ is chosen small enough
that the screened potential remains in its Coulombic short-distance regime,
$V(r)\simeq -Z/r$, while avoiding evaluation of the irregular
Riccati--Neumann function exactly at the origin. The running phase is therefore
initialized at $r_{\min}$ using the leading small-$r$ asymptotic form
\begin{align}
\delta_{\ell}(k;r_{\min})
\simeq
\frac{
2\mu Z\,k^{2\ell+1} r_{\min}^{2\ell+2}
}{
(2\ell+2)\left[(2\ell+1)!!\right]^2
}.
\label{eq:rmin}
\end{align}
The inner radius $r_{\min}$ is determined by inverting
Eq.~\eqref{eq:rmin} for a target initial phase magnitude
$|\delta_{\ell}(k;r_{\min})|=10^{-10}$~rad, subject to an additional
screening-length cap that ensures the starting point remains within the
Coulombic short-range regime. The actual initial phase is then evaluated
from Eq.~\eqref{eq:rmin} at the resulting $r_{\min}$.

To achieve the asymptotic limit $U(r)\to0$, the outer radial boundary must be
chosen only after the potential has become negligible compared with the free
radial kinetic term. This requirement is expressed as follows:
\begin{align}
|U(\rmax)|
&<
10^{-n}
\left|
k^2-\frac{\ell(\ell+1)}{\rmax^2}
\right|,
\nonumber\\
\rmax &\gg \frac{\sqrt{\ell(\ell+1)}}{k},
\label{eq:rmax}
\end{align}
where the production calculations use $n=9$. The second condition in
Eq.~\eqref{eq:rmax} is the dimensionally consistent form of the
turning-point condition, obtained from the root of
$k^2-\ell(\ell+1)/r^2$; in the printed form of Grandjouan and
Deutsch's Eq.~(30), it appears as
$\rmax\gg [\ell(\ell+1)]^{1/2}\,k$.
The radius $\rmax$ is not a
physical cutoff of the potential, but an outer asymptotic integration radius
for the phase equation. It is selected so that, for $r\ge\rmax$, the residual
effect of the screened interaction on the running phase is negligible. In this limit, the
running phase is effectively independent of the chosen endpoint, allowing the
asymptotic phase shift to be approximated by
\begin{align}
\delta_{\ell}(k)
&=
\lim_{r\to\infty}\delta_{\ell}(k;r)
\simeq
\delta_{\ell}(k;\rmax).
\label{eq:delta_outer_matching_radius}
\end{align}
For a model potential characterized by a parameter set
$\boldsymbol{\lambda}$, the phase function and hence the physical phase shift
inherit the corresponding parameter dependence through $U(r;\boldsymbol{\lambda})$.
The zero-temperature FSR may therefore be written as
\begin{align}
Z
&=
\lim_{L\to\infty}
s_L(\kf;\boldsymbol{\lambda}),
\nonumber\\
s_L(\kf;\boldsymbol{\lambda})
&=
\frac{2}{\pi}
\sum_{\ell=0}^{L}
(2\ell+1)\,
\delta_{\ell}(\kf;\boldsymbol{\lambda}).
\label{eq:static_friedel_sum_rule}
\end{align}
In practice, $L$ is finite and is chosen according to a truncation criterion
beyond which the remaining partial-wave contribution is negligible. For the
one-parameter Yukawa, hydrogenic, and Hulth\'en potentials,
$\boldsymbol{\lambda}=\alpha$, and the FSR reduces to the scalar nonlinear
root-finding problem
\begin{align}
F(Z,\rs;\alpha)
&\equiv
Z-
\frac{2}{\pi}
\sum_{\ell=0}^{L}
(2\ell+1)\,
\delta_{\ell}(\kf;\alpha)
=0.
\label{eq:friedel_residual_alpha}
\end{align}
For the two-parameter Apagyi--Nagy potential,
$\boldsymbol{\lambda}=(\alpha,\beta)$, with the mixing coefficient
$\mathcal W(\alpha,\beta;r_s)$ fixed by the cusp condition introduced in
Sec.~\ref{sec:model_potentials}. In that case, the FSR defines the allowed
locus
\begin{align}
F(Z,\rs;\alpha,\beta)
&\equiv
Z-
\frac{2}{\pi}
\sum_{\ell=0}^{L}
(2\ell+1)\,
\delta_{\ell}(\kf;\alpha,\beta)
=0,
\label{eq:friedel_residual_alpha_beta}
\end{align}
rather than uniquely determining both screening parameters by itself.

Thus, the same phase-shift constraint can be imposed consistently across the
different screened central potentials considered here, while retaining the
parameter structure specific to each model.

\subsubsection{Variable-amplitude equation}
\label{sec:vpa_amplitude}

The phase equation developed in Sec.~\ref{sec:vpa_phase} determines the
running phase $\delta_{\ell}(k;r)$ and, asymptotically, the physical scattering
phase shift. For the reconstruction of the corresponding radial solution,
however, the phase must be supplemented by a running amplitude. This
amplitude construction was developed in detail by Calogero
\cite{1967Calogero_VariablePhaseApproach}. Here we formulate it directly in
the notation of Grandjouan and Deutsch
\cite{1975GrandjouanPhaseShiftsStatic}, so that it is fully consistent with
the phase equation used above.

Following Calogero, Grandjouan and Deutsch represent the regular reduced radial solution in terms
of two running auxiliary scattering functions 
\begin{align}
    S_{\ell}(k; r) &= -k^{-1}\int_{0}^{r}dr' U(r')\hat{j}_{\ell}(kr')\psi_{\ell}(r'), \\
    C_{\ell}(k; r) &=1 -k^{-1}\int_{0}^{r}dr' U(r')\hat{n}_{\ell}(kr')\psi_{\ell}(r'),
    \label{eq:S_l_C_l}
\end{align}
such that
\begin{align}
\psi_{\ell}(k;r)
&=
C_{\ell}(k;r)\,\hat{j}_{\ell}(kr)
-
S_{\ell}(k;r)\,\hat{n}_{\ell}(kr),
\label{eq:vpa_CS_wavefunction}
\end{align}
with the origin conditions
\begin{align}
C_{\ell}(k;0)&=1,
&
S_{\ell}(k;0)&=0,
\label{eq:vpa_CS_origin}
\end{align}
where the same convention
\begin{align}
U(r)
&=
\frac{2\mu}{\hbar^2}V(r)
\end{align}
is used as in Sec.~\ref{sec:vpa_phase}.
Eqs.~\eqref{eq:S_l_C_l} immediately yield the corresponding first-order equations 
\begin{align}
\frac{\partial S_{\ell}(k;r)}{\partial r}
&=
-\frac{U(r)}{k}
\hat{j}_{\ell}(kr)\,
\psi_{\ell}(k;r),
\label{eq:vpa_S_equation}
\\
\frac{\partial C_{\ell}(k;r)}{\partial r}
&=
-\frac{U(r)}{k}
\hat{n}_{\ell}(kr)\,
\psi_{\ell}(k;r),
\label{eq:vpa_C_equation}
\end{align}

To separate the phase and amplitude information contained in
$C_l$ and $S_l$, we write
\begin{align}
C_{\ell}(k;r)
&=
e^{\gamma_{\ell}(k;r)}
\cos\delta_{\ell}(k;r),
\nonumber\\
S_{\ell}(k;r)
&=
e^{\gamma_{\ell}(k;r)}
\sin\delta_{\ell}(k;r),
\label{eq:vpa_CS_gamma_delta}
\end{align}
where $\gamma_{\ell}(k;r)$ is the {running logarithmic amplitude}. The
origin conditions in Eq.~\eqref{eq:vpa_CS_origin} therefore imply
\begin{align}
\delta_{\ell}(k;0)
&=0,
&
\gamma_{\ell}(k;0)
&=0.
\label{eq:vpa_gamma_origin}
\end{align}
Calogero's running amplitude function, denoted by $\alpha_l(r)$, corresponds
here to $e^{\gamma_{\ell}(k;r)}$. This notation is convenient because the symbol
$\alpha$ is already reserved for the screening parameter of the model
potentials.

For compactness, define the two phase-rotated free-wave combinations
\begin{align}
Q_{\ell}(k,r)
&=
\cos\delta_{\ell}(k;r)\,\hat{j}_{\ell}(kr)
-
\sin\delta_{\ell}(k;r)\,\hat{n}_{\ell}(kr),
\label{eq:vpa_Q_definition}
\\
P_{\ell}(k,r)
&=
\sin\delta_{\ell}(k;r)\,\hat{j}_{\ell}(kr)
+
\cos\delta_{\ell}(k;r)\,\hat{n}_{\ell}(kr).
\label{eq:vpa_P_definition}
\end{align}
Equation~\eqref{eq:vpa_CS_wavefunction} can then be written as
\begin{align}
\psi_{\ell}(k;r)
&=
e^{\gamma_{\ell}(k;r)}
Q_{\ell}(k,r).
\label{eq:vpa_wavefunction_gamma}
\end{align}
The running phase follows from
\begin{align}
\frac{\partial\delta_{\ell}(k;r)}{\partial r}
&=
\frac{
C_{\ell}\,\partial_r S_{\ell}
-
S_{\ell}\,\partial_r C_{\ell}
}{
C_{\ell}^2+S_{\ell}^2
},
\label{eq:vpa_delta_from_CS}
\end{align}
which, upon use of Eqs.~\eqref{eq:vpa_S_equation} and
\eqref{eq:vpa_C_equation}, reproduces the phase equation
\eqref{eq:phase_Eq},
\begin{align}
\frac{\partial\delta_{\ell}(k;r)}{\partial r}
&=
-\frac{U(r)}{k}
Q_{\ell}^2(k,r).
\label{eq:vpa_phase_Q}
\end{align}
The independent combination of the same two auxiliary equations determines
the running amplitude. Since
\begin{align}
e^{2\gamma_{\ell}(k;r)}
&=
C_{\ell}^2(k;r)+S_{\ell}^2(k;r),
\end{align}
differentiation gives
\begin{align}
\frac{\partial\gamma_{\ell}(k;r)}{\partial r}
&=
\frac{
C_{\ell}\,\partial_r C_{\ell}
+
S_{\ell}\,\partial_r S_{\ell}
}{
C_{\ell}^2+S_{\ell}^2
}.
\label{eq:vpa_gamma_from_CS}
\end{align}
Substituting Eqs.~\eqref{eq:vpa_S_equation} and
\eqref{eq:vpa_C_equation} into Eq.~\eqref{eq:vpa_gamma_from_CS}, and using
Eqs.~\eqref{eq:vpa_Q_definition}--\eqref{eq:vpa_P_definition}, yields
\begin{align}
\frac{\partial\gamma_{\ell}(k;r)}{\partial r}
&=
-\frac{U(r)}{k}
Q_{\ell}(k,r)P_{\ell}(k,r).
\label{eq:vpa_amplitude_equation}
\end{align}
Equation~\eqref{eq:vpa_amplitude_equation} is the variable-amplitude equation
used here. Together with Eq.~\eqref{eq:vpa_phase_Q}, it forms a coupled
first-order phase--amplitude representation of the same radial
Schr\"odinger problem.

For the $s$ wave, the Riccati--Bessel functions satisfy
\begin{align}
\hat{j}_0(kr)&=\sin(kr),
&
\hat{n}_0(kr)&=-\cos(kr),
\end{align}
so that
\begin{align}
Q_0(k,r)
&=
\sin\!\left[kr+\delta_0(k;r)\right],
\\
P_0(k,r)
&=
-\cos\!\left[kr+\delta_0(k;r)\right].
\end{align}
The phase and amplitude equations therefore reduce to
\begin{align}
\frac{\partial\delta_0(k;r)}{\partial r}
&=
-\frac{U(r)}{k}
\sin^2\!\left[kr+\delta_0(k;r)\right],
\label{eq:vpa_swave_phase}
\\
\frac{\partial\gamma_0(k;r)}{\partial r}
&=
\frac{U(r)}{2k}
\sin\!\left\{
2\left[kr+\delta_0(k;r)\right]
\right\}.
\label{eq:vpa_swave_amplitude}
\end{align}
As the screened interaction vanishes at large distance,
both running functions approach constant asymptotic values. In particular,
\begin{align}
\gamma_{\ell}(k)
&\equiv
\lim_{r\rightarrow\infty}
\gamma_{\ell}(k;r),
\label{eq:vpa_gamma_asymptotic}
\end{align}
so that the asymptotic amplitude of the origin-normalized regular solution is
\begin{align}
\mathcal A_{\ell}(k)
&=
e^{\gamma_{\ell}(k)}.
\label{eq:vpa_asymptotic_amplitude}
\end{align}
Thus the phase equation determines the asymptotic phase of the regular
solution, while the companion amplitude equation determines its asymptotic
normalization. These two quantities will be used together in the following
subsection to construct the quantal scattering states associated with the
model potential.

\subsubsection{Quantal induced electron density from VPA scattering and bound states}
\label{sec:vpa_density}

The phase--amplitude formulation developed in
Secs.~\ref{sec:vpa_phase} and \ref{sec:vpa_amplitude} provides the information
required to reconstruct the continuum scattering states generated by a given
model potential. At zero temperature, the induced electron density is obtained
from the occupied bound spectrum together with the occupied continuum up to
the Fermi wave number $k_{\rm F}$.

We write the total quantal VPA density response as
\begin{align}
\Delta n^{\rm QM--VPA}(r)
&=
\Delta n_{\rm b}^{\rm QM--VPA}(r)
+
\Delta n_{\rm sc}^{\rm QM--VPA}(r),
\label{eq:vpa_density_decomposition}
\end{align}
where the bound-state and scattering-state contributions are treated
separately.

For a normalized bound reduced radial function $u_{\nu l}(r)$,
\begin{align}
\int_0^\infty
|u_{\nu \ell}(r)|^2\,dr
&=1,
\label{eq:vpa_bound_normalization}
\end{align}
the corresponding orbital is
\begin{align}
\Psi_{\nu \ell m}(\mathbf r)
&=
\frac{u_{\nu \ell}(r)}{r}
Y_{\ell m}(\hat{\mathbf r}).
\end{align}
At zero temperature, every ordinary negative-energy bound level is fully
occupied. After summation over the magnetic quantum number and the two spin
states, the bound contribution is therefore
\begin{align}
\Delta n_{\rm b}^{\rm QM--VPA}(r)
&=
\frac{1}{2\pi}
\sum_{\nu,\ell}
(2\ell+1)
\left|
\frac{u_{\nu \ell}(r)}{r}
\right|^2.
\label{eq:vpa_bound_density}
\end{align}
The homogeneous electron gas has no bound-state contribution.

For the continuum, the origin-normalized regular solution obtained from the
VPA is, from Eq.~\eqref{eq:vpa_wavefunction_gamma},
\begin{align}
\psi_{\ell}(k;r)
&=
e^{\gamma_{\ell}(k;r)}
Q_{\ell}(k,r),
\end{align}
whereas its asymptotic amplitude is
$\mathcal A_{\ell}(k)=e^{\gamma_{\ell}(k)}$, with
$\gamma_{\ell}(k)=\gamma_{\ell}(k;\infty)$. The continuum radial function entering the
density must instead be normalized to unit sine amplitude at infinity.
Accordingly,
\begin{align}
\varphi_{k\ell}^{\rm QM--VPA}(r)
&=
\frac{
e^{\gamma_{\ell}(k;r)-\gamma_{\ell}(k)}
}{
r
}
Q_l(k,r)
\nonumber\\
&=
\frac{
e^{\gamma_{\ell}(k;r)-\gamma_{\ell}(k)}
}{
r
}
\Bigl[
\cos\delta_{\ell}(k;r)\,\hat{j}_{\ell}(kr)
\nonumber\\
&\qquad-
\sin\delta_{\ell}(k;r)\,\hat{n}_{\ell}(kr)
\Bigr].
\label{eq:vpa_normalized_scattering_wavefunction}
\end{align}
It satisfies the asymptotic normalization
\begin{align}
\varphi_{k \ell}^{\rm QM--VPA}(r)
&\underset{r\to\infty}{\longrightarrow}
\frac{1}{r}
\sin\left(
kr-\frac{\ell \pi}{2}+\delta_{\ell}(k)
\right).
\label{eq:vpa_scattering_asymptotic_normalization}
\end{align}
Thus the phase alone determines the relative sine--cosine mixture, whereas
the factor
$e^{\gamma_{\ell}(k;r)-\gamma_{\ell}(k)}$
supplies the radial amplitude normalization required for a local-density
calculation.

For the free electron gas,
\begin{align}
\varphi_{k \ell}^{(0)}(r)
&=
k j_{\ell}(kr),
\label{eq:vpa_free_radial_function}
\end{align}
and the identity
\begin{align}
\sum_{\ell=0}^{\infty}
(2\ell+1)j_{\ell}^2(kr)
&=1
\label{eq:spherical_bessel_completeness}
\end{align}
recovers the homogeneous density
\begin{align}
n_0
&=
\frac{1}{\pi^2}
\int_0^{k_{\rm F}}k^2\,dk
=
\frac{k_{\rm F}^3}{3\pi^2}.
\label{eq:vpa_free_density}
\end{align}
It is therefore convenient to define the exact partial-wave contribution to
the continuum density difference as
\begin{align}
a_{\ell}^{\rm QM}(k;r)
&\equiv
\left|
\varphi_{k \ell}^{\rm QM--VPA}(r)
\right|^2
-
k^2j_{\ell}^2(kr),
\label{eq:vpa_channel_density_kernel}
\end{align}
and the complete continuum kernel
\begin{align}
\Phi(k;r)
&\equiv
\frac{2}{\pi}
\sum_{\ell=0}^{\infty}
(2\ell+1)
a_{\ell}^{\rm QM}(k;r).
\label{eq:vpa_Phi_definition}
\end{align}
The zero-temperature scattering contribution then takes the compact form
\begin{align}
\Delta n_{\rm sc}^{\rm QM--VPA}(r)
&=
\frac{1}{2\pi}
\int_0^{k_{\rm F}}
\Phi(k;r)\,dk.
\label{eq:vpa_scattering_density_Phi}
\end{align}
Combining Eqs.~\eqref{eq:vpa_bound_density} and
\eqref{eq:vpa_scattering_density_Phi} gives
\begin{align}
\Delta n^{\rm QM--VPA}(r)
&=
\frac{1}{2\pi}
\sum_{\nu,\ell}
(2\ell+1)
\left|
\frac{u_{\nu \ell}(r)}{r}
\right|^2
\nonumber\\
&\quad+
\frac{1}{2\pi}
\int_0^{k_{\rm F}}
\Phi(k;r)\,dk.
\label{eq:vpa_total_density_zeroT}
\end{align}
Equation~\eqref{eq:vpa_total_density_zeroT} is the quantal density response
generated by the chosen model potential. The same potential therefore enters
both the bound spectrum and the VPA continuum states.

At the origin, regular bound and continuum radial functions behave as
$r^l$, so all $l>0$ contributions vanish and only the $s$ wave survives.
For the continuum,
\begin{align}
Q_0(k,r)
&=
\sin\left[kr+\delta_0(k;r)\right],
\end{align}
and since $\delta_0(k;0)=0$,
\begin{align}
\lim_{r\to0}
\varphi_{k0}^{\rm QM--VPA}(r)
&=
k\,e^{-\gamma_0(k)}.
\label{eq:vpa_contact_scattering_wavefunction}
\end{align}
Consequently,
\begin{align}
\Phi(k;0)
&=
\frac{2}{\pi}
k^2
\left[
e^{-2\gamma_0(k)}-1
\right],
\label{eq:vpa_contact_Phi}
\end{align}
and the scattering contribution to the contact density becomes
\begin{align}
\Delta n_{\rm sc}^{\rm QM--VPA}(0)
&=
\frac{1}{\pi^2}
\int_0^{k_{\rm F}}
k^2
\left[
e^{-2\gamma_0(k)}-1
\right]dk.
\label{eq:vpa_contact_scattering_density}
\end{align}
Likewise, only $s$-wave bound states contribute at the origin,
\begin{align}
\Delta n_{\rm b}^{\rm QM--VPA}(0)
&=
\frac{1}{2\pi}
\sum_{\nu}
\left|
\lim_{r\to0}
\frac{u_{\nu 0}(r)}{r}
\right|^2
\nonumber\\
&=
\frac{1}{2\pi}
\sum_{\nu}
\left|
u_{\nu0}'(0)
\right|^2.
\label{eq:vpa_contact_bound_density}
\end{align}
The complete quantal contact density is therefore
\begin{align}
\Delta n^{\rm QM--VPA}(0)
&=
\frac{1}{2\pi}
\sum_{\nu}
\left|
u_{\nu0}'(0)
\right|^2
\nonumber\\
&\quad+
\frac{1}{\pi^2}
\int_0^{k_{\rm F}}
k^2
\left[
e^{-2\gamma_0(k)}-1
\right]dk.
\label{eq:vpa_contact_density}
\end{align}
Thus, although only the $s$ wave contributes exactly at contact, the general
finite-radius density in Eq.~\eqref{eq:vpa_total_density_zeroT} contains the
complete occupied bound and continuum spectra. In the following subsection,
this density is used to evaluate the corresponding density-derived contact
Hartree potential.

\subsubsection{Density-derived contact Hartree potential
$V_{\rm H}^{\rm QM--VPA}(0)$}
\label{sec:vpa_VH0}

Once the quantal induced density
$\Delta n^{\rm QM--VPA}(r)$ has been constructed from the bound and
continuum states of the model potential, the corresponding contact Hartree
potential is obtained directly from the spherical Hartree integral,
\begin{align}
V_{\rm H}^{\rm QM--VPA}(0)
&=
4\pi
\int_0^\infty
r\,
\Delta n^{\rm QM--VPA}(r)\,dr.
\label{eq:vpa_VH0_definition}
\end{align}
Using the decomposition in
Eq.~\eqref{eq:vpa_density_decomposition}, this may be written as
\begin{align}
V_{\rm H}^{\rm QM--VPA}(0)
&=
V_{{\rm H,b}}^{\rm QM--VPA}(0)
+
V_{{\rm H,sc}}^{\rm QM--VPA}(0),
\label{eq:vpa_VH0_decomposition}
\end{align}
where
\begin{align}
V_{{\rm H,b}}^{\rm QM--VPA}(0)
&=
4\pi
\int_0^\infty
r\,
\Delta n_{\rm b}^{\rm QM--VPA}(r)\,dr,
\\
V_{{\rm H,sc}}^{\rm QM--VPA}(0)
&=
4\pi
\int_0^\infty
r\,
\Delta n_{\rm sc}^{\rm QM--VPA}(r)\,dr.
\end{align}
This quantity is determined by the full radial density response generated
by the model potential and is therefore distinct from the Poisson-based
contact quantity $V_{\rm H}^{\rm Poisson}(0)$ introduced in
Sec.~\ref{sec:model_potentials}. No identification between the regular
short-range coefficient of the analytic model potential and
$V_{\rm H}^{\rm QM--VPA}(0)$ is assumed.

\section{Results}
\label{sec:results}

The results are organized around three successive questions. First, we establish the extent to which linear-response theory (LRT) reproduces the screening cloud generated by a proton in the uniform electron gas, considering in turn the spatially resolved induced density, the contact electron density, and the contact Hartree potential. The nonlinear reference calculations in this comparison are obtained within self-consistent DFT using specified approximate exchange--correlation functionals; they are therefore referred to below as DFT-LDA or DFT-PBE reference calculations rather than as exact many-body benchmarks. Second, we examine Friedel-constrained screened model potentials within the variable-phase approach (VPA), including the convergence of the Friedel sum and the determination of their screening parameters. Finally, we compare the density-derived contact quantities generated by these model potentials with the nonlinear DFT references in order to determine which aspects of the short-range screening cloud are, and are not, fixed by the Friedel constraint.

\subsection{Linear-response versus nonlinear DFT screening}
\label{sec:LRT_vs_DFT}

We begin with a proton, $Z=1$, embedded in the zero-temperature uniform electron gas. Unless otherwise stated, the local-field-corrected LRT results shown in the main text use the Kaplan--Kukkonen (KK) density local-field factor. A direct comparison with the Corradini--Del Sole--Onida--Palummo (CDOP) parametrization is given in Appendix~\ref{sec:LFC_Comparison}. The two parametrizations give essentially indistinguishable results for the contact observables considered here: at $\rs=2.07$, their RPA+LFC values of $n(0)/n_0$ differ by only approximately $6.5\times10^{-3}$, while the maximum difference in $V_{\rm H}(0)$ over their common range $2\leq \rs\leq6$ is approximately $1.6\times10^{-3}$~a.u., below $0.2\%$ in relative magnitude. We therefore use KK as the representative local-field correction in the figures below.
\begin{figure}[htbp!]
    \centering
    \includegraphics[width=\columnwidth]{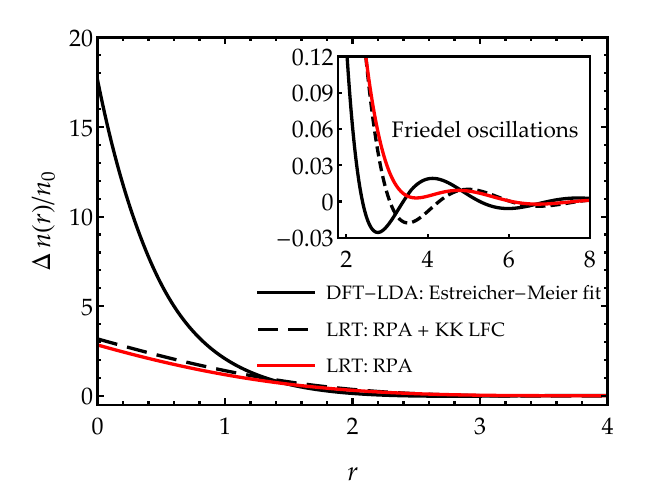}
    \caption{Normalized induced electron density $\Delta n(r)/n_0$ around a statically screened proton at $\rs=2.07$. The Estreicher--Meier parametrization of the self-consistent DFT-LDA density is compared with RPA and RPA+KK linear response. Both LRT treatments substantially underestimate the nonlinear charge accumulation in the near-proton region; the KK local-field correction increases the RPA response but does not remove this short-range discrepancy. The inset resolves the much smaller Friedel-oscillatory tail at larger distances.}
\label{fig:dn0_vs_r_DFT_vs_LRT}
\end{figure}
%
\subsubsection{Breakdown of linear response for the induced electron density}
\label{sec:LRT_DFT_density_profile}
Figure~\ref{fig:dn0_vs_r_DFT_vs_LRT} compares the normalized induced density $\Delta n(r)/n_0$ at $\rs=2.07$, representative of the electron density in Al. The nonlinear curve is the Estreicher--Meier parametrization of the self-consistent DFT-LDA induced density, Eq.~\eqref{eq:EM-dnr}, while the LRT curves are obtained from Eq.~\eqref{eq:dn_LRT} using the RPA dielectric function, Eq.~\eqref{eq:eps_RPA}, and the test-charge dielectric function with the KK local-field correction, Eq.~\eqref{eq:eps_LFC}.

The principal discrepancy occurs in the immediate vicinity of the proton. Both RPA and RPA+KK substantially underestimate the nonlinear accumulation of electronic charge in the near field. The KK correction increases the induced density relative to RPA, but the change is small compared with the separation between either LRT curve and the DFT-LDA-based result. Thus, incorporating exchange--correlation effects through a static local-field factor does not by itself remove the failure of the weak-perturbation approximation in the region where the bare Coulomb field is strongest.

At larger distances, the absolute induced density becomes much smaller and the screening cloud develops the expected Friedel-oscillatory structure. The inset in Fig.~\ref{fig:dn0_vs_r_DFT_vs_LRT} resolves this tail on an expanded scale. All three descriptions exhibit oscillatory screening, whereas their largest quantitative disagreement is concentrated in the nonlinear near-field response. This distinction between short-range charge accumulation and the long-range oscillatory tail will become important below when considering $V_{\rm H}(0)$, which is a radial moment of the entire induced density rather than a strictly local quantity.
\begin{figure}[htbp!]
    \centering
    \includegraphics[width=\columnwidth]{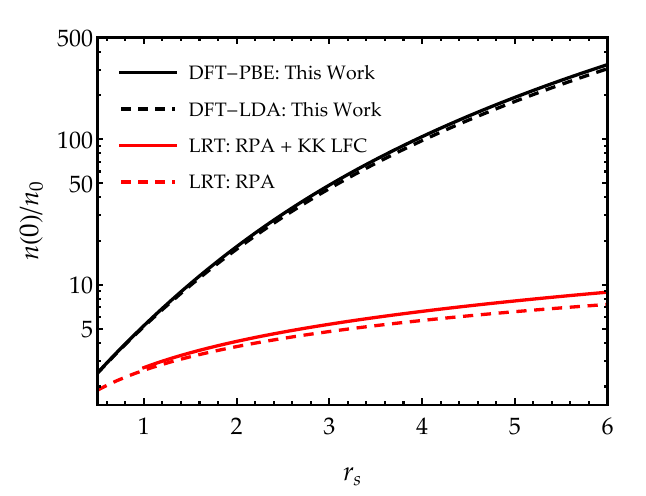}
    \caption{Normalized electron density at the proton position, $n(0)/n_0$, as a function of $\rs$. Self-consistent DFT-LDA and DFT-PBE results are shown over $0.5\leq\rs\leq6$ and compared with RPA over the same range and RPA+KK over $1\leq\rs\leq6$. The logarithmic ordinate emphasizes both the large nonlinear enhancement of the contact density throughout the metallic-density regime and the progressive narrowing of the DFT--LRT separation toward the high-density end of the calculated interval. The difference between the two semilocal DFT functionals is comparatively small on the scale of the DFT--LRT separation.}
\label{fig:n0_vs_rs_DFT_vs_LRT}
\end{figure}
%
\subsubsection{Contact electron density from LRT and DFT}
\label{sec:LRT_DFT_contact_density}
The near-field discrepancy in Fig.~\ref{fig:dn0_vs_r_DFT_vs_LRT} can be quantified directly through the contact density. Figure~\ref{fig:n0_vs_rs_DFT_vs_LRT} shows
$n(0)/n_0=1+\Delta n(0)/n_0$ as a function of $\rs$. The RPA values are obtained from Eq.~\eqref{eq:dn0_LRT} over $0.5\leq\rs\leq6$, while the KK-corrected results are shown over $1\leq\rs\leq6$. The nonlinear calculations comprise our self-consistent DFT-LDA and DFT-PBE results on the common $0.5\leq\rs\leq6$ grid.

The separation between linear and nonlinear screening is large throughout the metallic-density regime. For example, at $\rs=2$, the present DFT-LDA and DFT-PBE calculations give $n(0)/n_0\simeq17.54$ and $18.27$, respectively, whereas RPA and RPA+KK give approximately $3.76$ and $4.08$. The local-field correction therefore modifies the LRT contact density quantitatively but leaves the much larger nonlinear enhancement unresolved.

The extension to $\rs=0.5$ also reveals the expected trend toward closer agreement at increasing electron density. At $\rs=0.5$, the DFT-LDA and DFT-PBE values have decreased to approximately $2.470$ and $2.478$, while RPA gives approximately $1.887$. Thus the DFT--LRT separation narrows strongly toward the high-density end of the calculated interval, although a finite difference remains at $\rs=0.5$. We therefore interpret Fig.~\ref{fig:n0_vs_rs_DFT_vs_LRT} as demonstrating a trend toward the weak-coupling regime, not as a numerical realization of the strict $\rs\rightarrow0$ limit.

Within the two semilocal functionals considered here, PBE yields a systematically larger contact density than LDA, but their separation remains much smaller than the DFT--LRT separation over most of the displayed interval. This statement is deliberately restricted to the LDA/PBE comparison and should not be interpreted as a quantitative estimate of the full exchange--correlation or many-body uncertainty. The direction of the LDA--PBE difference is qualitatively consistent with the earlier first-gradient-correction study of Jena and Singwi~\cite{1978JenaElectronicStructureHydrogen}, which found an enhancement of the near-proton electron density relative to LDA. Their gradient correction predates PBE, however, and is used here only as qualitative historical context rather than as a quantitative PBE comparison.

Independent nonlinear calculations provide further perspective on the contact density. Table~I of Gondzik and Stachowiak~\cite{1985GondzikScreeningPositiveParticles} compares the total density $n(0)$ obtained from their HNC--HF variational many-body calculation, the collective-boson treatment of Arponen and Pajanne~\cite{1979ArponenElectronLiquidCollective}, and the self-consistent DFT-LDA calculations of Almbladh \emph{et al.}~\cite{1976AlmbladhScreeningProtonElectron}. At $\rs=2$, the three reported values are $0.461$, $0.594$, and $0.522~a_0^{-3}$, respectively; at $\rs=6$ they are $0.319$, $0.343$, and $0.335~a_0^{-3}$. The historical DFT-LDA values therefore lie between the two many-body estimates at both densities. Our present LDA values are correspondingly $n(0)\simeq0.523$ and $0.337~a_0^{-3}$ at $\rs=2$ and $6$, respectively, providing a useful consistency check on the present self-consistent calculation without elevating DFT-LDA to an exact benchmark.

The quantum Monte Carlo study of Sugiyama, Terray, and Alder~\cite{1988SugiyamaQuantumMonteCarlo} provides a further, although less quantitative, comparison. Their published electron--proton pair distribution functions were not tabulated with sufficient resolution to extract a reliable $n(0)$, and their calculations employed finite 54-electron cells. Nevertheless, the reported $g_{\rm e-p}(r)$ exhibits both a strong electron pile-up around the proton and Friedel-like oscillations about the uniform background. These results support the qualitative conclusion from Fig.~\ref{fig:n0_vs_rs_DFT_vs_LRT} that the near-proton response is strongly nonlinear.
\begin{figure}[htbp!]
    \centering
    \includegraphics[width=\columnwidth]{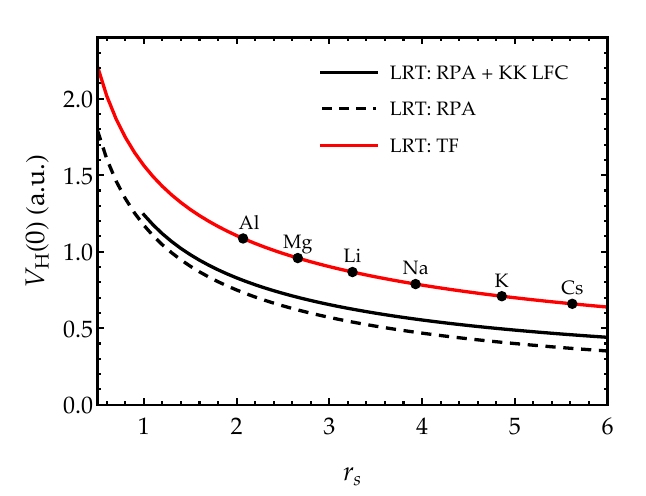}
    \caption{Linear-response contact Hartree potential $V_{\rm H}(0)$ as a function of $\rs$. Thomas--Fermi (TF) and RPA results are shown over $0.5\leq\rs\leq6$, while the RPA+KK result is shown over $1\leq\rs\leq6$. Representative simple metals are indicated on the TF curve at their corresponding density parameters. The KK local-field correction increases the RPA result, whereas the Thomas--Fermi value remains larger because of the approximate short-wavelength behavior of the TF dielectric response.}
\label{fig:VH0_linear_models}
\end{figure}
%
\subsubsection{Contact Hartree potential from LRT and DFT}
\label{sec:res_VH0}
The contact density considered above is a strictly local observable. By contrast, the contact Hartree potential $V_{\rm H}(0)$ is determined by the radial moment of the full induced density through Eq.~\eqref{eq:VH0_r}. Consequently, two screening descriptions can differ strongly in $\Delta n(0)$ without exhibiting the same relative difference in $V_{\rm H}(0)$, and long-range density rearrangements can contribute to the latter even though they have no effect on the value of the density exactly at the proton.

We first establish the LRT scale in Fig.~\ref{fig:VH0_linear_models}. The Thomas--Fermi and RPA results are shown over $0.5\leq\rs\leq6$, while the RPA+KK result is restricted to $1\leq\rs\leq6$. The RPA and RPA+KK curves follow from Eq.~\eqref{eq:VH0_LRT} with the dielectric functions in Eqs.~\eqref{eq:eps_RPA} and \eqref{eq:eps_LFC}, respectively, while the Thomas--Fermi result is given by the analytic expression in Eq.~\eqref{eq:VH0_TF}.

Over the displayed interval,
\begin{equation*}
V_{\rm H}^{\rm TF}(0)
>
V_{\rm H}^{\rm RPA+KK}(0)
>
V_{\rm H}^{\rm RPA}(0).
\end{equation*}
The KK local-field correction therefore raises the RPA value of the contact Hartree potential, but by substantially less than the separation between RPA and Thomas--Fermi theory. The elevated Thomas--Fermi result reflects the incorrect short-wavelength behavior of the Thomas--Fermi dielectric response. Although the corresponding Thomas--Fermi contact density is divergent, $V_{\rm H}^{\rm TF}(0)$ remains finite because the Hartree contact is an $r$-weighted moment of the complete screening cloud rather than the density evaluated at a single point. For this reason, Thomas--Fermi theory is retained here as a simple analytic reference, whereas RPA provides the more appropriate full-wave-vector LRT comparison with the nonlinear calculations below.

\begin{figure}[htbp!]
    \centering
    \includegraphics[width=\columnwidth]{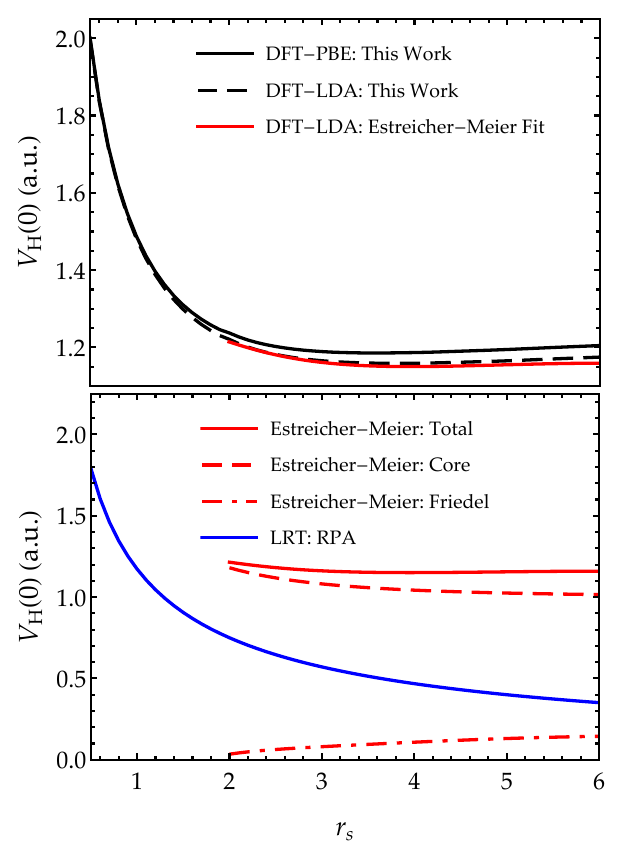}
    \caption{Contact Hartree potential $V_{\rm H}(0)$ as a function of $\rs$. Upper panel: self-consistent DFT-PBE and DFT-LDA reference calculations over $0.5\leq\rs\leq6$, together with the Estreicher--Meier parametrization of the DFT-LDA induced density over $2\leq\rs\leq6$. Lower panel: decomposition of the Estreicher--Meier result into its core [Eq.~\eqref{eq:VH0_LDA_core}] and Friedel [Eq.~\eqref{eq:VH0_LDA_friedel}] contributions, compared with the RPA linear-response result. The core term dominates the Estreicher--Meier Hartree moment, but the Friedel contribution grows with $\rs$ and becomes a non-negligible fraction of the total. In the metallic-density regime, the nonlinear DFT values remain substantially above RPA, whereas the DFT--RPA separation decreases toward the high-density end of the calculated interval. The DFT curves are reference calculations within the stated approximate exchange--correlation functionals and are not presented as exact many-body benchmarks.}
\label{fig:VH0_DFT_vs_linear}
\end{figure}

Figure~\ref{fig:VH0_DFT_vs_linear} then compares the nonlinear DFT reference calculations with RPA and resolves the spatial contributions to the Estreicher--Meier LDA-based result. The upper panel shows our self-consistent DFT-LDA and DFT-PBE values over $0.5\leq\rs\leq6$, together with the contact Hartree potential obtained from the Estreicher--Meier LDA density parametrization over its displayed range $2\leq\rs\leq6$. The lower panel decomposes the Estreicher--Meier result into the core and Friedel contributions defined in Eqs.~\eqref{eq:VH0_LDA_core} and \eqref{eq:VH0_LDA_friedel}, and compares their sum with the RPA result.
The DFT-LDA and DFT-PBE curves exhibit nearly the same overall density dependence. Their difference increases gradually with $\rs$, but remains modest for these two semilocal functionals. At $\rs=5$, for example, the present calculations give
\begin{align*}
V_{\rm H}^{\rm LDA}(0) &= 1.1654~\mathrm{a.u.}, &
V_{\rm H}^{\rm PBE}(0) &= 1.1947~\mathrm{a.u.},
\end{align*}
a difference of approximately $2.5\%$. This comparison should be interpreted narrowly: it shows that the difference between the two semilocal functionals considered here is modest. It does not provide a quantitative estimate of the full many-body uncertainty in $V_{\rm H}(0)$. In particular, no thermodynamic-limit wavefunction benchmark for this contact Hartree quantity is presently available. Accordingly, the DFT-LDA and DFT-PBE curves in Fig.~\ref{fig:VH0_DFT_vs_linear} are used as approximate nonlinear DFT reference calculations, not as exact benchmarks.

Within that qualification, the contrast with LRT is substantial in the metallic-density regime. At $\rs=5$, RPA gives $V_{\rm H}^{\rm RPA}(0)\simeq0.400$~a.u., compared with approximately $1.17$--$1.19$~a.u. from LDA and PBE. Thus, for the particular LDA/PBE comparison made here, the DFT--RPA separation is much larger than the LDA--PBE spread. Toward higher density the difference narrows: at $\rs=0.5$, the LDA, PBE, and RPA values are approximately $1.999$, $2.002$, and $1.792$~a.u., respectively. The residual difference at the highest density considered remains finite, so the calculation again establishes a trend toward closer linear and nonlinear response rather than a numerical demonstration of the strict $\rs\rightarrow0$ limit.

The Estreicher--Meier curve provides an independent representation of the LDA screening cloud because it is constructed by inserting their analytic LDA density parametrization into the Hartree moment. Over $2\leq\rs\leq6$, its resulting $V_{\rm H}(0)$ remains within approximately $1.4\%$ of our direct self-consistent LDA result. Since the Estreicher--Meier expression is an analytic fit to an LDA density rather than a distinct exchange--correlation approximation, the small residual difference between the two LDA-based curves is not interpreted as exchange--correlation-functional dependence. Appendix~\ref{sec:EM_cusp} shows that the EM radial fit also has a small, density-dependent deficit in its local cusp slope, decreasing from $5.80\%$ at $\rs=2$ to $0.33\%$ at $\rs=6$. That slope deficit cannot be translated into a corresponding error in $V_{\rm H}(0)$: the latter samples the full radial density, and the EM--DFT-LDA Hartree residual does not follow the decreasing cusp-deficit trend.

The lower panel of Fig.~\ref{fig:VH0_DFT_vs_linear} illustrates why $V_{\rm H}(0)$ should not be regarded as a purely contact-local observable. The Estreicher--Meier core term provides the dominant contribution, but the Friedel contribution grows systematically with $\rs$: it increases from approximately $0.034$~a.u. at $\rs=2$ to approximately $0.144$~a.u. at $\rs=6$, corresponding to roughly $2.8\%$ and $12.4\%$, respectively, of the total Estreicher--Meier contact Hartree potential. Long-range oscillatory screening therefore makes a measurable contribution to the $r$-weighted Hartree moment even though the contact density itself is controlled by the near-proton region.

The Estreicher--Meier radial Hartree integral was evaluated using adaptive quadrature for the core/finite-range contribution together with oscillatory quadrature for the asymptotic Friedel tail. Varying the finite-to-asymptotic matching point over $10\leq x_{\rm match}\leq160$ at representative densities changed $V_{\rm H}^{\rm Friedel}(0)$ by less than $2\times10^{-12}$~a.u.; tightening the production quadrature tolerance from $10^{-11}$ to $10^{-12}$ changed it by less than $6\times10^{-13}$~a.u. The core-plus-Friedel decomposition in Fig.~\ref{fig:VH0_DFT_vs_linear} is therefore numerically converged well beyond the scale visible in the figure.

A further independent historical check on the absolute nonlinear DFT
scale is provided by Popovi\'c \textit{et al.}~
\cite{1976PopovicTheoryHeatSolution}.  They treated the screening of a
proton in a uniform electron gas at the mean densities appropriate to
Al and Mg using the Kohn--Sham density-functional framework with a
local Hedin--Lundqvist exchange--correlation approximation and an
approximately self-consistent nonlinear construction.  Their tabulated
electron--proton interaction energy, $V_{\rm int}/Z_p$, is related to
the present contact Hartree convention by
$V_{\rm H}(0)=-V_{\rm int}/Z_p$.  For the physical proton,
their exchange--correlation calculation therefore gives
$V_{\rm H}(0)=1.186$~a.u. at $\rs=2.064$ (the Al electron density)
and $1.157$~a.u. at $\rs=2.642$ (the Mg electron density).
Linear interpolation of the present DFT-LDA results to the same
densities gives $1.214$ and $1.176$~a.u., respectively, so that the
historical values differ from the present LDA calculation by only
approximately $2.3\%$ and $1.6\%$.  Given the different numerical
implementation, the historical local exchange--correlation
parametrization, and the approximate self-consistency of the earlier
calculation, this close agreement provides an independent consistency
check on the absolute magnitude of the density-derived contact Hartree
potential.  Popovi\'c \textit{et al.} also reported nonlinear
Hartree-only values of $1.016$ and $0.937$~a.u. at the same two
densities, illustrating the appreciable enhancement of the contact
Hartree response produced by the inclusion of local
exchange--correlation effects.  These historical results are therefore
used here as an independent DFT-LDA consistency check rather than as an
exact many-body benchmark.

Arponen and Pajanne~\cite{1979ArponenElectronLiquidCollective} also reported the induced electrostatic screening potential directly. Their collective-boson result indicates somewhat stronger induced screening than the DFT-LDA-based Estreicher--Meier reconstruction, while remaining on the same overall nonlinear scale and exhibiting a similar radial structure. This comparison, together with the contact-density results discussed in Sec.~\ref{sec:LRT_DFT_contact_density}, provides an independent many-body consistency check on the magnitude of nonlinear proton screening. It does not remove the functional and many-body uncertainty associated with the DFT reference calculations.

\subsection{VPA for model potentials: Friedel sum rule}
\label{sec:vpa_fsr_results}

Having established the nonlinear density and contact response within
DFT, we now turn to a reduced scattering description based on the
screened model potentials introduced in Sec.~\ref{sec:model_potentials}.
The variable-phase approach provides the continuum phase shifts
required by the static Friedel sum rule (FSR), while an independent
negative-energy calculation determines the discrete spectrum when
bound states are present.  We first validate these two numerical
ingredients and their consistency through Levinson's theorem.  We
then quantify the numerical convergence of the exact-$T=0$ FSR
calculation and determine the FSR-constrained screening parameter for
the Yukawa, hydrogenic, and Hulth\'en potentials over the full density
interval considered in this work.  The extension to two-parameter
potentials is considered in Sec.~\ref{sec:two_parameter_fsr}.

\subsubsection{Phase-shift and bound-state validation}
\label{sec:phase_bound_validation}

Before solving the Friedel condition for the model-potential
parameters, we validate the two spectral ingredients entering the
calculation: the continuum phase shifts and the discrete bound-state
spectrum.  This is particularly important near a bound-state
threshold, where the absolute branch of the phase shift must remain
consistent with the number of states that have crossed into the
discrete spectrum.  As introduced in Sec.~\ref{sec:FSR}, we define
\begin{align}
s(k)=\frac{2}{\pi}
\sum_{\ell=0}^{\infty}(2\ell+1)\delta_{\ell}(k),
\label{eq:s_validation}
\end{align}
while Levinson's theorem gives
\begin{align}
\delta_{\ell}(0)=n_{\ell}^{b}\pi,
\qquad
s(0)=
2\sum_{\ell=0}^{\infty}(2\ell+1)n_{\ell}^{b}
\equiv N_{\rm cap}.
\label{eq:levinson_validation}
\end{align}
Here $n_{\ell}^{b}$ denotes the number of radial bound states in the
$\ell$ channel, and $N_{\rm cap}$ is their spin- and
orbital-degeneracy-weighted spectral capacity.  The latter is a
state-counting quantity and should not be confused with the actual
many-electron occupation of those states.

Among the validation tests performed for the phase-shift solver, a
particularly stringent combined benchmark is provided by the classic
calculations of Rogers and co-workers for the attractive static
screened Coulomb (Yukawa) potential
\cite{1970RogersBoundEigenstatesStatic,1971RogersPhaseShiftsStatic}.
Rogers parametrizes the potential by the screening length $D$, which
for $Z=1$ is related to the present inverse screening parameter by
$D=1/\alpha$. His weighted phase-shift sum $G_B^{-}(k)/\pi$ corresponds
to $s(k)/2$ in the convention of Eq.~\eqref{eq:s_validation}.
For the fixed-proton convention used here, Rogers' dimensionless
momentum is numerically equal to the momentum variable $k$ used in
the present calculation.

\begin{figure}[htbp!]
    \centering
    \includegraphics[width=\columnwidth]
    {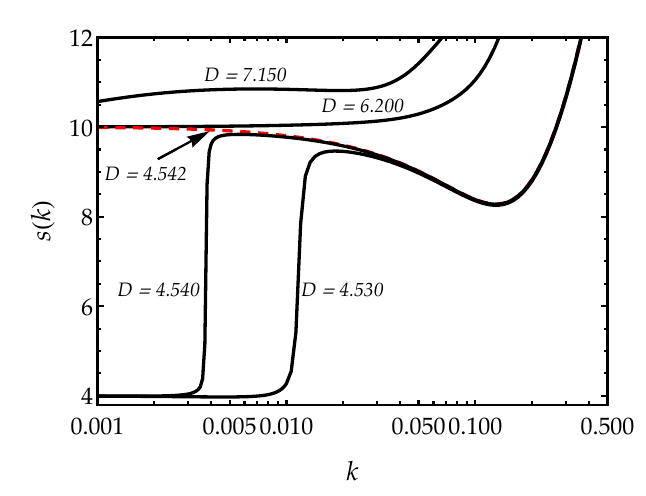}
    \caption{Weighted phase-shift sum $s(k)$ for the attractive
    Yukawa potential at selected screening lengths from the benchmark
    of Rogers \cite{1971RogersPhaseShiftsStatic}. The plotted quantity $s(k)$ corresponds to twice the weighted phase-shift sum reported by Rogers. The red
    dashed curve, $D=4.542$, lies immediately on the bound side of
    the narrow $p$-wave threshold between $D=4.540$ and $D=4.542$.
    The corresponding Levinson limits change from $s(0)=4$ at
    $D=4.540$ to $s(0)=10$ at $D=4.542$, consistently with the
    appearance of one additional $\ell=1$ radial bound state.}
\label{fig:sk_vs_k_Yukawa_Rogers_Benchmark}
\end{figure}

Figure~\ref{fig:sk_vs_k_Yukawa_Rogers_Benchmark} shows $s(k)$ for five
representative screening lengths surrounding the shallow $p$-state
threshold identified by Rogers.  The quantum-mechanical phase shifts
are obtained from the variable-phase equation using the LSODA solver.
For this historical benchmark, the high-$\ell$ tail of the phase sum
is completed using the independently tested QM/Born-matching
procedure; the production exact-$T=0$ Friedel calculations in
Sec.~\ref{sec:fsr_convergence_alpha} instead use the separately
audited adaptive pure-QM $L_{\max}$ criterion.

The pair $D=4.540$ and $D=4.542$ provides a substantially more
demanding test than a comparison performed far from a spectral
threshold.  The independent negative-energy calculation gives
\begin{align*}
D=4.540:
\qquad
(n_0^b,n_1^b)=(2,0),
\qquad
N_{\rm cap}=4,
\end{align*}
whereas
\begin{align*}
D=4.542:
\qquad
(n_0^b,n_1^b)=(2,1),
\qquad
N_{\rm cap}=10.
\end{align*}
Equation~\eqref{eq:levinson_validation} therefore requires
\begin{align}
N_{\rm cap}:4\longrightarrow10,
\qquad
\Delta s(0)=2(2\ell+1)=6
\quad(\ell=1),
\label{eq:Rogers_threshold_jump}
\end{align}
precisely the change approached by the two calculated curves in
Fig.~\ref{fig:sk_vs_k_Yukawa_Rogers_Benchmark}.  Thus, the same
narrow threshold is resolved independently from the positive-energy
scattering calculation and from the negative-energy spectrum.  This
agreement simultaneously tests the absolute magnitude and branch of
the low-energy phase shifts, their degeneracy-weighted partial-wave
sum, the bound-state count, and the Levinson bookkeeping.  The
remaining curves reproduce the characteristic momentum dependence
reported by Rogers and are also consistent with the independent
variable-phase calculations of Grandjouan and Deutsch
\cite{1975GrandjouanPhaseShiftsStatic}.

The bound-state solver was additionally benchmarked against
independent analytic or published spectral results for each of the
three screened-Coulomb forms used below.  We summarize only the
principal checks in Table~\ref{tab:model_bound_validation}; their
purpose here is to establish the reliability of the discrete-spectrum
classification entering the Friedel/Levinson accounting rather than
to introduce a separate model-potential validation study.

\begin{table*}[tp]
\centering
\caption{Selected independent bound-state benchmarks for the
screened model potentials.  Energy differences are given in Hartree
where applicable.}
\label{tab:model_bound_validation}
\small
\renewcommand{\arraystretch}{1.15}
\begin{tabular}{p{0.14\linewidth} p{0.41\linewidth} p{0.37\linewidth}}
\hline
Potential & Independent benchmark & Present numerical check \\
\hline

Yukawa &
Rogers bound-state spectrum
\cite{1970RogersBoundEigenstatesStatic}, together with
bound/unbound-state classifications spanning the tabulated
screening strengths &
All 17 tabulated bound-state energies and 31 corresponding
unbound/blank-state classifications are reproduced; the maximum
absolute energy difference is approximately
$2\times10^{-5}$ Ha. \\

Hulth\'en &
Exact analytic $s$-state spectrum,
$E_{ns}=-\frac{1}{2}(n^{-1}-n\alpha/2)^2$, together with
non-$s$ numerical eigenvalues of Varshni
\cite{1990VarshniEigenenergiesOscillatorStrengths} &
The maximum error is $1.92\times10^{-9}$ Ha for the exact
$s$ states and $5.01\times10^{-8}$ Ha for the Varshni
non-$s$ states. \\

Hydrogenic &
Critical screening strength for disappearance of the $1s$ bound
state reported by Apagyi and Nagy
\cite{1988ApagyiNagyTwoParameter} &
The independently determined threshold is
$\alpha_{1s}^{c}=1.821058756042$ in the present
convention.  Since the screening parameter of
Ref.~\cite{1988ApagyiNagyTwoParameter} satisfies
$\alpha=2\alpha_F$, this corresponds to
$\alpha_F^{c}=0.910529378$, consistent with their published
critical value $\alpha_F^{c}\simeq0.91$. \\

\hline
\end{tabular}
\end{table*}

The validation of the continuum phase shifts is not limited to the
Rogers comparison shown in
Fig.~\ref{fig:sk_vs_k_Yukawa_Rogers_Benchmark}.  During development
of the variable-phase implementation, quantal Yukawa phase shifts
were compared with independent published results over a broad range
of electron momenta and angular-momentum channels, including the
calculations of Grandjouan and Deutsch
\cite{1975GrandjouanPhaseShiftsStatic}, Arista
\cite{2002AristaEnergyLossIons}, and Arista and Sigmund
\cite{2007Arista-Sigmund}.  These comparisons reproduce the
published phase-shift trends and magnitudes.  The Rogers benchmark is
shown here because it provides an especially compact end-to-end test:
the same physical problem probes the low-energy phase branch,
degeneracy-weighted angular summation, a shallow bound-state
threshold, and the associated Levinson count.

Together, the continuum and discrete-spectrum tests establish the
numerical ingredients required for the subsequent Friedel analysis:
reliable quantum-mechanical phase shifts, controlled angular-momentum
summation, and consistent identification and counting of bound
states.

\subsubsection{FSR convergence and self-consistent screening parameters}
\label{sec:fsr_convergence_alpha}

For each one-parameter potential
$\nu\in\{\mathrm{Yuk},\mathrm{Hyd},\mathrm{Hult}\}$, the screening
parameter is determined independently from the exact-$T=0$ static
Friedel condition
\begin{align}
F_\nu(\alpha;r_s)
\equiv
Z-
s_{L_{\max}}^\nu(k_{\rm F};\alpha)
=0,
\label{eq:FSR_model_residual}
\end{align}
with
\begin{align}
s_{L_{\max}}^\nu(k_{\rm F};\alpha)
=
\frac{2}{\pi}
\sum_{\ell=0}^{L_{\max}}
(2\ell+1)\delta_{\ell}^{\nu}(k_{\rm F};\alpha).
\label{eq:FSR_finite_sum_results}
\end{align}
We denote the resulting root by
$\alpha_{\rm FSR}^{\nu}(r_s)$.  The term
``self-consistent screening parameter'' is used here only in this
restricted sense: $\alpha_{\rm FSR}^{\nu}$ is the parameter that
makes the chosen analytic potential satisfy the Friedel constraint.
It does not imply Hartree or Kohn--Sham self-consistency between that
analytic potential and the density it generates.

Before carrying out the density sweep, the exact-$T=0$ Friedel
pipeline was subjected to a top-down numerical-error analysis.  The
relevant accuracy is the uncertainty induced in the final root, not
merely the local error of an individual phase shift.  Perturbations
of the calculated Friedel residual were therefore propagated through
the local root conditioning according to
\begin{align}
\Delta\alpha_{\rm FSR}
\simeq
-\frac{\Delta F}
{\partial F/\partial\alpha}.
\label{eq:FSR_conditioning}
\end{align}
For the Yukawa potential, a 51-state conditioning scan over
$1\le r_s\le6$ gives
\begin{align}
1.1715
\lesssim
\left|
\frac{\partial F}{\partial\alpha}
\right|
\lesssim
1.5800,
\label{eq:FSR_slope_range}
\end{align}
showing that no nearly flat or poorly conditioned Friedel root occurs
over the production interval.  The numerical controls were then
frozen to satisfy the adopted target
\begin{align}
|\Delta\alpha_{\rm FSR}|
\le
10^{-5}\,.
\label{eq:alpha_accuracy_target}
\end{align}
Representative sensitivities from this audit are summarized in
Table~\ref{tab:FSR_numerical_error}.  The listed
$\Delta\alpha_{\rm FSR}$ values are empirical changes relative to
tighter numerical references and are intended as convergence
diagnostics rather than rigorous mathematical error bounds.

\begin{table*}[tp]
\centering
\caption{Representative numerical-convergence budget used to freeze
the exact-$T=0$ Friedel pipeline before the 51-state production
sweeps.  The sensitivity tests were performed for the Yukawa
potential at representative densities and propagated to the fitted
screening parameter through Eq.~\eqref{eq:FSR_conditioning}.}
\label{tab:FSR_numerical_error}
\small
\renewcommand{\arraystretch}{1.15}
\begin{tabular}{p{0.32\linewidth} p{0.31\linewidth} p{0.23\linewidth}}
\hline
Numerical control & Production setting & Max. tested
$|\Delta\alpha_{\rm FSR}|$ \\
\hline

LSODA phase-equation tolerances &
$(r_{\rm tol},a_{\rm tol})=(10^{-9},10^{-12})$ &
$\sim 1.0\times10^{-8}$ \\

Near-origin / phase-resolution control &
$10^{-9}$ &
$1.2\times10^{-8}$ \\

Adaptive outer radial endpoint &
$\epsilon_{\rm tail}=10^{-9}$,
$c_{\rm tp}=1.5$ &
$4.8\times10^{-7}$ \\

Adaptive angular tail &
$\tau_L=10^{-6}$,
$L_{\min}=3$,
$N_{\rm tail}=4$ &
$5.4\times10^{-7}$ \\

Brent root location &
$(r_{\rm tol},x_{\rm tol})=(10^{-8},10^{-10})$ &
$\lesssim3\times10^{-8}$ \\

\hline
\end{tabular}
\end{table*}

The angular cutoff is not assigned from a fixed empirical formula.
Instead, the weighted partial-wave contributions $(2\ell+1)\delta_\ell(k_{\rm F};\alpha)$ are evaluated successively in $\ell$. The effective $L_{\max}$ is accepted only after the last
significant contribution is followed by $N_{\rm tail}=4$ consecutive channels satisfying the prescribed angular threshold. If this certified tail is not reached before the current angular ceiling, the ceiling is enlarged and the test is repeated. The radial endpoint is determined similarly from the decay of the screened interaction, with
an additional safety distance from the classical turning point. These adaptive criteria prevent the production calculation from relying on a single fixed radial box or a preassigned partial-wave cutoff.

With all numerical controls frozen, the screening parameter was then
recomputed independently for
\begin{align*}
r_s
=
1.0,1.1,\ldots,6.0
\end{align*}
for each of the Yukawa, hydrogenic, and Hulth\'en potentials.  Thus,
51 density states were evaluated for each model, corresponding to
153 independent Friedel roots in total. No numerical tolerance was
retuned during these production sweeps. The maximum realized FSR
residuals were
$3.57\times10^{-9}$,
$4.00\times10^{-9}$, and
$3.62\times10^{-9}$ for the Yukawa, hydrogenic, and Hulth\'en
calculations, respectively.  The largest adaptive angular cutoffs
encountered over the same grids were
$L_{\max}=18$, $14$, and $13$, respectively.
\begin{figure}[htbp!]
\centering
\includegraphics[width=\columnwidth]
{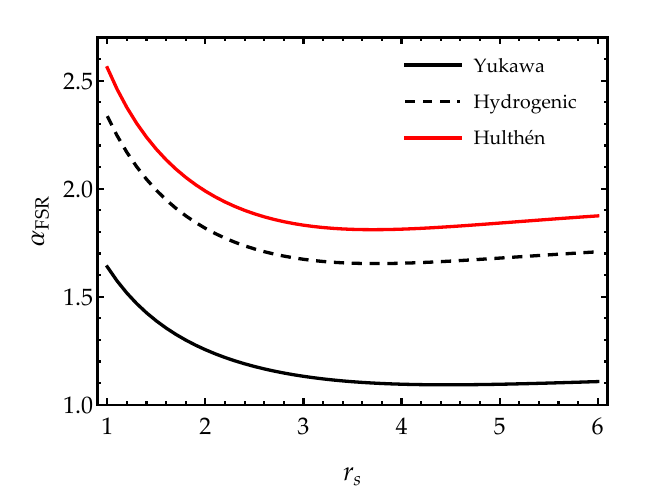}
\caption{Exact-$T=0$ Friedel-constrained screening parameters
$\alpha_{\rm FSR}$ for a proton impurity, $Z=1$, obtained
independently for the Yukawa, hydrogenic, and Hulth\'en model
potentials at 51 density points,
$r_s=1.0,1.1,\ldots,6.0$.  Each curve is obtained by solving
Eq.~\eqref{eq:FSR_model_residual} using the same frozen numerical
policy.  Although all three potentials satisfy the same total
Friedel condition, the fitted screening parameter remains strongly
dependent on the analytic form of the model potential.}
\label{fig:AlphaFSR_vs_rs_AllPots}
\end{figure}
Figure~\ref{fig:AlphaFSR_vs_rs_AllPots} shows a common qualitative
density dependence for the three analytic forms. Starting from the
high-density end, $\alpha_{\rm FSR}$ decreases rapidly, passes
through a broad shallow minimum, and then increases weakly toward
$r_s=6$.  The absolute values nevertheless remain clearly
separated throughout the interval: the Hulth\'en screening parameter
is largest, the hydrogenic value is intermediate, and the Yukawa
value is smallest.  This ordering should not be interpreted as an
ordering of a unique physical screening length.  The parameter
$\alpha$ enters three different analytic screening functions, and its
numerical value is therefore intrinsically model dependent.

More fundamentally, satisfaction of the same total FSR does not
make the underlying scattering problems identical.  At the fitted
parameter each model obeys
\begin{align}
\frac{2}{\pi}
\sum_{\ell}
(2\ell+1)
\delta_\ell^\nu
\left(k_{\rm F};\alpha_{\rm FSR}^{\nu}\right)
=
Z,
\label{eq:FSR_global_constraint}
\end{align}
but Eq.~\eqref{eq:FSR_global_constraint} fixes only one
degeneracy-weighted scalar combination of the phase shifts.  It does
not require the individual
$\delta_\ell^\nu(k_{\rm F})$ to coincide channel by channel, nor does it
uniquely determine the radial form or short-range structure of the
screened interaction.  We therefore do not infer model independence
of the fitted potentials merely from their common Friedel sum.

It is noteworthy that the VPA-based $\alpha_{\rm FSR}$ findings in Figure~\ref{fig:AlphaFSR_vs_rs_AllPots} are in agreement with those obtained from the numerical solutions of the Schr\"odinger equation for these model potentials reported earlier by Arista \cite{2006Arista_ChargeStates}, by Nersisyan and Fern\'andez-Varea \cite{2013NersisyanSecondorderBornApproximation}, and by Vincent and co-workers \cite{2008VincentDescriptionMigratingProton}.

The one-parameter calculations thus demonstrate both the utility and
the limitation of the FSR constraint.  For each chosen analytic form,
the FSR selects a well-defined and numerically stable screening
parameter, but substantial model freedom remains in quantities not
fixed by the total phase-shift sum.  This naturally motivates asking
whether an additional short-range condition can reduce that freedom.
We therefore turn next to two-parameter screened potentials, beginning
with the Apagyi--Nagy construction in which the Friedel condition is
combined with the electron--proton cusp constraint.

\subsubsection{Two-parameter potentials: Friedel/cusp-constrained
Apagyi--Nagy model}
\label{sec:two_parameter_fsr}
As discussed in Sec.~\ref{sec:model_potentials}, the
Apagyi--Nagy construction introduces two range parameters
$\alpha$ and $\beta$, while the mixing coefficient
$\mathcal{W}(\alpha,\beta;r_s)$ is fixed analytically by the
electron--proton cusp condition.  Imposing the Friedel sum rule
therefore supplies one further scalar constraint on the two remaining
range parameters.  At fixed $r_s$, however, this does not select a
unique pair $(\alpha,\beta)$; rather, it defines a continuous locus
$\alpha(\beta;r_s)$ of Friedel/cusp-constrained potentials.  This is
the family displayed in Fig.~1 of Apagyi and Nagy
\cite{1988ApagyiNagyTwoParameter}.  The two exact constraints are
therefore complementary but do not, by themselves, remove the
remaining one-parameter freedom of the model.

To construct a definite set of potentials for the present
calculations, we follow the representative choice used by Apagyi and
Nagy in their Table~1 and fix
$\alpha=0.7$.  The second range parameter $\beta$ is then
determined independently at each density from the exact-$T=0$ FSR
using the VPA developed above.  Because the model is symmetric under
$\alpha\leftrightarrow\beta$, we restrict the calculation to the
ordered physical branch $\alpha<\beta$ and require the cusp-derived
mixing coefficient to satisfy $0\leq\mathcal{W}\leq1$.

For each of the nine density parameters
$r_s=1.0,1.4,1.5,1.9,2.0,3.0,4.0,5.0,$ and $6.0$, the search on
this physical branch yields a single certified FSR root at fixed
$\alpha=0.7$.  Here ``single'' refers to the intersection
of the FSR/cusp locus with the chosen $\alpha=0.7$ slice; it should
not be interpreted as uniqueness of the full two-parameter
Friedel/cusp construction.  The maximum FSR residual among the nine
states is below $3\times10^{-9}$, and the adaptive partial-wave
cutoff varies from $L_{\max}=25$ at $r_s=1$ to $L_{\max}=8$ at
$r_s=6$.  The resulting parameters are listed in
Table~\ref{tab:AN_FSR_parameters}.
\begin{table}[htbp!]
\centering
\caption{Friedel/cusp-constrained Apagyi--Nagy parameters for the
representative branch $\alpha=0.7$.  The present values of
$\beta_{\rm FSR}$ are obtained independently from the exact-$T=0$
VPA calculation.  The final column gives the corresponding values
tabulated by Apagyi and Nagy \cite{1988ApagyiNagyTwoParameter} for
comparison.}
\label{tab:AN_FSR_parameters}
\small
\renewcommand{\arraystretch}{1.10}
\begin{tabular}{ccc}
\hline
$r_s$ &
Present $\beta_{\rm FSR}$ &
Ref.~\cite{1988ApagyiNagyTwoParameter} \\
\hline
1.0 & 1.353478 & 1.365 \\
1.4 & 1.210771 & 1.225 \\
1.5 & 1.191005 & 1.205 \\
1.9 & 1.145877 & 1.155 \\
2.0 & 1.140347 & 1.150 \\
3.0 & 1.132477 & 1.135 \\
4.0 & 1.128779 & 1.130 \\
5.0 & 1.110604 & 1.110 \\
6.0 & 1.092708 & 1.090 \\
\hline
\end{tabular}
\end{table}
The small differences between the present FSR-constrained parameters
and the values reported by Apagyi and Nagy cannot be attributed solely
to the tabulation precision of the latter.  The historical phase shifts
were obtained using Lad\'anyi's finite-basis variational method
\cite{1988ApagyiNagyTwoParameter}. Although this method can yield
highly accurate scattering results when appropriately converged, its
accuracy depends on the basis size, nonlinear scale parameters, and
associated stability and error diagnostics. These numerical details
were not reported in the original screening calculation. We therefore regard differences in the underlying phase-shift calculations as a plausible source of the residual discrepancy, without attempting to assign it quantitatively.

The nine states labeled by $r_s$ values in Table~\ref{tab:AN_FSR_parameters} and their corresponding parameter pairs define the prescribed Apagyi--Nagy
potentials used in the subsequent quantal-density calculations. The
comparison also emphasizes that the additional cusp constraint does
not convert the Friedel rule into a unique two-parameter fit: the
choice $\alpha=0.7$ selects one representative member of
the allowed Friedel/cusp family at each density.

\subsection{VPA for model potentials: Quantal electron density}
\label{sec:vpa_QM_dn}

The Friedel sum rule constrains the integrated scattering response of a
screened impurity potential, but it does not uniquely determine the
real-space density generated by that potential. We therefore proceed
beyond the phase-shift analysis and examine the quantal induced
electron density constructed directly from the occupied bound and
continuum states of the FSR-constrained Yukawa, Hulth\'en, and
hydrogenic potentials.

The required formalism was developed in
Sec.~\ref{sec:vpa_density}. In particular, the QM-VPA density separates
into the bound and scattering contributions defined in
Eq.~\eqref{eq:vpa_density_decomposition}, with the continuum density
obtained from the momentum-resolved kernel
$\Phi(k;r)$ defined through
Eqs.~\eqref{eq:vpa_channel_density_kernel} and
\eqref{eq:vpa_Phi_definition} and integrated over the occupied Fermi
sea according to Eq.~\eqref{eq:vpa_scattering_density_Phi}.
The resulting total density is given by
Eq.~\eqref{eq:vpa_total_density_zeroT}. We emphasize throughout that
this quantal density is distinct from any charge density obtained by
applying Poisson's equation directly to the analytic form of a model
potential.

\subsubsection{Radial profile: bound and scattering contributions}
\label{sec:vpa_dn_r_bound_scatt}

We first consider the representative state $r_s=3$, for which all three
model potentials support a shallow $1s$ bound state while retaining
clear differences in their short-range screening structure. Each
potential is evaluated using its own zero-temperature FSR-constrained
screening parameter.

Before examining the final induced density, it is useful to inspect one
of its principal continuum ingredients. As shown in
Sec.~\ref{sec:vpa_density}, the scattering density is obtained by
integrating the continuum kernel $\Phi(k;r)$ over all occupied momenta,
Eq.~\eqref{eq:vpa_scattering_density_Phi}. The behavior of this kernel
at the upper boundary $k=k_{\rm F}$ is especially informative because the
long-range response of a degenerate Fermi gas is controlled by the
sharp occupation boundary at the Fermi surface.

\begin{figure}[htbp!]
\centering
\includegraphics[width=\columnwidth]
{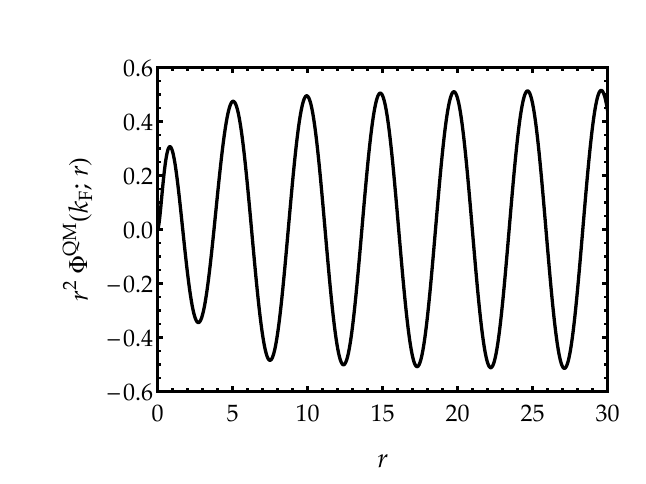}
\caption{
Fermi-edge continuum kernel for the FSR-constrained Yukawa potential at
$r_s=3$, displayed as $r^2\Phi(k_{\rm F};r)$ to remove the dominant radial
decay and expose the long-range oscillatory structure. The kernel is
one momentum-resolved ingredient of the scattering density defined in
Eq.~\eqref{eq:vpa_scattering_density_Phi}, not the density itself.
The median outer zero-crossing separation is
$2.454579\,a_0$, compared with
$\pi/(2k_{\rm F})=2.455446\,a_0$ for a $2k_{\rm F}$ oscillation. Thus a monotonic,
exponentially screened Yukawa potential generates continuum scattering
states carrying the characteristic Friedel wavelength.
}
\label{fig:vpa_phi_kf_yuk_rs3}
\end{figure}

Figure~\ref{fig:vpa_phi_kf_yuk_rs3} therefore shows
$r^2\Phi(k_{\rm F};r)$ for the FSR-constrained Yukawa potential at $r_s=3$.
The Yukawa case provides a particularly transparent test because the
underlying screened potential is monotonic and exponentially decaying.
Consequently, an oscillatory long-range structure in the continuum
kernel cannot be attributed to oscillations already present in the
input potential.

After a short-range transient, $r^2\Phi(k_{\rm F};r)$ approaches a nearly
constant-amplitude oscillation. At $r_s=3$,
$k_{\rm F}=0.63971943$, giving the characteristic Friedel period
and adjacent-zero spacing
\begin{align*}
\lambda_{2k_{\rm F}}
&=
\frac{\pi}{k_{\rm F}}
=
4.910891\,a_0,
\\
\Delta r_{\rm zero}^{(2k_{\rm F})}
&=
\frac{\pi}{2k_{\rm F}}
=
2.455446\,a_0.
\end{align*}
The median adjacent-zero separation measured from the numerical
continuum kernel in the outer region is
$2.454579\,a_0$, corresponding to
\begin{align*}
\frac{\Delta r_{\rm zero}^{\rm QM}}
     {\pi/(2k_{\rm F})}
=
0.999647.
\end{align*}
Thus the Fermi-edge continuum kernel reproduces the expected $2k_{\rm F}$
oscillation scale to approximately $3.5\times10^{-4}$ in relative
terms.

The origin of this behavior is also evident from the scattering-state
asymptotics. Equation~\eqref{eq:vpa_scattering_asymptotic_normalization}
shows that each partial wave approaches a free spherical wave shifted
by the full scattering phase $\delta_\ell(k)$. Squaring this wave and
subtracting its free-electron counterpart, as prescribed by
Eq.~\eqref{eq:vpa_channel_density_kernel}, generates an oscillatory
density contribution whose radial phase depends explicitly on the
finite scattering phase shift. The subsequent integration over
occupied momenta in Eq.~\eqref{eq:vpa_scattering_density_Phi} selects
the Fermi-edge contribution asymptotically, producing the familiar
three-dimensional Friedel form
\begin{align*}
\Delta n_{\rm sc}^{\rm QM--VPA}(r)
\sim
\frac{A_F}{r^3}
\cos\!\left(2k_{\rm F} r+\phi_F\right),
\qquad r\rightarrow\infty .
\end{align*}
The result in Fig.~\ref{fig:vpa_phi_kf_yuk_rs3} therefore supplies a
real-space scattering interpretation of the same Fermi-surface physics
that appears as the $q=2k_{\rm F}$ nonanalyticity in the response function:
the characteristic wavelength is imposed by the Fermi surface, even
though the screened impurity potential itself is nonoscillatory.

It is important, however, to distinguish the period of a
Friedel oscillation from its phase and amplitude. The $2k_{\rm F}$
period is already present in linear response because the Lindhard
response contains the Fermi-surface nonanalyticity at $q=2k_{\rm F}$.
Correctly reproducing this wavelength therefore does not by itself
establish that the screening response is nonlinear. The detailed
phase of the Friedel tail contains additional information about the
scattering from the impurity.

This distinction was emphasized explicitly in the early nonlinear
screening calculation of Almbladh \emph{et al.}\
\cite{1976AlmbladhScreeningProtonElectron}. For a proton in an electron
gas at $r_s=4$, they found that linear response not only
underestimated the near-proton density but also gave an inadequate
description of the Friedel oscillations. In particular, they noted
that linear response produced essentially zero Friedel phase, whereas
their nonlinear self-consistent solution displaced the oscillations by
approximately $90^\circ$ over the metallic-density range. They further
emphasized that this phase is physically relevant because it controls
the interaction between screened impurities.

The same qualitative distinction is visible in the present results.
The inset of Fig.~\ref{fig:dn0_vs_r_DFT_vs_LRT} shows that RPA and
RPA+KK linear response reproduce an oscillatory tail but that their
zero crossings and extrema are displaced from those of the nonlinear
DFT-LDA density. The inclusion of a static local-field correction
changes the linear-response curve quantitatively but does not change
its perturbative order in the external field. Thus the existence and
wavelength of the Friedel oscillation are already contained in LRT,
while its detailed phase need not agree with a fully nonlinear
calculation.

\begin{figure}[htbp!]
\centering
\includegraphics[width=\columnwidth]
{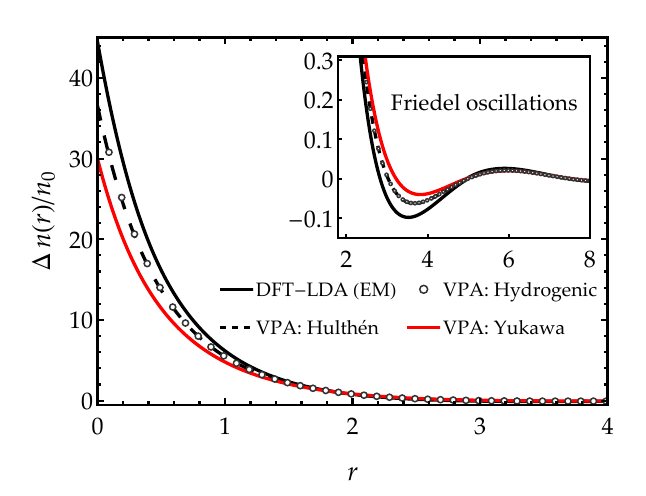}
\caption{
Normalized total induced electron density at $r_s=3$.
The nonlinear DFT-LDA density represented by the Estreicher--Meier
parametrization is compared with the QM-VPA densities generated by the
FSR-constrained Yukawa, Hulth\'en, and hydrogenic potentials. The
Hulth\'en and hydrogenic profiles are nearly coincident in the
near-field region, while the Yukawa response is smaller. The inset
resolves the Friedel tail and shows that the three QM-VPA oscillations
are nearly phase aligned with the nonlinear DFT-LDA result. This
contrasts with the larger DFT--LRT phase displacement visible in
Fig.~\ref{fig:dn0_vs_r_DFT_vs_LRT}: LRT contains the $2k_{\rm F}$ Friedel
wavelength, whereas the nonperturbative scattering calculation also
retains the finite phase shifts that control the detailed real-space
phase of the oscillation.
}
\label{fig:vpa_dn_total_rs3}
\end{figure}

The situation is markedly different when the FSR-constrained model
potentials are treated by the full QM-VPA scattering calculation.
Figure~\ref{fig:vpa_dn_total_rs3} compares their total quantal densities
with the nonlinear DFT-LDA density represented by the
Estreicher--Meier parametrization. The DFT-LDA curve is used as a
self-consistent nonlinear comparison rather than as an exact
many-body benchmark. All densities are normalized by the homogeneous
density $n_0$.
In the near field, DFT-LDA produces the strongest electron
accumulation, the Hulth\'en and hydrogenic QM-VPA densities are nearly
coincident, and the Yukawa density is smaller. The agreement is not
limited, however, to this overall radial decay. The inset shows that
the Friedel oscillations generated by all three QM-VPA model
potentials are also nearly phase aligned with the DFT-LDA oscillation:
the principal zero crossings and extrema occur at closely similar
radial positions even though their amplitudes are not identical.

This behavior is significant in view of the contrast with linear
response. The QM-VPA calculation does not linearize the scattering
from the prescribed model potential. Once that potential has been
specified, its continuum wavefunctions and the phase shifts entering
Eq.~\eqref{eq:vpa_scattering_asymptotic_normalization} are obtained
nonperturbatively. Equivalently, the scattering solution retains the
multiple-scattering content that would appear order by order in a
Born expansion, rather than retaining only the first-order density
response to the perturbing field. Self-consistent DFT and QM-VPA are
not equivalent theories---DFT determines its effective Kohn--Sham
potential self-consistently from the density, whereas QM-VPA solves the
scattering problem for a prescribed FSR-constrained model
potential---but both retain nonperturbative single-particle scattering
physics that is absent from LRT.

The near phase agreement between QM-VPA and DFT-LDA can therefore be
viewed as a manifestation of this common nonlinear/nonperturbative
scattering content. The Fermi surface fixes the characteristic
$2k_{\rm F}$ wavelength, as demonstrated directly in
Fig.~\ref{fig:vpa_phi_kf_yuk_rs3}, while the finite scattering phase
shifts determine how that oscillation is displaced in real space.
Linear response captures the former but, for a proton-strength
perturbation in the metallic-density regime, can miss the latter.
This interpretation is consistent with the nonlinear Friedel-phase
shift identified by Almbladh \emph{et al.}\
\cite{1976AlmbladhScreeningProtonElectron}.

The phase agreement should not be attributed to the Friedel sum rule
alone. The FSR constrains a weighted sum of the asymptotic phase shifts
at $k_{\rm F}$, whereas the local density depends on the complete set of
continuum wavefunctions and their radial amplitudes and phases.
Consequently, satisfaction of the FSR is necessary for the global
screening charge but does not uniquely determine either the amplitude
or the phase of the real-space density oscillations. The close
DFT--QM-VPA phase correspondence in
Fig.~\ref{fig:vpa_dn_total_rs3} is therefore additional information
provided by the quantal scattering calculation.
\begin{figure}[htbp!]
\centering
\includegraphics[width=\columnwidth]
{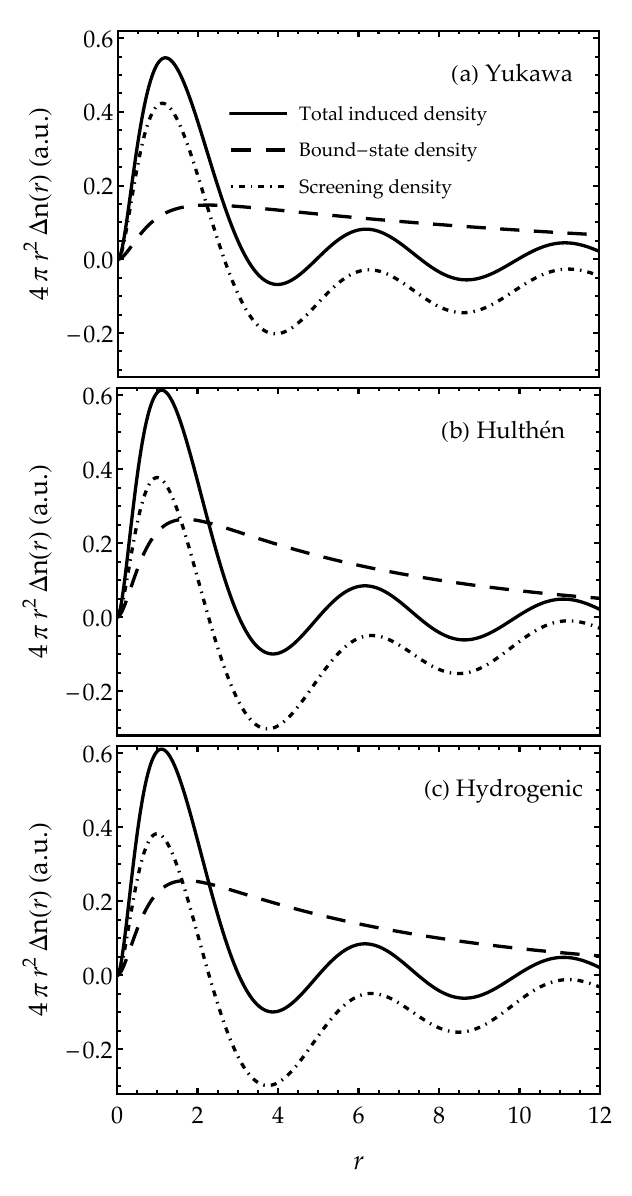}
\caption{
Bound, scattering, and total contributions to the QM-VPA induced
electron density at $r_s=3$, shown through
$4\pi r^2\Delta n(r)$ for (a) Yukawa, (b) Hulth\'en, and
(c) hydrogenic FSR-constrained potentials. The bound contribution is
positive and nonoscillatory, whereas the scattering contribution
changes sign and carries the long-range Friedel oscillations. The
Hulth\'en and hydrogenic decompositions are very similar, while the
Yukawa potential exhibits a weaker bound contribution and a different
continuum compensation. The figure directly connects the oscillatory
total densities of Fig.~\ref{fig:vpa_dn_total_rs3} with the continuum
kernel examined in Fig.~\ref{fig:vpa_phi_kf_yuk_rs3}.
}
\label{fig:vpa_dn_bound_scatt_rs3}
\end{figure}
The total density does not by itself reveal how the screening cloud is
partitioned between localized and continuum states. We therefore
separate the two contributions already defined in
Eq.~\eqref{eq:vpa_density_decomposition} and display the radial
screening-charge integrands
$4\pi r^2\Delta n_{\rm b}^{\rm QM--VPA}(r)$ and
$4\pi r^2\Delta n_{\rm sc}^{\rm QM--VPA}(r)$ together with their sum.
Their radial integrals give the corresponding accumulated charges,
\begin{align*}
Q_x(R)
=
4\pi\int_0^R
r^2\Delta n_x(r)\,dr,
\qquad
x=\{{\rm b},{\rm sc},{\rm total}\}.
\end{align*}

Figure~\ref{fig:vpa_dn_bound_scatt_rs3} shows the decomposition for all
three model potentials. The bound contribution is positive and
nonoscillatory, as expected from the squared modulus of the localized
bound-state wavefunction. The scattering contribution behaves
qualitatively differently. It is positive close to the proton, changes
sign at larger radius, and develops the alternating structure that
survives in the Friedel tail of the total density. Thus the
oscillations displayed in Fig.~\ref{fig:vpa_dn_total_rs3} can be
identified directly with the continuum sector whose Fermi-edge
structure was examined in Fig.~\ref{fig:vpa_phi_kf_yuk_rs3}.

The decomposition also exposes model dependence that is less evident
from the total density alone. The Hulth\'en and hydrogenic potentials
produce not merely similar total densities, but remarkably similar
bound and scattering contributions separately. Their close total
profiles therefore arise from a similar partition of the screening
response between localized and continuum states. The Yukawa potential,
by contrast, produces a weaker bound contribution and a correspondingly
different continuum compensation.

This result provides another direct demonstration that the FSR does
not uniquely determine the local density. Different model potentials
can satisfy the same unit-charge Friedel constraint while distributing
that charge differently between bound and scattering states and while
producing different near-field densities.

\begin{figure}[htbp!]
\centering
\includegraphics[width=\columnwidth]
{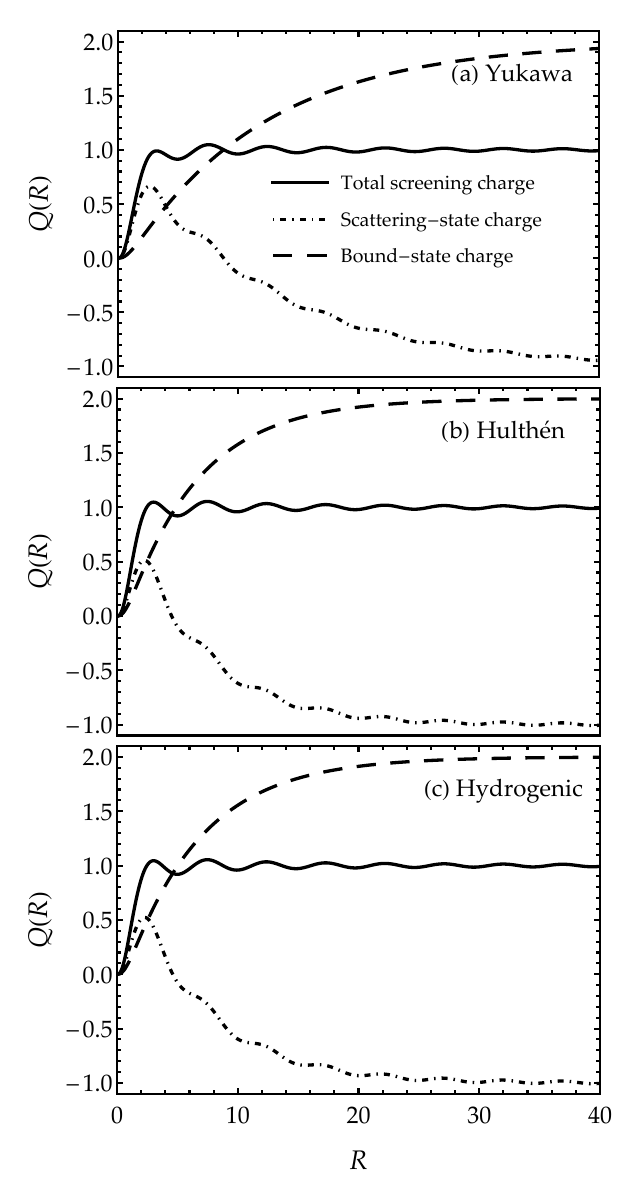}
\caption{
Accumulated bound, scattering, and total screening charges at $r_s=3$
for (a) Yukawa, (b) Hulth\'en, and (c) hydrogenic FSR-constrained
potentials. With one doubly occupied $1s$ level,
$Q_{\rm b}(R)\rightarrow2$ and
$Q_{\rm sc}(R)\rightarrow-1$, leaving the required total screening
charge $Q_{\rm total}(R)\rightarrow1$. The total charge becomes close
to unity on a substantially shorter radial scale than that required
for the separate bound and scattering components to reach their
asymptotic limits, demonstrating their strong real-space
compensation. Horizontal guide lines indicate the asymptotic values
$2$, $1$, and $-1$.
}
\label{fig:vpa_Q_bound_scatt_rs3}
\end{figure}

The bound--continuum charge accounting becomes still more transparent
after radial integration. Figure~\ref{fig:vpa_Q_bound_scatt_rs3}
shows the accumulated charges $Q_{\rm b}(R)$,
$Q_{\rm sc}(R)$, and $Q_{\rm total}(R)$ obtained directly from the
stored QM-VPA density components. At $r_s=3$, each of the three model
potentials supports one doubly occupied $1s$ level. The corresponding
Levinson/Friedel bookkeeping therefore requires
\begin{align*}
Q_{\rm b}(\infty)\rightarrow2,
\qquad
Q_{\rm sc}(\infty)\rightarrow-1,
\qquad
Q_{\rm total}(\infty)\rightarrow1.
\end{align*}
The numerical results recover precisely this structure.

The total screening charge rises rapidly toward unity and then performs
small, decaying Friedel oscillations about the required value. The
individual bound and scattering charges evolve on a considerably
longer radial scale toward their respective $+2$ and $-1$ asymptotes.
On the full $80\,a_0$ production domain, all three potentials recover
the asymptotic $2-1=1$ charge balance to within a few parts in
$10^{-3}$ or better.

The separation of these radial scales is physically important. The
existence of a spatially extended shallow bound state does not imply
that the net screening cloud extends over the same distance. The
positive bound-state charge accumulated at a given radius is
simultaneously compensated by the continuum response, allowing
$Q_{\rm total}(R)$ to approach the required unit screening charge well
before $Q_{\rm b}(R)$ and $Q_{\rm sc}(R)$ have individually saturated.
This is the same bound--continuum compensation emphasized in earlier
self-consistent jellium calculations of hydrogen impurities
\cite{1976AlmbladhScreeningProtonElectron,
1978JenaElectronicStructureHydrogen}.

The model dependence again follows the pattern observed in
Fig.~\ref{fig:vpa_dn_bound_scatt_rs3}. Hulth\'en and hydrogenic
screening display very similar accumulation profiles, whereas the
Yukawa bound charge approaches its asymptotic value more slowly.
Nevertheless, the continuum contribution compensates the bound sector
so that all three potentials recover the same unit total screening
charge required by the FSR.

Taken together, Figures~\ref{fig:vpa_phi_kf_yuk_rs3}--
\ref{fig:vpa_Q_bound_scatt_rs3} provide a connected microscopic
picture of the QM-VPA screening density. The Fermi-edge continuum
kernel first establishes the correct $2k_{\rm F}$ spatial scale. Integration
over the occupied continuum then produces a total quantal density whose
Friedel oscillations are not only of the expected wavelength but are
also nearly phase aligned with the nonlinear DFT-LDA density. The
bound--scattering decomposition identifies these oscillations with the
continuum sector, and the accumulated charges finally recover the
Levinson/Friedel $2-1=1$ bookkeeping.

This sequence also separates two aspects of Friedel physics that should
not be conflated. The oscillation wavelength is fixed by the Fermi
surface and is already accessible within LRT. The detailed phase and
amplitude are sensitive to the strength of impurity scattering and
therefore to nonlinear or nonperturbative response. The present
QM-VPA calculation captures the latter through the full scattering
wavefunctions of the prescribed model potential, while self-consistent
DFT captures it through the nonlinear Kohn--Sham screening problem.
Their close Friedel-phase correspondence, contrasted with the phase
displacement found in LRT and historically emphasized by Almbladh
\emph{et al.}, provides a useful real-space signature of the nonlinear
screening physics retained by the model-potential VPA treatment.

Finally, these results reinforce the distinction between three objects:
the analytic screened potential, any Poisson-derived density associated
with that analytic form, and the quantal density generated from its
occupied eigenstates. Only the last of these,
$\Delta n^{\rm QM--VPA}(r)$, contains the complete bound and continuum
wavefunction information used for the density-derived contact
quantities considered below.

\subsubsection{Contact induced density: bound and scattering contributions}
\label{sec:vpa_dn_0_bound_scatt}

We next turn from the finite-radius structure of the screening cloud to
its strictly local value at the proton.  The contact limit provides a
particularly sensitive test of the model-potential construction because
the Friedel sum rule and the contact density constrain different aspects
of the screening solution.  The FSR fixes the integrated scattering
response through the degeneracy-weighted phase-shift sum, whereas
$\Delta n(0)$ depends on the amplitudes of the occupied quantum states
directly at the origin.

As shown in Sec.~\ref{sec:vpa_density}, regular bound and continuum
radial functions behave as $r^\ell$ near the origin, so all
$\ell>0$ contributions vanish at $r=0$ and only the $s$ wave survives.
The scattering contribution is therefore given by
Eq.~\eqref{eq:vpa_contact_scattering_density}, while the bound-state
contribution is given by Eq.~\eqref{eq:vpa_contact_bound_density}.
Their sum gives the complete QM-VPA contact induced density,
Eq.~\eqref{eq:vpa_contact_density}.  Consequently, the contact density
probes information that is not determined by the total Friedel charge:
different model potentials can satisfy the same unit-charge FSR while
producing different $s$-wave amplitudes at the proton.

\begin{figure}[htbp!]
\centering

\includegraphics[width=\columnwidth]
{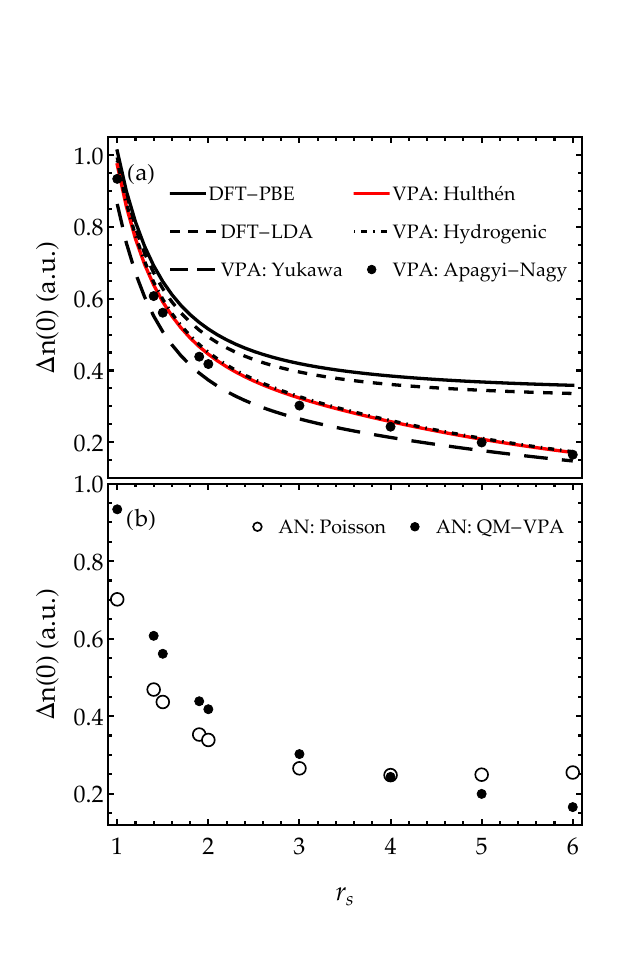}



\caption{
Contact induced electron density at the proton.
\textbf{(a)} Total QM-VPA contact density
$\Delta n^{\rm QM--VPA}(0)$ for the FSR-constrained Yukawa,
Hulth\'en, and hydrogenic potentials over
$1\leq r_s\leq6$, together with the nine
Friedel/cusp-constrained Apagyi--Nagy states of
Table~\ref{tab:AN_FSR_parameters}.  Self-consistent DFT-LDA and
DFT-PBE results are shown for nonlinear comparison.  The Hulth\'en
and hydrogenic results remain nearly coincident, whereas the Yukawa
contact density is systematically smaller; the separation between the
model-potential and DFT contact densities increases toward larger
$r_s$.
\textbf{(b)} Contact density associated with the same nine
Friedel/cusp-constrained Apagyi--Nagy potentials obtained from two
different constructions: the Poisson-density ansatz of
Eq.~\eqref{eq:dn_poisson_AN} and the QM-VPA density generated by the
occupied eigenstates of the corresponding potential.  The two results
approach one another near $r_s\simeq4$ and reverse their ordering
across this region, demonstrating that the Poisson-based density used
in constructing an analytic screened potential is not, in general, the
quantal density subsequently generated by that potential.
}
\label{fig:vpa_dn0_allmodels}
\end{figure}

Figure~\ref{fig:vpa_dn0_allmodels}(a) compares the total quantal contact
density $\Delta n^{\rm QM--VPA}(0)$ obtained from the FSR-constrained
Yukawa, Hulth\'en, and hydrogenic potentials over the 51-state interval
$1\leq r_s\leq6$. The nine Friedel/cusp-constrained
Apagyi--Nagy potentials defined in
Sec.~\ref{sec:two_parameter_fsr} and
Table~\ref{tab:AN_FSR_parameters} are included as discrete points.
For comparison, the self-consistent DFT-LDA and DFT-PBE contact
densities are shown on the same absolute-density scale.  As elsewhere
in this work, these DFT results are used as nonlinear self-consistent
comparisons and not as exact many-body benchmarks.

All of the QM-VPA contact densities decrease as the electron gas becomes
more dilute, but the quantitative result remains sensitive to the
analytic form of the FSR-constrained potential.  The Hulth\'en and
hydrogenic curves are nearly coincident over the full interval,
consistent with the close similarity of their finite-radius density
profiles in Sec.~\ref{sec:vpa_dn_r_bound_scatt}.  The Yukawa contact
density is systematically smaller.  At $r_s=1$, for example, the
Yukawa, Hulth\'en, and hydrogenic values are approximately
$0.864$, $0.973$, and $0.984\,a_0^{-3}$, respectively, while the
corresponding DFT-LDA and DFT-PBE values are approximately
$0.994$ and $1.011\,a_0^{-3}$.  Thus the Hulth\'en and hydrogenic
models lie close to the nonlinear DFT values at the high-density end
of the model-potential interval, whereas the Yukawa model already gives
a visibly smaller contact response.

The separation becomes substantially larger toward lower electron
density.  At $r_s=6$, the Yukawa, Hulth\'en, and hydrogenic QM-VPA
contact densities have decreased to approximately
$0.148$, $0.172$, and $0.174\,a_0^{-3}$, respectively, compared with
approximately $0.336$ and $0.359\,a_0^{-3}$ from DFT-LDA and DFT-PBE.
The agreement of the different model potentials with the unit Friedel
charge therefore does not imply corresponding agreement in the local
contact amplitude.  The distinction becomes increasingly visible as
$r_s$ increases.
The two-parameter Apagyi--Nagy results provide an additional test of
this conclusion.  Their nine QM-VPA contact densities lie within, or
close to, the range spanned by the one-parameter model potentials and
follow the same overall density dependence.  In particular, imposing
the exact electron--proton cusp condition on the auxiliary Poisson density in addition to the FSR does
not force the independently reconstructed quantal contact density onto the self-consistent DFT
curves. The one-parameter FSR-only models already satisfy the exact quantum cusp automatically (Appendix~\ref{app:qm_vpa_cusp}); the additional Apagyi--Nagy condition instead restricts the analytic potential through its prescribed Poisson-based density ansatz. Since the cusp fixes a derivative relative to a contact value rather than the contact value itself, neither construction uniquely determines $\Delta n^{\rm QM--VPA}(0)$.

A more direct consequence of the Apagyi--Nagy construction is shown in
Fig.~\ref{fig:vpa_dn0_allmodels}(b).  As discussed in
Sec.~\ref{sec:model_potentials}, this model was constructed by first
postulating the normalized two-exponential density
$\Delta n_{\rm AN}^{\rm Poisson}(r)$ in
Eq.~\eqref{eq:dn_poisson_AN}, with the mixing coefficient
$\mathcal{W}(\alpha,\beta;r_s)$ fixed by the cusp condition through
Eq.~\eqref{eq:AN_weight}, and then obtaining the corresponding
screened interaction from Poisson's equation,
Eq.~\eqref{eq:AN_potential}.  Once this potential is treated instead
as a quantum scattering Hamiltonian, however, the occupied bound and
continuum eigenstates generate the distinct QM-VPA density of
Eq.~\eqref{eq:vpa_total_density_zeroT} and, at the origin, the contact
value of Eq.~\eqref{eq:vpa_contact_density}.

Figure~\ref{fig:vpa_dn0_allmodels}(b) compares these two contact
densities for exactly the same nine Friedel/cusp-constrained
Apagyi--Nagy potentials.  The distinction is substantial and strongly
density dependent.  At $r_s=1$, the Poisson ansatz gives
$\Delta n_{\rm AN}^{\rm Poisson}(0)\simeq0.701\,a_0^{-3}$, whereas the
quantal calculation gives
$\Delta n_{\rm AN}^{\rm QM--VPA}(0)\simeq0.934\,a_0^{-3}$.
The difference decreases with increasing $r_s$, and the two values
become nearly equal around $r_s\simeq4$, where they are approximately
$0.248$ and $0.243\,a_0^{-3}$, respectively.  Beyond this region the
ordering reverses: at $r_s=6$ the Poisson value is approximately
$0.255\,a_0^{-3}$ while the QM-VPA value has decreased to approximately
$0.165\,a_0^{-3}$.
\begin{figure}[htbp!]
\centering
\includegraphics[width=\columnwidth]
{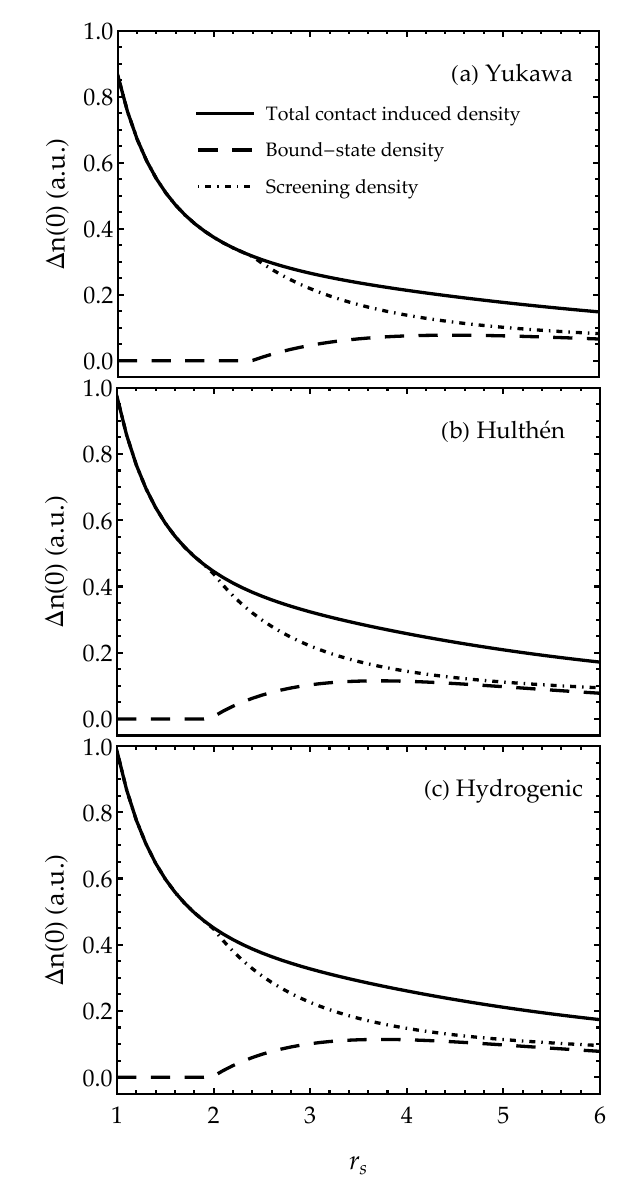}
\caption{
Total, bound-state, and scattering-state QM-VPA contact induced
densities as functions of $r_s$ for the FSR-constrained
\textbf{(a)} Yukawa, \textbf{(b)} Hulth\'en, and
\textbf{(c)} hydrogenic potentials.  The decomposition is that of
Eq.~\eqref{eq:vpa_contact_density}, with the scattering and bound
contributions given by
Eqs.~\eqref{eq:vpa_contact_scattering_density} and
\eqref{eq:vpa_contact_bound_density}, respectively.
At high electron density the contact response is entirely continuum
like.  After the shallow $1s$ state enters the discrete spectrum, a
bound-state contribution appears while the continuum contribution is
correspondingly reduced.  The total contact density evolves smoothly
through this redistribution.  Hulth\'en and hydrogenic screening give
closely similar component-level decompositions, whereas the Yukawa
potential yields a smaller contact response.  Even at $r_s=6$, the
bound state contributes only about $45\%$ of the total contact density,
so the continuum remains quantitatively important at the proton.
}
\label{fig:vpa_dn0_components}
\end{figure}
The reversal of the ordering is particularly instructive.  It shows
that the difference between the two constructions cannot be represented
by a simple density-independent rescaling of the contact amplitude.
Rather, the Poisson ansatz and the occupied-state QM-VPA calculation
represent genuinely different density responses.  The same screened
potential can therefore satisfy the imposed normalization, cusp, and
Friedel constraints without reproducing, through its quantum
eigenstates, the particular density from which its analytic form was
originally generated.  This provides a direct numerical illustration
of the distinction emphasized throughout the present analysis between
an analytic screened potential, a Poisson-derived density associated
with that potential, and the quantal density generated by its occupied
states.

To identify the microscopic origin of the model dependence seen in
Fig.~\ref{fig:vpa_dn0_allmodels}(a), we next resolve the total QM-VPA
contact density into its bound- and scattering-state contributions.
Figure~\ref{fig:vpa_dn0_components} shows this decomposition separately
for the Yukawa, Hulth\'en, and hydrogenic potentials over the same
51-state interval.  The use of three vertically stacked panels avoids
superposing nine closely spaced curves on a single set of axes and
makes the evolution of the two spectral sectors directly visible.

At the high-density end of the interval, where no ordinary negative-energy
bound state is present, the total contact response is generated entirely
by continuum scattering states.  As $r_s$ increases and a shallow
$1s$ level enters the discrete spectrum, a nonzero bound-state
contribution appears.  The scattering contribution continues to
decrease with increasing $r_s$, while an increasing share of the
contact response is transferred to the bound sector.  No corresponding
discontinuity is apparent in the total contact density: the formation
of the bound state is accompanied by a compensating rearrangement of
the continuum contribution.

This behavior is the contact counterpart of the bound--continuum
compensation observed in the radial densities and accumulated charges
of Sec.~\ref{sec:vpa_dn_r_bound_scatt}.  The distinction is important:
the appearance of a discrete level does not simply add a new positive
density on top of an otherwise unchanged continuum response.  Instead,
the occupied spectrum reorganizes between localized and scattering
sectors, while their sum determines the physical induced density.

The model dependence is also resolved more clearly by this
decomposition.  The Hulth\'en and hydrogenic potentials generate
closely similar bound contributions and closely similar scattering
contributions separately, not merely similar total contact densities.
The Yukawa potential produces a smaller contact response in both
sectors over much of the interval.  Thus the difference among the
one-parameter models cannot be attributed solely to the presence or
absence of the shallow bound state.  It reflects the complete
short-range $s$-wave response of both the localized and continuum
wavefunctions to the particular FSR-constrained potential.

At the dilute end of the calculated interval, the bound and continuum
contributions remain comparable rather than reducing to a purely
localized contact response.  At $r_s=6$, the bound contribution
accounts for approximately $44.5\%$, $45.2\%$, and $44.9\%$ of the
total contact density for Yukawa, Hulth\'en, and hydrogenic screening,
respectively.  A substantial continuum contribution therefore remains
at the proton even after the $1s$ state is well established.  This
again emphasizes that the local contact response is a property of the
complete occupied spectrum rather than of the bound state alone.

This decomposition also clarifies why the contact density contains
information that is absent from the Friedel sum alone.  The FSR fixes
the asymptotic, degeneracy-weighted phase-shift sum required for unit
screening, whereas
Eqs.~\eqref{eq:vpa_contact_scattering_density} and
\eqref{eq:vpa_contact_bound_density} depend on the $s$-wave amplitudes
at the origin.  The global charge constraint and the local contact
amplitude are therefore independent characteristics of the screening
solution.  Having characterized both the radial structure and its
contact limit, we next return to the spatially accumulated screening
charge and use it to define a characteristic screening-cloud radius
$R_{\rm VPA}$.

\subsubsection{Screening-cloud radius: $R_{\rm VPA}$}
\label{sec:vpa_r_vpa}

The radial densities discussed above contain substantially more
information than can be represented by a single screening length.
Nevertheless, it is useful to assign a characteristic spatial scale to
the induced screening cloud in a manner that can be applied consistently
to different density constructions.  For this purpose, we adopt the
accumulated-charge definition introduced by Perrot
\cite{1982PerrotTemperatureDependent}.  Unlike the screening parameter
of a prescribed analytic model potential, this construction depends
only on the induced electron density and can therefore be applied on the
same footing to a QM-VPA density or to a nonlinear self-consistent
density such as Perrot's local-density-functional result.
\begin{figure}[htbp!]
\centering
\includegraphics[width=\columnwidth]
{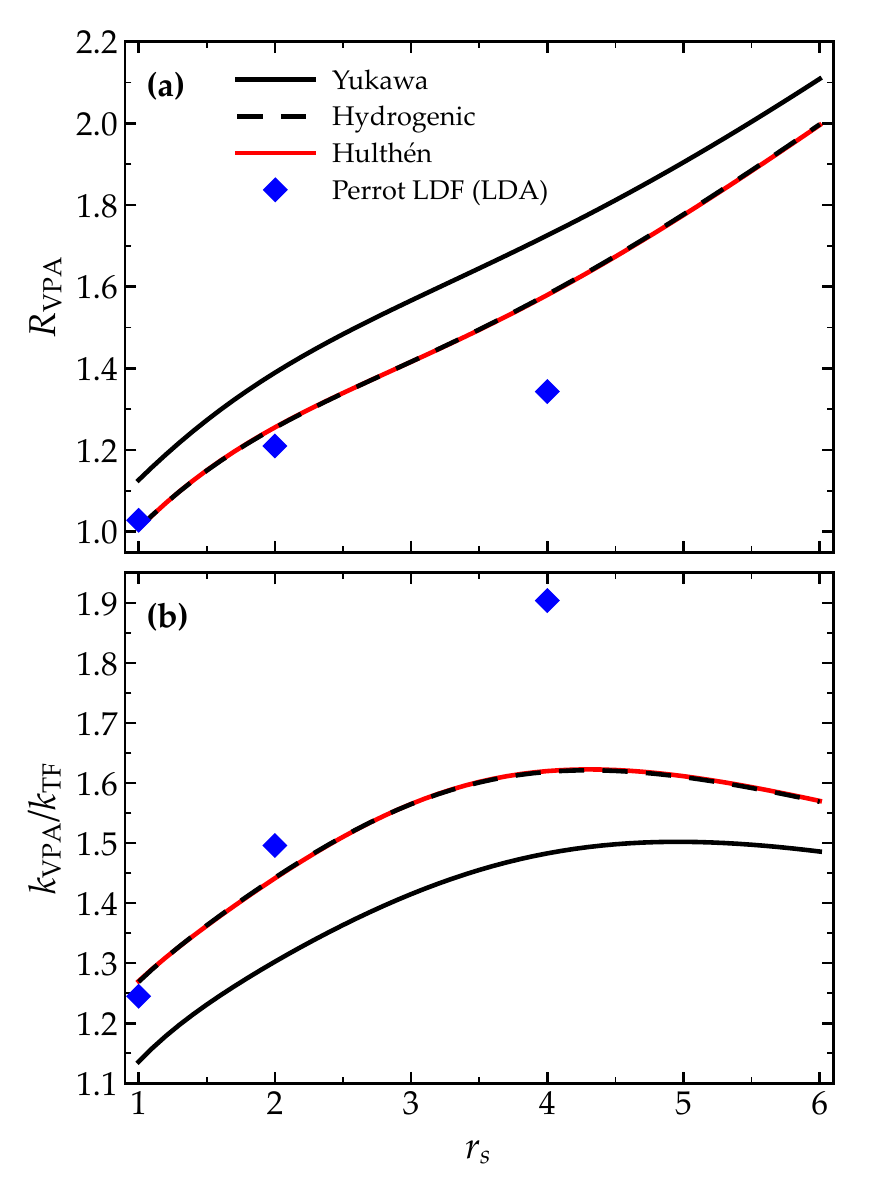}
\caption{
Density-derived characteristic screening-cloud size for a proton in
the zero-temperature electron gas.
\textbf{(a)} Screening radius $R_{\rm VPA}$ defined by
Eq.~\eqref{eq:RVPA_definition} for the FSR-constrained Yukawa,
Hulth\'en, and hydrogenic model potentials over the 51-state grid
$r_s=1.0,1.1,\ldots,6.0$.  The three discrete symbols show the
corresponding radii $R_{\rm LDF}=2/k_{\rm LDF}$ obtained from Perrot's
zero-temperature nonlinear local-density-functional results at
$r_s=1$, $2$, and $4$.
\textbf{(b)} Corresponding density-derived inverse screening scale
normalized by the Thomas--Fermi momentum,
$k_{\rm VPA}/k_{\rm TF}$, together with Perrot's reported
$k_{\rm LDF}/k_{\rm TF}$ values.  The Hulth\'en and hydrogenic QM-VPA
results are nearly coincident throughout the density interval, whereas
the Yukawa screening cloud is systematically more extended.  The
increasing separation from Perrot's self-consistent LDF result toward
larger $r_s$ is consistent with the growing quantitative importance of
exchange--correlation and self-consistency effects that are absent from
the prescribed FSR-constrained model potentials.
}
\label{fig:vpa_rvpa_allpots}
\end{figure}
Perrot's construction is referenced to the linearized
Thomas--Fermi screening cloud.  In the notation used throughout the
present work, its inverse screening length is the Thomas--Fermi momentum
$k_{\rm TF}$ already introduced below Eq.~\eqref{eq:eps_RPA}, with
$k_{\rm TF}^2=(4/\pi)k_{\rm F}$.  The corresponding Yukawa-type screening
density has the accumulated displaced charge
\begin{align*}
\frac{Q_{\rm TF}(R)}{Z}
=
1-\left(1+k_{\rm TF}R\right)e^{-k_{\rm TF}R}.
\end{align*}
A sphere of radius $2/k_{\rm TF}$ therefore contains the fraction
\begin{align*}
\frac{Q_{\rm TF}(2/k_{\rm TF})}{Z}
=
1-3e^{-2}
=
0.593994\ldots
\end{align*}
of the total screening charge.  Perrot used this same enclosed-charge
fraction to assign an effective screening radius to a general nonlinear
density.

For the present QM-VPA densities, the accumulated total charge
$Q_{\rm total}(R)$ was introduced in
Sec.~\ref{sec:vpa_dn_r_bound_scatt}.  We therefore define the
characteristic screening-cloud radius by
\begin{align}
Q_{\rm total}(R_{\rm VPA})
&=
\left(1-3e^{-2}\right)Z,
\nonumber\\
k_{\rm VPA}
&=
\frac{2}{R_{\rm VPA}} .
\label{eq:RVPA_definition}
\end{align}
When Friedel oscillations produce later recrossings of the prescribed
charge fraction, $R_{\rm VPA}$ is defined by the first outward crossing.
The ratio $k_{\rm VPA}/k_{\rm TF}$ then measures the compactness of the
fully quantal screening cloud relative to the linearized TF reference
at the same electron density.  Values above unity correspond to a
screening cloud that reaches the fixed $59.4\%$ enclosed-charge
fraction at a smaller radius than the TF cloud.

This interpretation also makes clear that $k_{\rm VPA}$ should not be
identified with the model-potential parameter $\alpha_{\rm FSR}$.
As discussed in Sec.~\ref{sec:vpa_fsr_results} and illustrated in
Fig.~\ref{fig:AlphaFSR_vs_rs_AllPots}, $\alpha_{\rm FSR}$ is determined
by imposing the global Friedel constraint on a chosen analytic
potential family.  By contrast, $k_{\rm VPA}$ is extracted afterward
from the quantal density generated by the occupied bound and continuum
states of that potential.  Thus, $\alpha_{\rm FSR}$ characterizes the
chosen scattering potential, whereas $k_{\rm VPA}$ is a
density-derived effective inverse screening length.

Figure~\ref{fig:vpa_rvpa_allpots}(a) shows $R_{\rm VPA}$ for the
FSR-constrained Yukawa, Hulth\'en, and hydrogenic potentials over the
full 51-state interval $r_s=1.0,1.1,\ldots,6.0$.  The absolute
screening radius increases smoothly with $r_s$ for all three models,
as expected when the background electron gas is diluted.  The
Hulth\'en and hydrogenic results are nearly indistinguishable over the
entire interval: their values of $R_{\rm VPA}$ differ by less than
approximately $0.1\%$ over all 51 states.  At $r_s=1$, for example,
they give $R_{\rm VPA}=1.0073$ and $1.0082\,a_0$, respectively, while
at $r_s=6$ the corresponding values are $1.9958$ and
$1.9975\,a_0$.  The Yukawa screening cloud is systematically more
extended, with $R_{\rm VPA}=1.1264\,a_0$ at $r_s=1$ and
$2.1090\,a_0$ at $r_s=6$.

The near coincidence of the Hulth\'en and hydrogenic results is not an
isolated feature of $R_{\rm VPA}$.  It continues the post-FSR
similarity already found for the screened interactions, the finite-radius
QM-VPA densities in Fig.~\ref{fig:vpa_dn_total_rs3}, and the contact
densities in Fig.~\ref{fig:vpa_dn0_allmodels}.  Once constrained by the
same Friedel condition, these two potential families therefore generate
remarkably similar results for several distinct density-derived
observables.  This similarity should not, however, be interpreted as
establishing model independence.  The Yukawa results remain clearly
separated from the Hulth\'en/hydrogenic pair, while the two-parameter
Apagyi--Nagy results discussed above provide an additional example in
which satisfying the Friedel condition, even together with the exact
cusp imposed on its auxiliary Poisson-density ansatz, does not uniquely determine the quantal density
response.  The present result therefore demonstrates strong robustness
between two particular model families rather than uniqueness under the
FSR constraint.

The normalized inverse screening scale in
Fig.~\ref{fig:vpa_rvpa_allpots}(b) provides a complementary view.
For the Hulth\'en and hydrogenic potentials,
$k_{\rm VPA}/k_{\rm TF}$ increases from approximately $1.27$ at
$r_s=1$ to a broad maximum of approximately $1.62$ near
$r_s\simeq4$--$5$, before decreasing slightly toward approximately
$1.57$ at $r_s=6$.  The Yukawa curve follows the same qualitative
behavior at lower values, rising from approximately $1.14$ at
$r_s=1$ to about $1.50$ at its broad maximum before decreasing weakly
toward the dilute end.  Thus, according to this common
enclosed-charge criterion, all three QM-VPA screening clouds are more
compact than the linearized TF reference throughout the interval
studied, although their relative enhancement over TF is not monotonic
with $r_s$.

For comparison, Fig.~\ref{fig:vpa_rvpa_allpots} also includes Perrot's
zero-temperature nonlinear local-density-functional results at
$r_s=1$, $2$, and $4$
\cite{1982PerrotTemperatureDependent}.  The lower panel shows his
reported $k_{\rm LDF}/k_{\rm TF}$ values directly, while the
corresponding upper-panel radii are obtained from the same construction,
$R_{\rm LDF}=2/k_{\rm LDF}$.  At $r_s=1$, the Hulth\'en and
hydrogenic QM-VPA radii are already close to the nonlinear LDF result,
whereas the Yukawa cloud is somewhat more extended.  The separation
increases with $r_s$.  At $r_s=4$, Perrot's LDF result gives
$k_{\rm LDF}/k_{\rm TF}=1.904$, whereas the corresponding QM-VPA
values are approximately $1.62$ for the Hulth\'en/hydrogenic pair and
$1.48$ for Yukawa screening.

This growing difference relative to the self-consistent LDF result is
consistent with the broader trend already found for the contact
density.  As $r_s$ increases, exchange--correlation effects become
progressively more important in the electron gas, whereas the
prescribed one-parameter model potentials contain neither an explicit
exchange--correlation potential nor a self-consistent density
dependence.  The increasing separation from Perrot's result is
therefore consistent with the growing importance of physics absent
from the simple model-potential description.  We do not attribute the
entire difference uniquely to exchange--correlation, since the QM-VPA
and LDF calculations also differ more generally in their construction
and degree of self-consistency.  Perrot's LDF result is consequently
used here as a nonlinear self-consistent comparison rather than as an
exact many-body benchmark.

The screening radius and the contact density characterize different
aspects of the same induced electron cloud.  The former is determined
by an accumulated, $r^2$-weighted charge criterion and therefore
measures its spatial extent, whereas the contact density probes the
strict short-range limit.  The contact Hartree potential considered
next provides a third and more directly electrostatic measure:
according to the exact identity in Eq.~\eqref{eq:VH0_r},
$V_{\rm H}(0)$ is the first radial moment of the full induced density.
It therefore combines information from the near-field accumulation,
the redistribution between bound and continuum states, and the
extended screening cloud in a single density-derived quantity.

\subsection{VPA for model potentials: Quantal density-derived contact Hartree potential}
\label{sec:vpa_QM_Vh0}

We finally turn to the density-derived contact Hartree potential of the
model-potential screening clouds.  As defined in
Eq.~\eqref{eq:vpa_VH0_definition}, this quantity is obtained from the
complete QM-VPA induced density generated by the occupied bound and
continuum states of the prescribed screened interaction.  It is therefore
distinct from the Poisson-based contact Hartree potential introduced in
Sec.~\ref{sec:model_potentials}.  The distinction is precisely that
between the two constructions summarized in
Eqs.~\eqref{eq:poisson_model_chain} and
\eqref{eq:quantal_model_chain}: the regular short-range coefficient of an
analytic screened potential is the contact Hartree potential of the
Poisson density associated with that potential, whereas
$V_{\rm H}^{\rm QM--VPA}(0)$ is the first radial moment of the
independently generated quantum density.  No equality between these two
quantities is assumed.  We first examine potentials constructed using
global charge screening and, for Apagyi--Nagy, a cusp imposed on an auxiliary Poisson-derived density,
and then contrast them with a qualitatively different class of
two-parameter potentials whose construction also incorporates feedback
from the density generated by the trial potential.
\begin{figure}[htbp!]
\centering
\includegraphics[width=\columnwidth]
{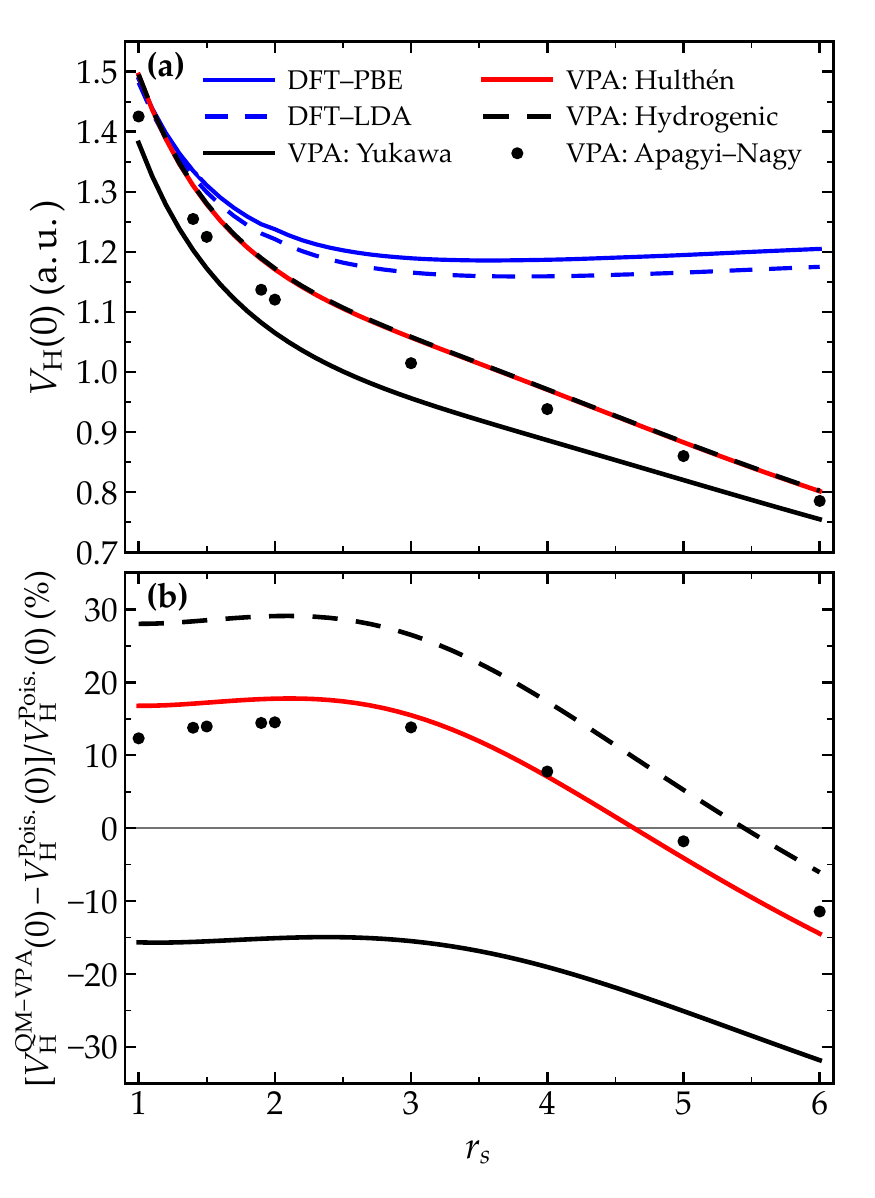}
\caption{
Density-derived contact Hartree potential for a proton in the
zero-temperature electron gas.
\textbf{(a)} Total
$V_{\rm H}^{\rm QM--VPA}(0)$ obtained from the quantal induced density
of the FSR-constrained Yukawa, Hulth\'en, and hydrogenic potentials over
the 51-state grid $r_s=1.0,1.1,\ldots,6.0$, together with the nine
Friedel/cusp-constrained Apagyi--Nagy states of
Table~\ref{tab:AN_FSR_parameters}.  Self-consistent DFT-LDA and
DFT-PBE results are shown as nonlinear reference calculations.  The
Hulth\'en and hydrogenic QM-VPA results are nearly coincident over the
full interval, whereas the Yukawa result is systematically smaller;
the separation between all prescribed model-potential results and the
DFT references increases toward larger $r_s$.
\textbf{(b)} Signed relative difference between the density-derived
QM-VPA contact Hartree potential and the corresponding Poisson-based
contact $V_{\rm H}^{\rm Poisson}(0)$ associated with the same analytic
potential, as defined in Sec.~\ref{sec:model_potentials} and summarized
in Table~\ref{tab:poisson_model_densities}.  Positive values indicate
that the QM-VPA contact exceeds the Poisson-based value, and negative
values indicate the opposite ordering.  For the Apagyi--Nagy model,
$V_{\rm H}^{\rm Poisson}(0)$ is calculated here from its analytic
Poisson-density ansatz through Eq.~\eqref{eq:VH0_poisson_AN}; it is not
a contact-Hartree quantity reported in the original Apagyi--Nagy work.
The strongly model- and density-dependent differences demonstrate that
the regular short-range coefficient of a Friedel-constrained screened
potential cannot in general be identified with the contact Hartree
potential of the quantal density generated by that potential.
}
\label{fig:vpa_vh0_allmodels}
\end{figure}
\subsubsection{FSR- and cusp-constrained model potentials: nonuniqueness of the density-derived contact Hartree potential}
\label{sec:FSR_cup_for_vh0}

Figure~\ref{fig:vpa_vh0_allmodels}(a) shows
$V_{\rm H}^{\rm QM--VPA}(0)$ for the FSR-constrained Yukawa,
Hulth\'en, and hydrogenic potentials over the complete 51-state grid
$1\leq r_s\leq6$.  The nine Friedel/cusp-constrained
Apagyi--Nagy potentials defined in
Sec.~\ref{sec:two_parameter_fsr} and
Table~\ref{tab:AN_FSR_parameters} are included as discrete points.
The self-consistent DFT-LDA and DFT-PBE results are shown on the same
scale for nonlinear comparison.  As in Sec.~\ref{sec:LRT_vs_DFT},
these DFT calculations are used as approximate nonlinear reference
calculations and not as exact many-body benchmarks.

A striking feature of Fig.~\ref{fig:vpa_vh0_allmodels}(a) is the
near coincidence of the Hulth\'en and hydrogenic results over the
entire density interval.  Their maximum absolute separation over the
51-state grid is only approximately $1.5\times10^{-3}$~a.u., and their
maximum symmetric relative difference is approximately $0.13\%$.
This extends the close Hulth\'en--hydrogenic correspondence already
found for the finite-radius induced density, the contact density, and
the density-derived screening radius.  The agreement should not,
however, be interpreted as a consequence of the Friedel sum rule alone:
the Yukawa result remains clearly separated from this pair despite
satisfying the same unit-charge Friedel constraint.

At the high-density end, the Hulth\'en and hydrogenic contact Hartree
potentials are already close to the nonlinear DFT values.  At
$r_s=1$, they are approximately $1.495$~a.u., compared with
$1.482$ and $1.488$~a.u. from DFT-LDA and DFT-PBE, respectively.
The Yukawa result is smaller, approximately $1.382$~a.u.  The
separation from the DFT calculations then grows systematically as the
electron gas becomes more dilute.  At $r_s=2$, the Hulth\'en and
hydrogenic results are approximately $4\%$ below DFT-LDA, whereas the
Yukawa value is approximately $13\%$ lower.  By $r_s=4$, the
corresponding deviations are approximately $16\%$ for the
Hulth\'en/hydrogenic pair and $24\%$ for Yukawa.  At $r_s=6$, the
Hulth\'en and hydrogenic values have decreased to approximately
$0.801$ and $0.802$~a.u., and the Yukawa result to approximately
$0.755$~a.u., compared with $1.175$ and $1.205$~a.u. from DFT-LDA
and DFT-PBE.

The increasing separation toward larger $r_s$ is consistent with the
trend found above for the contact density and the screening-cloud
radius.  The prescribed FSR-constrained model potentials do not contain
an explicit exchange--correlation potential and are not updated
self-consistently from the density that their quantum states generate.
Both differences become increasingly important when comparing with a
self-consistent DFT calculation as the electron gas is diluted.  We do
not attribute the full discrepancy uniquely to exchange--correlation,
since the two approaches differ more generally in their construction
and degree of self-consistency.  The comparison instead shows directly
that enforcing the correct Friedel charge on a prescribed scattering
potential is not sufficient to reproduce the nonlinear DFT contact
Hartree response quantitatively over the full density interval.

The nine Apagyi--Nagy points provide a complementary test involving an
additional exact short-range constraint.  These calculations use the
representative $\alpha=0.7\,$ branch specified in
Sec.~\ref{sec:two_parameter_fsr}; they should therefore be regarded as
one selected sequence from the continuous Friedel/cusp-constrained
family rather than as a unique two-parameter solution.  Their
QM-VPA contact Hartree potentials follow the same overall decrease with
$r_s$ as the one-parameter models, ranging from approximately
$1.425$~a.u. at $r_s=1$ to $0.785$~a.u. at $r_s=6$.  Imposing the
electron--proton cusp on the auxiliary Apagyi--Nagy Poisson-density ansatz in addition to the FSR thus does not
force the quantum-density-derived contact Hartree potential onto the
self-consistent DFT curves.  This is consistent with the roles of the
two constraints established above: the FSR fixes the scalar
degeneracy-weighted phase-shift combination in
Eq.~\eqref{eq:FSR_global_constraint}, while the cusp condition constrains
the local derivative of the density relative to its value.  Neither
condition specifies the complete radial density whose first moment
determines $V_{\rm H}(0)$.

The distinction between the Poisson and quantal constructions is made
explicit in Fig.~\ref{fig:vpa_vh0_allmodels}(b).  For each potential,
the figure shows the signed relative difference between
$V_{\rm H}^{\rm QM--VPA}(0)$ and the corresponding Poisson-based
contact Hartree potential summarized in
Table~\ref{tab:poisson_model_densities}.  The result is both substantial
and strongly model dependent.  For Yukawa screening,
$V_{\rm H}^{\rm QM--VPA}(0)$ remains below its Poisson-based counterpart
throughout the interval, with the difference changing from approximately
$-16\%$ at $r_s=1$ to approximately $-32\%$ at $r_s=6$.  The
Hulth\'en result instead begins approximately $17\%$ above its
Poisson value, reaches a maximum difference of about $18\%$ near
$r_s\simeq2$, and changes sign near $r_s\simeq4.6$--$4.7$ before
reaching approximately $-14\%$ at $r_s=6$.  The hydrogenic discrepancy
is larger at high density, approximately $28$--$29\%$ for
$r_s\simeq1$--$2$, and changes sign only near
$r_s\simeq5.4$--$5.5$.

The Apagyi--Nagy points exhibit the same qualitative lesson.  Here the
Poisson-based contact is calculated in the present work from the
analytic density ansatz through Eq.~\eqref{eq:VH0_poisson_AN}; it is
not a contact-Hartree quantity reported by Apagyi and Nagy.  For the
same nine prescribed Friedel/cusp-constrained potentials, the QM-VPA
result exceeds the Poisson value by approximately $12\%$ at $r_s=1$
and $15\%$ at $r_s=2$, the difference decreases to about $8\%$ at
$r_s=4$, and its sign reverses between $r_s=4$ and $5$, reaching
approximately $-11\%$ at $r_s=6$.  Thus the difference between the
Poisson and quantal constructions cannot be represented by a universal
or even a model-independent multiplicative correction.

The Hulth\'en--hydrogenic comparison makes this point particularly
clear. Their quantal contact Hartree potentials are nearly
indistinguishable, yet their deviations from their respective
Poisson-based contacts differ by more than ten percentage points over
much of the interval.  Conversely, the common FSR normalization does
not bring the Yukawa result onto the same density-derived contact.
The regular short-range coefficient of an analytic screened potential
therefore contains definite information about the Poisson density
associated with that analytic form, as established by
Eqs.~\eqref{eq:VH0_poisson_general} and
\eqref{eq:model_short_range_poisson}, but it does not determine the
contact Hartree potential of the quantum density generated when the
same potential is used as a scattering Hamiltonian.

The total $V_{\rm H}^{\rm QM--VPA}(0)$ curves also remain smooth as
bound-state weight emerges in the dilute part of the density interval.
This behavior is consistent with the bound--continuum compensation
already observed for the radial and contact densities and with the
decomposition in Eq.~\eqref{eq:vpa_VH0_decomposition}: formation of a
discrete state redistributes spectral weight between the bound and
continuum sectors rather than adding an independent contribution to an
otherwise unchanged screening cloud.  For this reason, the principal
comparison in Fig.~\ref{fig:vpa_vh0_allmodels} is made using the total
density-derived Hartree contact.

Taken together with the contact-density and screening-radius results of
the preceding subsection, Fig.~\ref{fig:vpa_vh0_allmodels} clarifies
the information content of the model-potential constraints.  The
Friedel sum rule fixes the total displaced charge but leaves substantial
freedom in the radial distribution of that charge.  The exact cusp imposed on the auxiliary density in the Apagyi--Nagy construction
adds local information about that ansatz, but does not determine the contact
amplitude or full screening cloud of the reconstructed quantum density.  The contact density,
$R_{\rm VPA}$, and $V_{\rm H}(0)$ probe, respectively, the strict
short-range amplitude, an accumulated $r^2$-weighted measure of spatial
extent, and the $r$-weighted electrostatic moment of the complete
density.  Their combined behavior therefore shows why agreement with
the FSR---even when supplemented by the exact cusp condition---does not
uniquely determine the nonlinear density-derived contact Hartree
potential.  This nonuniqueness should not, however, be interpreted as
an intrinsic failure of a reduced scattering-potential description.
Rather, it identifies the information that is missing when the model
potential is fixed only by global charge screening and local
short-range constraints.  The next subsection tests a different
two-parameter construction in which the generated density itself enters
the determination of the effective potential.

\subsubsection{Density-informed approximately self-consistent model potentials: recovery of the contact Hartree response}
\label{sec:Popovic_potential_vh0}

The preceding results raise a more constructive question: can a compact
screened potential reproduce the density-derived contact Hartree
response if information from the density generated by that potential is
used in fixing its parameters?  A particularly useful historical test
is provided by the nonlinear proton-screening calculation of
Popovi\'c \textit{et al.}~\cite{1974PopovicNonlinearSelfConsistentTheory, 1976PopovicTheoryHeatSolution}. Their starting point was the density-functional single-particle problem in which the effective potential was written as the sum of the total electrostatic potential and a local exchange--correlation contribution. They considered both a nonlinear Hartree approximation, obtained by omitting the exchange--correlation term, and a nonlinear calculation
including a local Hedin--Lundqvist exchange--correlation potential.

A direct one-parameter self-consistency iteration was found not to
converge while maintaining the Friedel condition.  Popovi\'c
\textit{et al.} therefore introduced a flexible two-parameter trial
potential and determined its parameters so that the Friedel sum rule was
satisfied not only by the trial potential itself but also by the new
effective potential regenerated from the induced density produced by
that trial potential.  The latter step is the essential distinction
from the FSR-only and FSR/cusp constructions considered above: the
generated density enters explicitly through the density-to-potential
feedback used to impose the second Friedel condition.  The resulting
scheme was described by Popovi\'c \textit{et al.} as an approximate
self-consistent solution.

For a physical proton, their selected trial potential can be written
\begin{align}
V_{\rm Pop}(r;\alpha,\beta)
&=
-\frac{1}{r}\exp\!\left(-\alpha r^{\beta}\right).
\label{eq:Popovic_trial_potential}
\end{align}
This form also belongs to the class of scattering potentials admissible
for the VPA construction of Sec.~\ref{sec:vpa_phase}.  For
$\alpha>0$ and $\beta>0$, its short-distance expansion is
\begin{align}
V_{\rm Pop}(r;\alpha,\beta)
&=
-\frac{1}{r}
+\alpha r^{\beta-1}
+\mathcal{O}\!\left(r^{2\beta-1}\right),
\qquad r\rightarrow0,
\label{eq:Popovic_short_range}
\end{align}
so the leading singularity is Coulombic, corresponding to $m=1<2$ in
Eq.~\eqref{eq:origin_regular_condition}.  If $\beta<1$, the next term
is itself singular but remains weaker than $1/r$; if $\beta>1$, it
vanishes at the origin.  At large distance,
$rV_{\rm Pop}(r)=-\exp(-\alpha r^\beta)\rightarrow0$, so the potential
is short ranged and satisfies the asymptotic screening requirement as
well.  Thus the same bound-state and continuum VPA machinery used for
the screened Coulomb potentials above can be applied without altering
the scattering boundary conditions.

Popovi\'c \textit{et al.} tabulated the electron--proton interaction
quantity $V_{\rm int}/Z_p$ for the Al and Mg electron-gas densities.
With the positive-magnitude convention for $V_{\rm H}(0)$ used in the
present work, their quantity is related to the contact Hartree potential
by
\begin{align}
V_{\rm H}(0)
&=
-\frac{V_{\rm int}}{Z_p},
\label{eq:Popovic_Vint_to_VH}
\end{align}
because $V_{\rm int}=-Z_p\int d^3r\,\Delta n(r)/r$. For the physical
proton, $Z_p=1$, their nonlinear Hartree calculation gives
$V_{\rm H}(0)=1.016$~a.u. at the Al density
$r_s=2.064$ and $0.937$~a.u. at the Mg density
$r_s=2.642$.  When the local exchange--correlation potential is
included in their approximate self-consistency construction, the
corresponding Hartree contacts of the exchange--correlation-modified
densities become $1.186$ and $1.157$~a.u., respectively.  The latter
numbers remain Hartree contacts: the exchange--correlation potential is
not added to $V_{\rm H}(0)$, but instead modifies the density from which
the Hartree moment is evaluated.

We tested both constructions using the published Popovi\'c parameters
directly, without refitting them or re-solving the FSR.  For the
Hartree-only branch the tabulated parameters are
$(\alpha,\beta)=(1.273,0.954)$ for Al and
$(1.197,0.943)$ for Mg.  Propagating these fixed potentials through the
same zero-temperature QM-VPA density pipeline used above gives
$V_{\rm H}^{\rm QM--VPA}(0)=1.015556$~a.u. for Al and
$0.937263$~a.u. for Mg, differing from the historical values by only
$-0.0437\%$ and $+0.0281\%$, respectively.  The calculation also
reproduces the qualitative bound-state structure reported for the
Hartree approximation: no bound state is found at the Al density,
whereas a very shallow doubly occupied $s$ state is present at the Mg
density.

An even more stringent reconstruction is obtained from the
exchange--correlation branch.  Using the corresponding published
parameters $(1.153,1.171)$ for Al and $(1.063,1.237)$ for Mg, while
introducing no explicit exchange--correlation functional into the
present VPA calculation, we obtain
$V_{\rm H}^{\rm QM--VPA}(0)=1.185735$ and $1.156837$~a.u.  These values
differ from the Popovi\'c results $1.186$ and $1.157$~a.u. by only
$-0.0223\%$ and $-0.0140\%$, respectively, well within the precision
of the historical tabulation.  The XC-branch potentials also generate
a shallow doubly occupied $s$ state at both densities, in qualitative
agreement with the bound-state pattern reported by Popovi\'c
\textit{et al.}

It is important to distinguish this reconstruction from an independent
exchange--correlation calculation.  The present QM-VPA pipeline contains
no exchange--correlation functional.  In the Hartree branch, the
agreement shows that the density-informed two-parameter trial potential
faithfully reproduces the contact moment of the approximately
self-consistent Hartree screening cloud.  In the exchange--correlation
branch, the exchange--correlation physics enters only implicitly
through the published values of $\alpha$ and $\beta$, which were
obtained from Popovi\'c \textit{et al.}'s exchange--correlation-inclusive
approximate self-consistency construction.  The resulting change of the
reconstructed contact, approximately $16.8\%$ for Al and $23.4\%$ for
Mg relative to the Hartree branch, therefore demonstrates how strongly
the density consequences of exchange and correlation can be encoded in
the effective potential even when no explicit $V_{\rm xc}$ is evaluated
during the subsequent VPA reconstruction.

These Popovi\'c tests are not independent many-body benchmarks in the
same sense as the DFT-LDA/PBE comparisons in
Fig.~\ref{fig:vpa_vh0_allmodels}: their tabulated interaction energies
were obtained from the displaced densities entering the same
approximately self-consistent construction from which the trial
potentials were determined.  Their significance is instead
methodological.  They show that a compact analytic potential can carry
the density information required to reproduce the nonlocal
$r$-weighted contact Hartree moment when its parameters have been
constrained using feedback from the generated density.  This conclusion
also shows that the improvement is not attributable simply to the use of
two adjustable parameters.  The Apagyi--Nagy potential is likewise a
two-parameter construction, but its FSR and auxiliary-density cusp constraints do not
uniquely determine the quantum-density-derived $V_{\rm H}(0)$.  In the Popovi\'c construction the
second condition is density informed: the density generated by the
trial potential is used to reconstruct a new effective potential, and
the Friedel condition is imposed on that regenerated potential as well.

The combined results therefore sharpen the central message of this
section.  The FSR fixes the total displaced charge, and the cusp
condition adds exact local derivative information (imposed on an auxiliary density in the Apagyi--Nagy construction), but neither determines the full
quantum screening density or its first radial moment.  A reduced scattering
potential can nevertheless reproduce the density-derived contact
Hartree response with high fidelity when information from the
approximately self-consistent density--potential feedback is encoded in
its parameters.  For contact-Hartree modeling, the decisive issue is
therefore not simply the number of parameters in the screened
interaction, but the physical information used to determine them.

\section{Discussion}
\label{sec:discussion}

\subsection{Bound states}
\label{subsec:bound_states_disc}

Bound states enter the present screening problem differently depending on which contact-Hartree construction is considered. For the exact $T=0$ Friedel constraint used to determine the parameters of a screened model potential, they do not require separate bookkeeping. As shown in Sec.~\ref{sec:FSR}, the chemical potential is pinned at $\mu=E_F>0$, all negative-energy bound levels are fully occupied, and Levinson's theorem reduces the combined bound-plus-continuum charge accounting to the total zero-temperature FSR evaluated at $k_{\rm F}$. Consequently, the FSR-constrained parameters entering the Poisson-based contact Hartree potential $V_{{\rm H},\nu}^{\rm Poisson}(0)$ of Eqs.~\eqref{eq:VH0_poisson_general} and \eqref{eq:model_short_range_poisson} can be determined without explicitly constructing the bound-state density.

The situation is different for the quantal density-derived contact Hartree potential. Once the model potential is treated as a one-electron scattering Hamiltonian, the full induced density contains both continuum and discrete spectral contributions. An occupied bound state therefore contributes through its spatial wavefunction to $\Delta n(r)$ and hence to the radial moment defining $V_{\rm H}^{\rm QM--VPA}(0)$. Levinson's theorem guarantees the appropriate global charge accounting, but it does not determine how the bound charge is distributed in space. The explicit treatment of bound states is therefore required here for internal consistency of the quantal density reconstruction, but the existence and properties of those states remain dependent on the particular spherical jellium or screened-potential Hamiltonian being used and should not be overinterpreted as evidence for corresponding localized impurity states in a real crystalline metal.

The absence of explicit bound-state dependence in determining the FSR-constrained model parameters is a special feature of the exact $T=0$ limit. In this limit, the occupied bound charge equals the bound-state capacity and the generalized finite-temperature FSR of Eq.~\eqref{eq:FSR_finiteT} reduces to the usual zero-temperature Friedel sum. At finite temperature this cancellation no longer occurs in general: the bound-state energies and their Fermi--Dirac occupations enter the generalized FSR explicitly. The screening parameters obtained from that relation, and therefore even the associated Poisson-based contact Hartree potential, can then depend on the discrete bound-state spectrum and its thermal occupation. A finite-temperature quantal density reconstruction would require, in addition, the corresponding thermally weighted bound-state contributions to $\Delta n(r)$.

The seminal early DFT calculations of a proton in jellium in the metallic-density range $\rs\in[2,6]$
\cite{1976AlmbladhScreeningProtonElectron,1977Zaremba,1978JenaElectronicStructureHydrogen}
find a shallow, doubly occupied, $s$-like bound state. More recent spin-polarized DFT and diffusion Monte Carlo calculations by Takada and co-workers
\cite{2015TakadaEmergenceKondoSinglet}
place the emergence of an isolated $H^-$ ion at much lower electron density, around $\rs\sim12.5$. Their energy-based argument is instructive: formation of a neutral H-like configuration becomes energetically plausible when $E_F=1.84/\rs^2$ becomes comparable with the hydrogen $1s$ binding energy, whereas formation of an isolated $H^-$ requires comparison with the much smaller electron affinity.

Estreicher--Meier
\cite{1983_Estreicher-Meier}
fitted the total $\Delta n(r)$ from the available LDA calculations, so their parametrization already contains the effects of the bound-state contribution present in those calculations. Their Figure~1 shows a maximum binding magnitude of approximately $|\epsilon_b|\simeq0.012$ around $\rs\simeq4.9$. An effective spatial scale for such a shallow state is
$a_b\sim1/k_b=1/\sqrt{2|\epsilon_b|}\simeq6.5\,a_0$,
which is comparable to the lattice constant of BCC lithium
\cite{2015Papaconstantopoulos}.
As noted by Meier
\cite{1975MeierElectronDensitiesCharged},
such spatially extended bound states expose an important limitation of the jellium description of a real metal. For a genuinely localized impurity state, its spatial extent must be considered relative to the positions of the host ions rather than only with respect to a homogeneous positive background. Jena and Singwi
\cite{1978JenaElectronicStructureHydrogen}
similarly argued that coupling of these shallow states to the ionic cores of the lattice, together with finite quasiparticle-lifetime broadening, could destabilize them, so that the corresponding jellium states may be more appropriately interpreted as resonant states.

Even within the jellium framework, however, the bound-state spectrum is model dependent: the existence of a discrete state, its threshold in $\rs$, its binding energy, and its spatial extent depend on the effective electron--proton interaction used to describe screening. The bound states obtained here should therefore be understood as spectral properties of the particular self-consistent or screened-potential Hamiltonians considered, rather than as universal properties of a proton in jellium. A further level of uncertainty arises when one asks whether such a state survives as a localized electronic state in a particular metal, where the ionic lattice and electronic band structure must be treated explicitly. Within each model, however, any occupied bound state must be included consistently in the quantal density and contact-Hartree reconstruction.

\subsection{The utility and limitations of constrained model-potential methods}
\label{sec:WHyVPA}

The present calculations show that constrained model potentials can perform remarkably well in an appropriate density regime. Figures~\ref{fig:vpa_dn0_allmodels} and \ref{fig:vpa_vh0_allmodels} show that, toward the high-density end of the interval studied here, the hydrogenic and Hulth\'en potentials in particular approach the nonlinear DFT-LDA and DFT-PBE results for both the contact density and the density-derived contact Hartree potential. For roughly $\rs\lesssim1.5$, the agreement is already close despite the simplicity of these one-parameter interactions and the absence of an explicitly self-consistent exchange--correlation potential in their construction. This high-density agreement should not, however, be interpreted as demonstrating that exchange and correlation can simply be neglected at the origin. Rather, it shows that in the increasingly high-density, weak-coupling regime, rather simple constrained interactions can reproduce the contact observables with progressively greater fidelity.

The situation changes systematically as $\rs$ increases. The separation between the prescribed model-potential results and the nonlinear DFT references grows for both $\Delta n(0)$ and $V_{\rm H}(0)$. This behavior exposes the limited information content of the Friedel constraint itself. The FSR fixes the total displaced screening charge, but it does not prescribe how that charge is distributed in radius. Potentials satisfying the same total-screening condition can therefore generate different contact amplitudes, different intermediate-range density profiles, and different values of the $r$-weighted Hartree moment. The increasing disagreement toward lower electron density is thus not surprising once the contact observables are reconstructed from the full quantal density rather than inferred from the analytic screening parameter.

The two-parameter Apagyi--Nagy construction provides a particularly useful test of whether adding exact short-range information to potential construction removes this ambiguity. Its cusp constraint is imposed on the auxiliary total-density ansatz underlying the Poisson construction and, together with the FSR, restricts the potential parameters. This should not be confused with the exact cusp of the independently reconstructed QM--VPA density: all of the Coulomb-singular model potentials possess that property automatically. For the FSR-only Yukawa, hydrogenic, and Hulth\'en models, Appendix~\ref{app:qm_vpa_cusp} verifies this numerically at $r_s=2$, with relative derivative deviations below $0.006\%$. Nevertheless, the QM--VPA results in Figs.~\ref{fig:vpa_dn0_allmodels} and \ref{fig:vpa_vh0_allmodels} do not collapse onto the nonlinear DFT curves; the Apagyi--Nagy contact-Hartree values can even lie below those of some one-parameter models. Thus an extra exact constraint in the analytic construction need not improve the density-derived contact response monotonically. The FSR constrains total screening, whereas the cusp constrains a local derivative relative to its contact amplitude; neither determines the complete radial screening cloud.

The density-informed two-parameter potentials of Popovi\'c and Stott and their subsequent extension by Popovi\'c \textit{et al.}~
\cite{1974PopovicNonlinearSelfConsistentTheory, 1976PopovicTheoryHeatSolution}
provide the complementary case. In their construction, the density generated by the trial potential enters the approximate density--potential feedback used to determine the potential parameters. When their published Hartree-only and exchange--correlation-informed parameters are propagated through the present QM--VPA density/Hartree pipeline without refitting, the reconstructed contact Hartree potentials reproduce the corresponding historical values at the few-$10^{-4}$ relative level. As emphasized in Sec.~\ref{sec:Popovic_potential_vh0}, this is not an independent many-body benchmark, since the historical interaction energies and the trial potentials arise from the same approximately self-consistent construction. Its significance is methodological: a compact analytic potential can encode enough information about the generated density to recover the nonlocal Hartree moment with high fidelity when that information has entered the determination of its parameters.

This comparison also provides an important validation of the present VPA implementation. The Popovi\'c--Stott parameters are taken directly from the historical calculation and are not refitted to the present density or Hartree results. The agreement therefore tests the complete reconstruction chain from a prescribed effective potential through its bound and continuum states, the resulting quantal density, and finally the radial Hartree moment. It further shows that the decisive issue is not simply whether a screened potential has one or two free parameters, but what physical information has been used to determine those parameters.

A complementary indication of why the reduced-potential description becomes more demanding toward lower electron density is provided by the spatial extent of the screening cloud itself. Figure~\ref{fig:vpa_rvpa_allpots} shows that the density-derived screening radius $R_{\rm VPA}$ increases systematically with $\rs$, so that an increasingly extended region of the induced density participates in the screening response. Moreover, the lower panel of Fig.~\ref{fig:VH0_DFT_vs_linear} shows that the Friedel component of the Estreicher--Meier DFT-LDA density makes an increasingly important contribution to the contact Hartree moment, rising from approximately $0.034$~a.u. at $\rs=2$ to approximately $0.144$~a.u. at $\rs=6$. Although the analytic screened potentials considered here are themselves monotonic and nonoscillatory, their occupied scattering states generate Friedel oscillations through the Fermi-surface structure of the continuum. The difficulty is therefore not the absence of Friedel oscillations from the QM--VPA density, but the limited freedom of a rigid, finitely constrained model potential to reproduce their self-consistent phase, amplitude, and accompanying long-range redistribution of charge as the screening cloud becomes more extended.

A particularly revealing historical perspective is provided by Estreicher and Meier~\cite{1983_Estreicher-Meier}. In Sec.~II of their work, they revisited the two-parameter Popovi\'c--Stott trial potential over the metallic-density range. They found that the construction gave ``surprisingly good results'' at high electron density, including those densities representing Al and Mg metals, but deteriorated as $\rs$ increased. They attributed this deterioration to the absence of repeated exchange--correlation feedback and, at lower densities, to the increasing screening length, which made the prescribed trial form a progressively poorer approximation to the self-consistent effective potential. Their Fig.~2 illustrates the consequence at $\rs=5.63$: the approximate density substantially overestimates the contact density and develops Friedel oscillations that are out of phase with the fully self-consistent result. The same work further notes that the influence of exchange and correlation becomes more pronounced as the density decreases, making full self-consistency progressively more demanding.

The parallel with the present results is noteworthy. Estreicher and Meier diagnosed the limitation through the radial density and contact density, whereas the present calculation extends the test to the fully quantal density-derived contact Hartree potential. Their Fig.~2 does not by itself establish the sign or magnitude of the error in $V_{\rm H}(0)$, because the latter is an $r$-weighted moment of the entire density, including the oscillatory region. It does, however, provide an independent historical demonstration of the same underlying limitation: a constrained analytic potential may reproduce the self-consistent solution very well at high density while becoming increasingly inadequate as the detailed spatial structure of the nonlinear screening cloud acquires greater importance.

None of this constitutes a limitation of the variable-phase approach itself. Within a prescribed central-potential problem, VPA is the scattering solver; the approximation enters earlier, when the self-consistent Kohn--Sham effective potential is replaced by a prescribed analytic form with a finite number of adjustable parameters. The quantitative limitations identified here should therefore be attributed to the restricted information encoded in the constrained model potentials, rather than to the VPA used to solve their scattering problem. For a specified central potential, the VPA provides the scattering phase shifts directly and, through the accompanying variable-amplitude equation used here, reconstructs the continuum wavefunctions required for the quantal induced density. The present phase-plus-amplitude formulation therefore remains useful precisely because it allows the consequences of a chosen effective interaction and of the physical constraints imposed upon it to be examined at the level of phase shifts, bound states, the full quantal induced density, and the resulting contact Hartree response.

\section{Conclusion}
\label{sec:conclusion}
In this work we have examined the nonlinear static screening of a positive charge in a homogeneous electron gas, with particular emphasis on the contact density $\Delta n(0)$ and the density-derived contact Hartree potential $V_{\rm H}(0)$. Although evaluated at the impurity position, $V_{\rm H}(0)$ is not a strictly local observable but an $r$-weighted moment of the full induced density. The calculations confirm that Thomas--Fermi, RPA, and local-field-corrected linear response substantially underestimate the nonlinear contact response of a proton throughout the metallic-density regime when compared with the present DFT-LDA and DFT-PBE reference calculations and with established self-consistent results. The Estreicher--Meier parametrization provides a compact representation of the DFT-LDA screening density and reproduces the corresponding contact Hartree potential closely, while also making explicit the non-negligible contribution of the Friedel-oscillatory tail to the Hartree moment. Its approximate induced-density cusp should be distinguished from the exact total-density cusp; the resulting local slope deficit alone does not quantify any error in the nonlocal Hartree moment (Appendix~\ref{sec:EM_cusp}). The difference between the LDA and PBE results is comparatively modest on the scale of the nonlinear enhancement over linear response. Likewise, for the two ground-state local-field-factor parametrizations examined here, Kaplan--Kukkonen and Corradini--Del Sole--Onida--Palummo, the differences in $\Delta n(0)$ and $V_{\rm H}(0)$ are negligible at the scale of the present comparisons. This latter observation is specific to these two parametrizations and the density range considered, and should not be interpreted as a general insensitivity of contact screening to local-field corrections.

Complementing these DFT and linear-response benchmarks, we used the phase-plus-amplitude VPA to reconstruct the full quantal density generated by prescribed screened model potentials and thereby distinguish the density-derived contact Hartree potential from the regular short-range coefficient of the analytic interaction. The Friedel-constrained Yukawa, hydrogenic, and Hulth\'en potentials satisfy the same global total-screening condition but generate different $\Delta n(0)$ and $V_{\rm H}(0)$, demonstrating that the FSR does not uniquely determine the contact response. The two-parameter Apagyi--Nagy construction strengthens this test: its additional exact cusp condition is imposed on an auxiliary total-density ansatz used to construct the potential via Poisson's equation, not on the independently reconstructed quantum density. Indeed, the Coulomb singularity of each screened potential ensures the exact quantum-density cusp automatically; this is verified numerically at $r_s=2$ for the three FSR-only models, with relative slope deviations below $0.006\%$ (Appendix~\ref{app:qm_vpa_cusp}). The additional Apagyi--Nagy construction constraint nevertheless does not determine the complete radial screening cloud or its contact Hartree moment. By contrast, when the published density-informed Popovi\'c--Stott potentials are propagated through the present QM--VPA density/Hartree reconstruction without refitting, the resulting Hartree-only and exchange--correlation-informed contact potentials reproduce the corresponding historical values at the few-$10^{-4}$ relative level. This comparison shows that constraints imposed on a reduced potential and observables reconstructed from that potential are different levels of information: the FSR constrains a global scattering quantity and the cusp constrains local short-range behavior, whereas accurate recovery of $V_{\rm H}(0)$ requires sufficient information about the full generated screening density.

The performance of the constrained model potentials is strongly density dependent. Toward the high-density end of the interval studied here, particularly for the hydrogenic and Hulth\'en forms, the QM--VPA contact density and density-derived contact Hartree potential approach the nonlinear DFT-LDA and DFT-PBE results, with already close agreement for roughly $\rs\lesssim1.5$. As $\rs$ increases, however, the separation from the DFT references grows. This behavior accompanies an increasing spatial extent of the screening cloud: the density-derived radius $R_{\rm VPA}$ increases systematically with $\rs$, while the Friedel contribution to the Estreicher--Meier contact Hartree moment grows from approximately $0.034$~a.u. at $\rs=2$ to approximately $0.144$~a.u. at $\rs=6$. Thus, toward lower electron density, an increasingly extended and structured region of the screening cloud contributes to $V_{\rm H}(0)$, making a rigid finitely constrained analytic potential progressively less able to reproduce the self-consistent response. This density dependence is consistent with the historical analysis of Estreicher and Meier, who likewise found the two-parameter Popovi\'c--Stott trial potential to perform well at high density but to deteriorate as the screening length and the importance of exchange--correlation feedback increased. These limitations belong to the prescribed model potentials, not to the variable-phase method itself: for a specified central interaction, the phase-plus-amplitude VPA provides the scattering phase shifts and normalized continuum states required to reconstruct its full quantal density and contact Hartree response.

The present results therefore establish a controlled one-center benchmark for nonlinear proton screening in jellium, including the induced density, scattering information, and the density-derived contact Hartree potential. In the context of electron-screening enhancement of low-energy nuclear reactions in metallic environments, $V_{\rm H}(0)$ provides a physically meaningful one-center measure of the electrostatic barrier lowering produced by the screening cloud of an isolated proton. It should not, however, be identified with the effective screening energy of an approaching proton pair. As the two positive centers approach one another, the electron gas undergoes a genuinely nonlinear two-center rearrangement that cannot in general be obtained by superposing two isolated screening clouds. Determining the corresponding electron-mediated proton--proton interaction therefore requires a genuine two-center calculation. The one-center quantities established here provide the baseline against which such a calculation can be formulated and tested.

\begin{acknowledgments}
The authors thank Professor Roger Falcone, Professor Gianluca Gregori, Dr. Kosta Yanev, and Mr. Kai-Jian Xiao for their feedback on this manuscript, and Mr. Wilson Wu for his support of this work. M.S. particularly thanks Prof. Giovanni Vignale for introducing the paper by Kaplan and Kukkonen on the local-field factor and for his valuable comments on the manuscript. A.X.C. particularly thanks Professor Alfred Wong for his original inspiration and encouragement to pursue this work. This research was funded by Alpha Ring International Limited.
\end{acknowledgments}

\appendix

\section{Parametrized LFC \texorpdfstring{$G_+(q)$}{G+(q)}}
\label{sec:Gplusq}

The static local-field correction or local-field factor for a uniform electron gas can be expressed in terms of the static density-density response function as \cite{2005GiulianiQuantumTheoryElectron} 
\begin{align*}
\chi^{-1}(q) = \chi_0^{-1}(q) - v_q \left[1- G_{+}(q; r_s)\right]
\end{align*}

\subsection{Parametrization due to Corradini--Del Sole--Onida--Palummo (CDOP)}

Moroni \emph{et al.} \cite{1995MoroniStaticResponseLocal} proposed a simple formula for $G_{+}(q)$ that reproduces the main quantitative features of the local-field factor obtained numerically by diffusion quantum Monte Carlo calculations in the metallic region $\rs \in [2, 10]$. In 1998, Corradini and co-workers expanded on the original formula by Moroni and co-workers and presented a more tractable analytical $G_{+}(q)$ based on Lorentzian and Gaussian functions that reads \cite{1998CorradiniAnalyticalExpressionsLocalfield}
\begin{align}
G_{+}(q; \rs) = C(\rs)Q^2+\frac{B(\rs)Q^2}{g(\rs)+Q^2}+\alpha(\rs) Q^4 e^{-\beta(\rs) Q^2}, 
\label{eq:Gq_Corradini}
\end{align}
where $Q = q/\kf$, $g(\rs)=B(\rs)/\left[A(\rs)-C(\rs)\right]$ and $\alpha(\rs) = \frac{1.5}{\rs^{1/4}}\frac{A(\rs)}{B(\rs)g(\rs)}$ and $\beta(\rs) = \frac{1.2}{B(\rs)g(\rs)}$ minimize the differences with the numerical $G(q)$ results obtained by Moroni \emph{et al.} $A(\rs)$, $B(\rs)$, and $C(\rs)$ are explicit functions of $\rs$ given by:
\begin{align}
A(\rs) & = \frac{1}{4}+\frac{\rs^2}{27\alpha_3^2}\left[2\epsilon^{\prime}_C(\rs)-\rs \epsilon^{''}_C(\rs)\right]\nonumber \\
B(\rs) & = \frac{1+a_1\rs ^{1/2}+a_2\rs^{3/2}}{3+b_1\rs^{1/2}+b_2\rs^{3/2}}\nonumber \\
C(\rs) & = -\frac{\pi \alpha_3 \rs}{2}\left[\epsilon_C(\rs)+\rs \epsilon^{\prime}_C(\rs)\right],
\label{eq:Corradini_ABC}
\end{align}
where $\alpha_3 = [4/(9\pi)]^{1/3}$ and  $\epsilon^{\prime}_C(\rs) = d\epsilon_C(\rs)/d\rs$, $\epsilon^{''}_C(\rs) = d^2\epsilon_C(\rs)/d\rs^2$ with $\epsilon_C(\rs)$ being the Perdew--Wang correlation energy per electron expressed in a.u. given by Eq.~\eqref{eq:Ecorr_PW}. The fitting parameters in the $B(\rs)$ expression are given by $a_1 = 2.15$, $a_2 = 0.435$, $b_1 = 1.57$, and $b_2 = 0.409$.

\subsection{Parametrization due to Kaplan--Kukkonen (KK)}

In a more recent paper, Kaplan and Kukkonen \cite{2023KaplanQMCconsistentStaticSpin} presented a parametrization that fits the older quantum Monte Carlo data by Moroni and co-workers \cite{1995MoroniStaticResponseLocal} better than the parametrization of Corradini and co-workers did, and it also fits closely to a set of recent data \cite{2019ChenCombinedVariationalDiagrammatic, 2021Kukkonen-Chen} satisfying exact asymptotic limits. The expression of the LFC by Kaplan--Kukkonen is given by:
\begin{align}
G_{+}(q; \rs) = & x^2 \left[A_{+}(\rs)+\alpha_{+}(\rs)x^4\right]\nonumber\\
& \times H(x^4/16; a_{3+}, a_{4+})\nonumber\\
& + \left[C(\rs)x^2+B_{+}(\rs)\right]\nonumber\\
& \times \left[1-H(x^4/16; a_{3+}, a_{4+})\right]
\label{eq:Gq_KK_main}\\
\alpha_{+}(\rs) & = a_{0+}+a_{1+}\exp(-a_{2+}\rs)
\label{eq:alpha_KK}
\end{align}
where $x = q/\kf$, $A_{+}(\rs), B_{+}(\rs), C(\rs)$ are, respectively, identical to those provided by Corradini \emph{et al.} in Eq.~\eqref{eq:Corradini_ABC}, and the smoothed step function
\begin{align}
H(y; \beta, \gamma) = \frac{(e^{\beta \gamma}-1)e^{-\beta y}}{1+(e^{\beta\gamma}-2)e^{-\beta y}}
\label{eq:Hfnc_KK}
\end{align}
represents a simple and reasonable transition from low-$q$ behavior of the QMC data to the large-$q$ asymptotics. This function is constructed to satisfy three limits: $H(0; \beta, \gamma)=1$, $H(\gamma; \beta, \gamma)=1/2$, and $H(\infty; \beta, \gamma)=0$. The $a_{i+}$ parameters are fitted to QMC data, and they are given in Table~\ref{tab:KK_parameters}.  
\begin{table}[htbp!]
\centering
\caption{Fit parameters $a_{i+}$ with $i = 0, 1, 2, 3, 4$ for the model LFCs of Eq.~\eqref{eq:Gq_KK_main}, together with the estimated parameter uncertainties.}
\label{tab:KK_parameters}
\begin{tabular}{c r@{\,$\pm$\,}l}
\toprule
Parameter & \multicolumn{2}{c}{Value} \\
\midrule
$a_{0+}$ & $-0.00451760$ & $0.002$ \\
$a_{1+}$ & $ 0.0155766$ & $0.002$ \\
$a_{2+}$ & $ 0.422624$ & $0.2$ \\
$a_{3+}$ & $ 3.516054$ & $0.5$ \\
$a_{4+}$ & $ 1.01583$ & $0.04$ \\
\bottomrule
\end{tabular}
\end{table}
%

\subsection{LFC Comparison}
\label{sec:LFC_Comparison}

\begin{figure}[htbp!]
    \centering
    \includegraphics[width=\columnwidth]{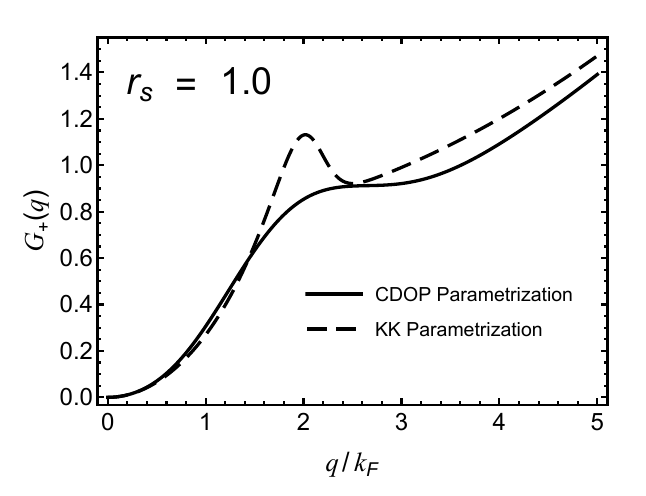}
    \caption{Static density local-field factor $G_+(q)$ of the unpolarized uniform electron gas at $r_s=1$. The solid black curve shows the Corradini--Del Sole--Onida--Palummo (CDOP) parametrization [Eq.~\eqref{eq:Gq_Corradini}], the dashed black curve shows the Kaplan--Kukkonen (KK) parametrization [Eq.~\eqref{eq:Gq_KK_main}]. The two analytic parametrizations share the same exact small-$q$ and large-$q$ constraints, but differ in their interpolation around $q\simeq2k_{\rm F}$. The KK fit exhibits a pronounced $2k_{\rm F}$ hump, whereas the CDOP form is smoother.
}
\label{fig:Gq_KK_CDOP_rs1}
\end{figure}

Figure~\ref{fig:Gq_KK_CDOP_rs1} compares three representations of the static density LFC of the spin-unpolarized electron gas $G_+(q)$ at $\rs=1$. The CDOP and KK parametrizations share the same exact small-$q$ and large-$q$ constraints, including the compressibility coefficient $A_+(r_s)$ and the large-$q$ coefficients $B_+(r_s)$ and $C(r_s)$, and both use the Perdew--Wang correlation energy [Eq.~\eqref{eq:Ecorr_PW}] as input. Their main difference is therefore not in the limiting behavior, but in the interpolation through the intermediate wave-vector regime. The CDOP form gives a smooth interpolation based on Lorentzian and Gaussian terms and was designed to reproduce the Moroni--Ceperley--Senatore ground-state QMC data \cite{1995MoroniStaticResponseLocal} while yielding an analytically Fourier-transformable exchange--correlation kernel. By contrast, the KK form uses a crossover construction fitted to more recent QMC data \cite{2021Kukkonen-Chen, 2019ChenCombinedVariationalDiagrammatic}, allowing a more pronounced structure near $q\simeq 2k_{\rm F}$. This distinction is evident in the figure: both parametrizations agree closely at small $q$, but the KK parametrization develops a clear hump around $q/k_{\rm F}\simeq2$, while the CDOP curve remains smoother.

\begin{figure}[htbp!]
    \centering
    \includegraphics[width=\columnwidth]{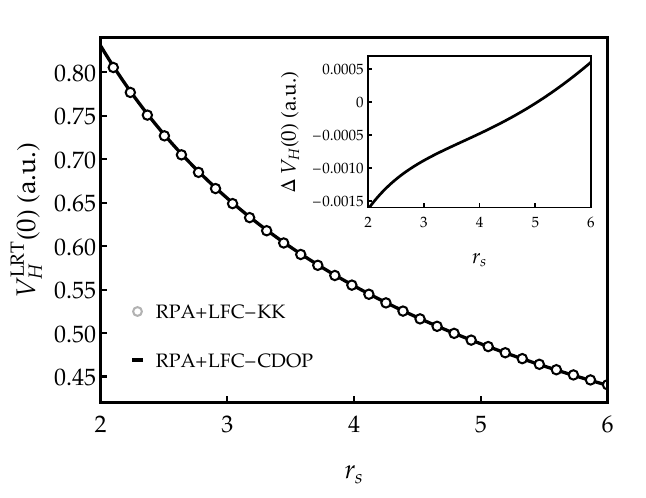}
\caption{Linear-response contact Hartree potential $V_H^{\rm LRT}(0)$ for a proton in the uniform electron gas computed within RPA+LFC using the Kaplan--Kukkonen (KK) and Corradini--Del Sole--Onida--Palummo (CDOP) density local-field factors. The two curves are visually indistinguishable on the scale of the main panel. The inset shows the resolved difference $\Delta V_{\rm H}(0)=V_{\rm H}^{\rm KK}(0)-V_{\rm H}^{\rm CDOP}(0)$, which remains below $\sim 1.65\times10^{-3}$~a.u. over $2\le r_s\le 6$. The smallness of this difference reflects the fact that the KK and CDOP parametrizations share the same exact small- and large-$q$ limits and differ mainly near the localized $q\simeq2k_{\rm F}$ crossover, while $V_{\rm H}^{\rm LRT}(0)$ is an integral over the full wave-vector range.}
\label{fig:VH0_KK_CDOP_LFC}
\end{figure}

Figure~\ref{fig:VH0_KK_CDOP_LFC} compares the contact Hartree potential $V_{\rm H}^{\rm LRT}(0)$ obtained within the test-charge dielectric formulation using the KK and CDOP density local-field factors. The two RPA+LFC curves are essentially indistinguishable over the metallic-density interval $2\le r_s\le 6$, decreasing monotonically from approximately $0.83$~a.u. to $0.44$~a.u. as the electron density is lowered. The inset resolves the small difference $\Delta V_{\rm H}(0)=V_{\rm H}^{\rm KK}(0)-V_{\rm H}^{\rm CDOP}(0)$, which remains below about $1.65\times 10^{-3}$~a.u. in magnitude over the full range. This near-equivalence is expected: the KK and CDOP parametrizations share the same exact small-$q$ and large-$q$ constraints, and their main difference is localized near the $q\simeq 2k_{\rm F}$ crossover, where the KK form exhibits a more pronounced hump. Since $V_{\rm H}^{\rm LRT}(0)$ is obtained by integrating the dielectric kernel over the entire wave-vector range, this localized difference in $G_+(q)$ is strongly averaged and produces a negligible correction to the contact Hartree potential. Thus, as for the induced density and on-top electron density, the choice between KK and CDOP does not affect the contact screening quantities at the level relevant for the present analysis.

%
\section{Estreicher--Meier's Friedel Term, \texorpdfstring{$f(2\kf r)$}{f(2kF r)}}
\label{sec:EM-FO}

Estreicher and Meier reported their parametrization of the induced
electron density around a proton in jellium in both their 1983 PRB article
\cite{1983_Estreicher-Meier} and a contemporaneous proceedings chapter
\cite{1983EstreicherParametrizationSelfConsistentlyCalculated}.
Here we follow the PRB parametrization of the Friedel-generating function
$f(x)$, where $x=2\kf r$. The function is piecewise, with each branch
expressed as a linear combination of Riccati--Bessel functions
$\widehat{j}_\ell(x)=xj_\ell(x)$, where $j_\ell$ denotes a spherical
Bessel function. The branches meet at $Z_2$, the radial position of the
second zero of $f(2\kf r)$:
\begin{align}
 Z_{2} = 1.52\,\rs + 0.462.
 \label{eq:Z2}
\end{align}
The corresponding chapter gives $Z_2=1.52\,\rs+0.464$ [its Eq.~(8)],
rather than the PRB intercept $0.462$. The published amplitude coefficients
are identical in the two sources. We retain the PRB value throughout;
the origin of the small intercept difference is not resolved here.

The two branches are parametrized as
\begin{align}
\label{eq:EM_f2kfr}
f(x)=
\begin{cases}
\begin{aligned}
&\dfrac{A_{0}}{x^{4}+1}\,\widehat{j}_{0}(x)\,(1-e^{-x})
\\
&+\dfrac{1}{x^{3}+1}
\sum_{\ell=1}^{3} A_{\ell}\,\widehat{j}_{\ell}(x),
\end{aligned}
& r < Z_{2},
\\[0.8em]
\dfrac{1}{x^{3}+1}
\sum_{\ell=2}^{5} B_{\ell}\,\widehat{j}_{\ell}(x),
& r > Z_{2}.
\end{cases}
\end{align}
The amplitudes $A_{\ell}$ (for $r<Z_{2}$) and $B_{\ell}$
(for $r>Z_{2}$) are smooth functions of $\rs$.
Estreicher and Meier tabulated these amplitudes in Table~I of
Ref.~\cite{1983_Estreicher-Meier} as polynomial fits in $\rs^{-1}$.
The coefficients are reproduced in Table~\ref{tab:EM_params}, using
\begin{align}
 A_{\ell}\;\text{or}\;B_{\ell}
 \;=\; a_{\ell}/\rs^{4} + b_{\ell}/\rs^{3} + c_{\ell}/\rs^{2}
 + d_{\ell}/\rs + e_{\ell}.
 \label{eq:AlBl-fit}
\end{align}

\begin{table}[htbp!]
 \centering
 \caption{Amplitudes $A_{\ell}$ and $B_{\ell}$ in the
 parametrization, as functions of $\rs$
 [Eq.~\eqref{eq:AlBl-fit}]. After Ref.~\cite{1983_Estreicher-Meier}.}
 \label{tab:EM_params}
 \begin{tabular}{c r r r r r}
  \toprule
   $\ell$ & $a_{\ell}$ & $b_{\ell}$ & $c_{\ell}$ & $d_{\ell}$ & $e_{\ell}$ \\
   \midrule
   \multicolumn{6}{l}{(a)\ \ $r < Z_{2}$:\ \ $A_{\ell} = a_{\ell}/\rs^{4}
   + b_{\ell}/\rs^{3} + c_{\ell}/\rs^{2} + d_{\ell}/\rs + e_{\ell}$} \\
   0 & $-9.879$  & $10.795$  & $-4.422$  & $0.696$  & $-0.018$ \\
   1 & $0.347$   & $2.257$   & $-1.711$  & $0.927$  & $-0.103$ \\
   2 & $14.900$  & $-20.780$ & $10.200$  & $-2.769$ & $0.233$  \\
   3 & $-15.040$ & $17.681$  & $-8.380$  & $1.946$  & $-0.156$ \\
   \midrule
   \multicolumn{6}{l}{(b)\ \ $r > Z_{2}$:\ \ $B_{\ell} = a_{\ell}/\rs^{4}
   + b_{\ell}/\rs^{3} + c_{\ell}/\rs^{2} + d_{\ell}/\rs + e_{\ell}$} \\
   2 & $-6.197$  & $5.882$   & $-1.256$  & $-0.379$ & $0.047$  \\
   3 & $-4.056$  & $6.326$   & $-6.186$  & $1.631$  & $-0.120$ \\
   4 & $2.388$   & $-6.313$  & $6.083$   & $-1.688$ & $0.122$  \\
   5 & $-16.430$ & $19.463$  & $-9.391$  & $1.820$  & $-0.114$ \\
   \bottomrule
 \end{tabular}
\end{table}

\subsection{Short-range cusp behavior of the EM parametrization}
\label{sec:EM_cusp}

The Riccati--Bessel functions obey
$\widehat{j}_\ell(x)=O(x^{\ell+1})$ as $x\to0$.
More specifically,
$\widehat{j}_0(x)(1-e^{-x})=x^2+O(x^3)$ and
$\widehat{j}_1(x)=x^2/3+O(x^4)$.
The inner branch of Eq.~\eqref{eq:EM_f2kfr} therefore gives
$f(x)=(A_0+A_1/3)x^2+O(x^3)$, so that
$f(0)=f'(0)=0$ (with the prime here denoting differentiation with
respect to $x$). The Friedel term consequently contributes neither to
$\Delta n_{\rm EM}(0)$ nor to its radial derivative at the origin.
In particular, the 40 polynomial coefficients in Table~\ref{tab:EM_params}
determine the parametrized Friedel contribution; the local contact slope
is instead fixed by the exponential core terms of the EM density fit.

For the proton ($Z=1$), the exact electron--proton cusp condition applies
to the total electron density $n(r)=n_0+\Delta n(r)$ in
Eq.~\eqref{eq:cusp} and reads
\begin{align}
 \Delta n'(0)=-2\,[n_0+\Delta n(0)],
 \qquad n_0=\frac{3}{4\pi\rs^3}.
 \label{eq:EM_exact_total_cusp}
\end{align}
In their proceedings chapter, Estreicher and Meier derive a cusp
condition for the electronic probability density from the near-origin
wavefunction behavior [their Eq.~(5)], but subsequently write
$\Delta n(r)\simeq\Delta n(0)e^{-2r}$ for the induced
density [their Eq.~(6)]. The exact cusp instead applies to the
total density $n(r)$, as also stated in Eq.~(1) of
Takada~\cite{2018TakadaOntopDensityNonlinear}. To first order in $r$,
it implies
\begin{align}
 \Delta n(r) = \Delta n(0)e^{-2r}-2n_0r+O(r^2),
 \qquad r\to0.
 \label{eq:EM_exact_induced_expansion}
\end{align}
The two exponential core terms in the full EM density
[Eq.~\eqref{eq:EM-dnr}] instead yield
$\Delta n'_{\rm EM}(0)=-2\Delta n_{\rm EM}(0)$.
This is an approximate induced-density cusp, not the exact
total-density relation in Eq.~\eqref{eq:EM_exact_total_cusp}.
The EM contact-density formula, Eq.~\eqref{eq:EM-dn0}, was fitted
to self-consistent LDA contact values. The approximate cusp was instead
built into the subsequent analytic radial representation,
Eq.~\eqref{eq:EM-dnr}; it was not an independent condition used to
determine the contact-density fit coefficients.
The fractional deficit in the magnitude of the cusp slope, evaluated
at the same EM contact density, is
\begin{align}
 \varepsilon_{\rm EM}(\rs)
 &=\frac{\bigl|-2[n_0+\Delta n_{\rm EM}(0)]
                 +2\Delta n_{\rm EM}(0)\bigr|}
              {2[n_0+\Delta n_{\rm EM}(0)]}
 \nonumber\\
 &=\frac{1}{1+\Delta n_{\rm EM}(0)/n_0}.
 \label{eq:EM_cusp_deficit}
\end{align}
Equivalently, the logarithmic derivative of the fitted total density
at the origin is
$n'_{\rm EM}(0)/n_{\rm EM}(0)=-2[1-\varepsilon_{\rm EM}(\rs)]$,
compared with the exact value $-2$.
Using the contact-density parametrization of Eq.~\eqref{eq:EM-dn0},
$\Delta n_{\rm EM}(0)=\pi^{-1}+\exp[-0.72-1.28\ln\rs
-0.385(\ln\rs)^2]$, gives the values in
Table~\ref{tab:EM_cusp_ratio}.
\begin{table}[htbp!]
 \centering
 \caption{Contact-density enhancement and relative cusp-slope deficit
 for the Estreicher--Meier fit. The discrepancy refers only to the
 local derivative, not to an error estimate for the full radial density
 or the contact Hartree potential.}
 \label{tab:EM_cusp_ratio}
 \begin{tabular}{c r r}
  \toprule
  $\rs$ & $\Delta n_{\rm EM}(0)/n_0$
         & $100\varepsilon_{\rm EM}$ (\%) \\
  \midrule
  2 & 16.249 & 5.80 \\
  3 & 44.477 & 2.20 \\
  4 & 95.892 & 1.03 \\
  5 & 178.648 & 0.56 \\
  6 & 300.913 & 0.33 \\
  \bottomrule
 \end{tabular}
\end{table}
Over $2\leq\rs\leq6$, the induced contact density substantially
exceeds the uniform background density, making the approximate
induced-density cusp progressively closer to the exact total-density
condition. The fractional slope deficit decreases from $5.80\%$ at
$\rs=2$ to $0.33\%$ at $\rs=6$. These values quantify only the
local-slope discrepancy; they neither bound the error of the
full radial fit nor provide an error estimate for $V_{\rm H}(0)$.

This distinction matters for the core--Friedel decomposition in
Eqs.~\eqref{eq:VH0_LDA_core} and \eqref{eq:VH0_LDA_friedel}.
Although the Friedel contribution has $f(0)=f'(0)=0$, its radial
moment generally contributes to
$V_{\rm H}(0)=4\pi\int_0^\infty r\,\Delta n(r)\,dr$.
Conversely, the approximate cusp is embedded in the exponential core,
which also contributes to this integral. Its possible influence through
the fitted radial profile cannot be excluded merely because the Hartree
integral contains no explicit contact derivative. In
Fig.~\ref{fig:VH0_DFT_vs_linear}, the small difference between the
EM-based and directly calculated DFT--LDA contact Hartree potentials
does not decrease with increasing $\rs$, unlike the EM cusp-slope
deficit. The contrasting trends do not identify the origin of the
Hartree difference or quantify any cusp-related contribution. In
particular, the $5.80\%$--$0.33\%$ local slope deficits cannot be
interpreted as corresponding errors in $V_{\rm H}(0)$ without a
separate analysis of the radial density difference.

\section{Short-Distance Cusp in Full-Wave-Vector Linear-Response Theory}
\label{sec:lrt_cusp}

The exact Coulomb cusp in Eq.~\eqref{eq:cusp} applies to the total
electron density, $n(r)=n_0+\Delta n(r)$, and requires
$n'(0^+)=-2Zn(0)$. Full-wave-vector linear-response theory (LRT)
does not reproduce this nonlinear identity at finite impurity charge;
however, contrary to what might be inferred by differentiating the
Fourier kernel directly at the origin, it predicts a nonzero
one-sided radial slope. We derive the corresponding first-order cusp
and explain the noncommutativity of the short-distance and
large-wave-vector limits.

\subsection{Large-wave-vector asymptote and the linearized cusp}

From Eq.~\eqref{eq:dn_LRT} and the definition of the
test-charge dielectric function in Eq.~\eqref{eq:eps_LFC},
the induced density can be written as
\begin{align}
\Delta n(r)&=\frac{1}{2\pi^2}\int_0^\infty
 q^2\Delta n(q)\,\frac{\sin(qr)}{qr}\,dq,
 \nonumber\\
 \Delta n(q)&=Z\left[1-\epsilon^{-1}(q)\right].
\label{eq:lrt_cusp_fourier}
\end{align}
The zero-temperature Lindhard function of Eq.~\eqref{eq:chi0_q} has the
large-$q$ expansion
\begin{align}
\chi_0(q)=-\frac{4n_0}{q^2}+O(q^{-4}),\qquad q\to\infty.
\label{eq:lrt_cusp_chi_tail}
\end{align}
In RPA, and also for the CDOP and KK static local-field factors
considered here, $G_+(q)=O(q^2)$ at large $q$ (Appendix~\ref{sec:Gplusq}).
Equation~\eqref{eq:chi_q} then implies
$\chi(q)=\chi_0(q)+O(q^{-4})$. Using
$\epsilon^{-1}(q)=1+(4\pi/q^2)\chi(q)$ gives
\begin{align}
\Delta n(q)=\frac{C}{q^4}+O(q^{-6}),
\qquad C=16\pi Zn_0.
\label{eq:lrt_cusp_nq_tail}
\end{align}
To determine the contact slope, it is convenient to subtract the
contact density before taking the small-$r$ limit:
\begin{align}
\Delta n(r)-\Delta n(0)
=\frac{1}{2\pi^2}\int_0^\infty dq\,q^2\Delta n(q)
\left[\frac{\sin(qr)}{qr}-1\right].
\label{eq:lrt_cusp_difference}
\end{align}
For a fixed large momentum $q_*$, the finite interval $0<q<q_*$
contributes only $O(r^2)$, whereas the $C/q^4$ high-$q$ tail yields
the leading linear term. With $t=qr$,
\begin{align}
\Delta n(r)-\Delta n(0)
&=\frac{Cr}{2\pi^2}\int_0^\infty
\frac{\sin t/t-1}{t^2}\,dt+O(r^2)\nonumber\\
&=-\frac{C}{8\pi}r+O(r^2)
=-2Zn_0r+O(r^2),
\label{eq:lrt_cusp_smallr}
\end{align}
where $\int_0^\infty(\sin t/t-1)t^{-2}\,dt=-\pi/4$.
Thus, for the full-$q$ RPA and the CDOP/KK-corrected LRT responses,
\begin{align}
\left.\frac{dn_{\rm LRT}}{dr}\right|_{0^+}=
-2Zn_0.
\label{eq:lrt_cusp_linearized}
\end{align}
This is the first-order term of the exact relation
$-2Z[n_0+\Delta n(0)]$. It does not imply that the LRT density
satisfies the full Kato cusp evaluated at its own enhanced
contact value.

\subsection{Noninterchangeable limits and an exact free-hydrogen illustration}

For each fixed $q$, the spherical Fourier kernel
$j_0(qr)=\sin(qr)/(qr)$ satisfies
$\partial_r j_0(qr)|_{r=0}=0$. The erroneous zero-slope inference
results from moving this derivative and the $r\to0^+$ limit
inside the integral in Eq.~\eqref{eq:lrt_cusp_fourier}.
The operation is not justified for the $q^{-4}$ tail:
$\int^\infty q^4\Delta n(q)\,dq$ diverges, so that the
Taylor series $j_0(qr)=1-q^2r^2/6+\cdots$ cannot be integrated
term by term. The short-distance contribution arises from a
wave-vector region $q\sim1/r$ that moves to infinity as $r\to0^+$.

The distinction is particularly transparent if the integral is
truncated at a finite upper wave vector $Q$. For
$\Delta n_Q(r)=(2\pi^2)^{-1}\int_0^Q q^2\Delta n(q)j_0(qr)\,dq$,
one obtains $\Delta n_Q'(0)=0$ for every finite $Q$.
For the full Lindhard-based response, however,
\begin{align}
\lim_{r\to0^+}\lim_{Q\to\infty}\Delta n_Q'(r)
&=-2Zn_0,\nonumber\\
\lim_{Q\to\infty}\lim_{r\to0^+}\Delta n_Q'(r)
&=0.
\label{eq:lrt_cusp_noncommuting}
\end{align}
Hence these two limiting operations do not commute.

A physical, exactly soluble illustration is provided by the
isolated hydrogen atom. For the neutral one-electron system ($Z=1$),
let $\alpha=2\,a_0^{-1}$, so that the exact $1s$ electron density and
the electrostatic potential of the nucleus plus its electron cloud are
\begin{align}
 n_{\rm H}(r)&=\frac{\alpha^3}{8\pi}e^{-\alpha r},
 &V_{\rm SC}^{\rm free-H}(r)
 &=\frac{1}{r}\left(1+\frac{\alpha r}{2}\right)e^{-\alpha r},
\label{eq:lrt_cusp_freeH_real}
\end{align}
where the latter is the known spherically averaged hydrogen potential
\cite{1999_jackson_classicalelectrodynamic}. This is an isolated
one-electron Coulomb problem, not a model of metallic LRT screening.
Following the main text, one can define a dielectric-style
ratio from the Fourier transform of the screened electrostatic
potential,
\begin{align}
 \epsilon_{\rm free-H}^{-1}(q)
 &=\frac{V_{\rm SC}^{\rm free-H}(q)}{4\pi/q^2}
 =\frac{q^2(q^2+2\alpha^2)}{(q^2+\alpha^2)^2},
 \nonumber\\
 1-\epsilon_{\rm free-H}^{-1}(q)
 &=\frac{\alpha^4}{(q^2+\alpha^2)^2}
 \sim\frac{\alpha^4}{q^4}\qquad(q\to\infty).
\label{eq:lrt_cusp_freeH_dielectric}
\end{align}
The ratio is an electrostatic bookkeeping identity for the known
one-electron density, not a microscopic dielectric function
of a translationally invariant electron gas. Nor does it identify
the Poisson density of a general fitted hydrogenic model potential
with the independently reconstructed QM--VPA density of that
potential (Sec.~2.3.1). Crucially, its corresponding density factor has the same $q^{-4}$ asymptotic power as the full-$q$
LRT response in Eq.~\eqref{eq:lrt_cusp_nq_tail}, although the
coefficient is different.

Indeed, the inverse Fourier integral can be evaluated exactly:
\begin{align}
 n_{\rm H}(r)
 &=\frac{1}{2\pi^2}\int_0^\infty dq\,q^2
 \left[1-\epsilon_{\rm free-H}^{-1}(q)\right]
 \frac{\sin(qr)}{qr}
 \nonumber\\
 &=\frac{\alpha^3}{8\pi}e^{-\alpha r}.
\label{eq:lrt_cusp_freeH_inverse}
\end{align}
Hence, although the derivative of the Fourier kernel vanishes at
$r=0$ for each finite $q$, the completed integral satisfies
\begin{align}
 n_{\rm H}'(0^+)
 &=-\frac{\alpha^4}{8\pi}
 =-\alpha n_{\rm H}(0)
 =-2n_{\rm H}(0),\nonumber\\
 &\qquad \alpha=2,\ Z=1.
\label{eq:lrt_cusp_freeH_slope}
\end{align}
In contrast, truncating the integral in
Eq.~\eqref{eq:lrt_cusp_freeH_inverse} at any finite $Q$ gives
$n_{{\rm H},Q}'(0)=0$. This known quantum density thus gives a
physical and exactly soluble demonstration of the same noncommuting
limits responsible for the LRT cusp. Here $n_0=0$; the hydrogen
slope obeys the full one-electron Kato cusp, whereas
Eq.~\eqref{eq:lrt_cusp_linearized} is the first-order cusp of a
perturbed uniform electron gas.

\subsection{Numerical consistency check and relation to the nonlinear cusp}

For the proton at $r_s=2.07$, the RPA radial density used in
Fig.~\ref{fig:dn0_vs_r_DFT_vs_LRT} gives
$\Delta n_{\rm RPA}(0)/n_0=2.8354015$.
Using this fixed contact value and fitting the first $20$
positive-radius points ($0<r\le0.37547\,a_0$) to
$\Delta n(r)/n_0=\Delta n(0)/n_0+s_1r+s_2r^2+s_3r^3$
without constraining $s_1$ gives
$s_1=-1.999637\,a_0^{-1}$, compared with
$-2\,a_0^{-1}$ from Eq.~\eqref{eq:lrt_cusp_linearized}.
This is a consistency check on the saved RPA profile, not a
separate proof of contact-derivative convergence: the numerical
contact value includes an analytic high-$q$ correction while the
finite-radius Fourier reconstruction uses a finite wave-vector
cutoff.

Finally, it is useful to distinguish the LRT cusp from the
approximate induced-density cusp of the Estreicher--Meier (EM)
parametrization discussed in Appendix~\ref{sec:EM_cusp}.
At $Z=1$, LRT retains the background term $-2n_0$,
whereas the EM fit has $\Delta n'_{\rm EM}(0)=-2\Delta n_{\rm EM}(0)$.
These are different partial realizations of the exact relation
$\Delta n'(0)=-2[n_0+\Delta n(0)]$; they must not be conflated
or added as independent density approximations. In particular,
the cusp constrains a local slope but determines neither the
contact density nor the nonlocal Hartree moment $V_{\rm H}(0)$.

\section{Solving the phase equation and Friedel pipeline}
\label{sec:friedel_pipeline}

For each choice of $\rs$, impurity charge $Z$, potential family $\nu$, and trial screening parameter $\alpha$, the pipeline evaluates the partial-wave phase shift $\delta_{\ell}(k_{\rm F};\alpha)$ by integrating Eq.~\eqref{eq:phase_Eq} over a finite radial interval,
\begin{align}
r_{\min}
\le r \le
r_{\max}.
\label{eq:radial_integration_window}
\end{align}
The integration is performed with LSODA through the \texttt{solve\_ivp} interface of the SciPy scientific-computing library. LSODA adaptively switches between nonstiff Adams and stiff backward-differentiation-formula modes, so the internal integration mesh is determined by local error control rather than by a manually fixed radial spacing \cite{1983_Petzold_LSODA,1983_Hindmarsh_odepack}. When running-phase profiles are not needed, the code stores only the endpoint information; this output choice does not affect LSODA's internal adaptive steps.

The inner radius is determined from the Coulombic small-$r$ phase in Eq.~\eqref{eq:rmin}. For a prescribed small phase scale $\epsilon_\phi$, the asymptotic startup radius is obtained from
\begin{align}
\left|
\delta_{\ell}^{(0)}(k;r_{\min}^{(0)})
\right|
&=
\epsilon_\phi,
\label{eq:rmin_phase_condition}
\end{align}
which gives
\begin{align}
r_{\min}^{(0)}
&=
\left[
\frac{
\epsilon_\phi(2\ell+2)\left[(2\ell+1)!!\right]^2
}{
2\mu |Z| k^{2\ell+1}
}
\right]^{1/(2\ell+2)}.
\label{eq:rmin_asymptotic_solution}
\end{align}
Because the expansion $V(r)\simeq -Z/r$ is valid only well inside the screening length, the actual starting point is additionally restricted by
\begin{align}
r_{\min}
&=
\min\left(
r_{\min}^{(0)},
\frac{\eta_{\rm in}}{\alpha}
\right),
\qquad
0<\eta_{\rm in}\ll 1.
\label{eq:rmin_screening_length_cap}
\end{align}
In the calculations reported here we use $\epsilon_\phi=10^{-10}$ and $\eta_{\rm in}=0.1$. The initial condition for Eq.~\eqref{eq:phase_Eq} is then the asymptotic value $\delta_{\ell}^{(0)}(k;r_{\min})$, rather than zero. This avoids starting the integration at the singular origin and keeps the Riccati--Neumann contribution in the phase equation numerically controlled.

The outer radius is determined from the same physical principle as Eq.~\eqref{eq:delta_outer_matching_radius}. Let
\begin{align}
r_{\rm tp}(\ell,k)
&=
\frac{\sqrt{\ell(\ell+1)}}{k}
\label{eq:centrifugal_turning_point}
\end{align}
denote the centrifugal turning radius of the free radial problem. The raw tail radius $R_{\rm tail}$ is defined as the smallest radius beyond the turning point for which
\begin{align}
|U(R_{\rm tail})|
&\le
\epsilon_{\rm tail}
\left|
k^2-\frac{\ell(\ell+1)}{R_{\rm tail}^2}
\right|,
\qquad
R_{\rm tail}>r_{\rm tp},
\label{eq:rtail_definition}
\end{align}
where $U(r)=2\mu V(r)$. The final integration endpoint also enforces a safety distance from the turning point,
\begin{align}
r_{\max}
&=
\max\left(
R_{\rm tail},
c_{\rm tp}\,r_{\rm tp}
\right),
\qquad
c_{\rm tp}>1.
\label{eq:rmax_final_definition}
\end{align}
The second condition prevents the endpoint from being placed too close to $r_{\rm tp}$, where the free radial kinetic term changes rapidly. In the calculations reported here we use $\epsilon_{\rm tail}=10^{-9}$ and $c_{\rm tp}=1.5$. For $\ell=0$, $r_{\rm tp}=0$, so the endpoint is determined solely by the tail-dominance condition. For the Yukawa $s$ wave, Eq.~\eqref{eq:rtail_definition} can be written in closed form on the principal Lambert branch,
\begin{align}
R_{\rm tail}^{(Y,\ell=0)}
&=
\frac{1}{\alpha}
W_0\!\left(
\frac{2\mu |Z|\alpha}{\epsilon_{\rm tail}k^2}
\right).
\label{eq:yukawa_swave_lambert_tail}
\end{align}
For $l>0$, and for the non-Yukawa potentials, $R_{\rm tail}$ is obtained numerically from Eq.~\eqref{eq:rtail_definition} using the potential evaluated directly in real space. In all cases, the endpoint is chosen so that the residual tail contribution to the phase equation is negligible on the prescribed scale, and hence
\begin{align}
\delta_{\ell}(k;\alpha)
&\simeq
\delta_{\ell}(k;r_{\max};\alpha).
\label{eq:phase_shift_endpoint_approximation}
\end{align}
Endpoint phases whose absolute value falls below the declared phase-resolution scale $\epsilon_\phi$ are set to zero. This phase-floor convention removes numerical endpoint noise from channels whose contribution to the Friedel sum rule is below the intended resolution.

The upper limit $L_{\max}$ in Eq.~\eqref{eq:static_friedel_sum_rule} is determined from the actual weighted partial-wave increments, not from a fixed empirical formula. Define
\begin{align}
d_{\ell}(k_{\rm F};\alpha)
&=
(2\ell+1)\delta_{\ell}(k_{\rm F};\alpha).
\label{eq:partial_wave_increment}
\end{align}
The code computes $\delta_{\ell}(k_{\rm F};\alpha)$ incrementally for $\ell=0,1,2,\ldots$ and tests the cutoff after each channel. Let $l_\ast$ denote the last angular momentum for which $|d_{\ell}(k_{\rm F};\alpha)|>\tau_L$. The effective cutoff is then
\begin{align}
L_{\max}
&=
\max(L_{\min},l_\ast),
\label{eq:lmax_definition}
\end{align}
provided that it is followed by a certified tail of negligible increments,
\begin{align}
|d_j(k_{\rm F};\alpha)|
&\le
\tau_L,
\qquad
j=L_{\max}+1,\ldots,L_{\max}+N_{\rm tail}.
\label{eq:lmax_tail_condition}
\end{align}
Here $\tau_L$ is the partial-wave tail tolerance, $L_{\min}$ is a small minimum retained cutoff, and $N_{\rm tail}$ is the number of consecutive sub-threshold channels required to certify convergence. The production calculations use $L_{\min}=3$ and $N_{\rm tail}=4$. Requiring several consecutive negligible terms prevents an accidental early cutoff near isolated dips or sign changes in $\delta_{\ell}$. If the tail is not certified before the current angular-momentum ceiling is reached, the ceiling is enlarged and the test is repeated.

With $L_{\max}$ determined in this way, the finite static Friedel sum is evaluated as
\begin{align}
s_{L_{\max}}(k_{\rm F};\alpha)
&=
\frac{2}{\pi}
\sum^{L_{\max}}_{\ell=0}
d_{\ell}(k_{\rm F};\alpha) \nonumber\\
&= \frac{2}{\pi}
\sum^{L_{\max}}_{\ell=0}
(2\ell+1)\delta_{\ell}(k_{\rm F};\alpha),
\label{eq:finite_static_friedel_sum}
\end{align}
and the self-consistent screening parameter is obtained from the positive root of
\begin{align}
F(\alpha;\rs,Z)
&=
Z-s_{L_{\max}}(k_{\rm F};\alpha)
=0.
\label{eq:finite_friedel_root_condition}
\end{align}
The root is found by a bracketed Brent procedure. The bracketing step enforces a sign change of $F(\alpha)$ on a positive interval of $\alpha$, while the Brent iteration provides a derivative-free scalar solve. The default Brent tolerances are chosen tighter than the Friedel residual target, and the final diagnostic evaluation records
\begin{align}
\Delta_{\rm FSR}
&=
\left|
Z-s_{L_{\max}}(k_{\rm F};\alpha_\ast)
\right|.
\label{eq:fsr_diagnostic_residual}
\end{align}
The production result is retained only after verifying that $\Delta_{\rm FSR}$ is consistent with the target residual scale $\tau_{\rm FSR}$ and is not limited by unresolved phase shifts, an uncertified partial-wave tail, or loose root-location tolerances.

This completes the static Friedel screening pipeline. For each model potential, the calculation maps
\begin{align}
(\rs,Z,\nu)
&\longmapsto
\alpha_\ast^{(\nu)}(\rs,Z),
\label{eq:screening_pipeline_mapping}
\end{align}
where $\alpha_\ast^{(\nu)}$ is the self-consistent screening parameter for the selected potential family. In the standard one-parameter calculations,
\begin{align}
\nu
&\in
\{\text{Yuk},\text{Hyd},\text{Hult}\},
\end{align}
corresponding to Yukawa, hydrogenic, and Hulth\'en forms, respectively. The same phase-equation and Friedel-sum-rule machinery is also used for fitted screened potentials, such as the Popovi\'c--Stott-type form, when their additional parameters are supplied externally. The physically relevant output is the corresponding self-consistent potential $V_\nu(r;\alpha_\ast^{(\nu)})$ and the associated Friedel-saturating phase-shift spectrum.

\section{Numerical Convergence of Variable-Phase Calculations}
\label{sec:vpa_convergence}

The variable-phase pipeline described in Appendix~\ref{sec:friedel_pipeline} was validated by verifying its numerical convergence with respect to the radial integration window and the partial-wave truncation. Two representative diagnostics are collected here.

\begin{figure}[htbp!]
    \centering
    \includegraphics[width=\columnwidth]{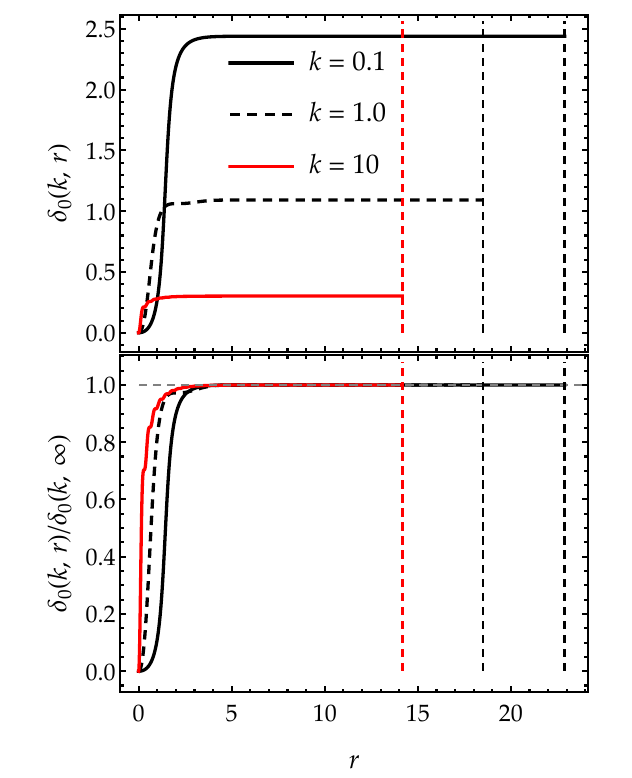}
    \caption{Running $s$-wave phase shift for the attractive Yukawa potential $V_{\text{Yuk}}(r)=-e^{-r}/r$, corresponding to $Z=1$ and $\alpha=1$, for $k=0.1$, $1$, and $10$ a.u. The upper panel shows the absolute running phase $\delta_0(k,r)$, whose asymptotic plateau defines the physical phase shift $\delta_0(k)=\lim_{r\to\infty}\delta_0(k,r)$. The lower panel shows the normalized running phase $\delta_0(k,r)/\delta_0(k,\infty)$, which measures the fraction of the final phase shift accumulated within radius $r$. The vertical dashed lines indicate the automatically selected $r_{\max}$ values for the three wave numbers, obtained from the large-distance tail-dominance criterion used to terminate the variable-phase integration [Eq.~\eqref{eq:rmax}]. The normalized curves show that the accumulated phase reaches its plateau well before the conservative $r_{\max}$ values. Thus, integration to $r_{\max}$ provides a robust asymptotic criterion, but it is not necessarily computationally optimal; most of the physically relevant phase accumulation occurs at substantially smaller radii.}
\label{fig:running_delta0_yukawa}
\end{figure}

As the first phase-shift result from our VPA pipeline, we present in Figure~\ref{fig:running_delta0_yukawa} the leading phase function $\delta_0(k, r)$ for a proton potential represented by a Yukawa potential with a typical screening parameter $\alpha=1$ for three momenta $k=0.1, 1.0, 10.0$. The figure shows the initial phase $\delta_0(k,\rmin) \simeq 0$ as postulated in Eq.~\eqref{eq:phase_origin_condition}, how the phase function $\delta_0(k, r)$ is accumulated as $r$ increases and saturates to its physical value as required by Eq.~\eqref{eq:phase_shift_saturation_definition}. The lower and upper limits of integration of the phase equation are determined, respectively, by $\rmin = 3.2\times 10^{-5}, 1.0 \times 10^{-5}, 3.2\times 10^{-6}$ [Eq.~\eqref{eq:rmin_asymptotic_solution}] and $\rmax = 22.9, 18.5, 14.2$ [Eq.~\eqref{eq:yukawa_swave_lambert_tail}] corresponding to $k=0.1, 1.0, 10.0$, respectively.

\begin{figure}[htbp!]
    \centering
    \includegraphics[width=\columnwidth]{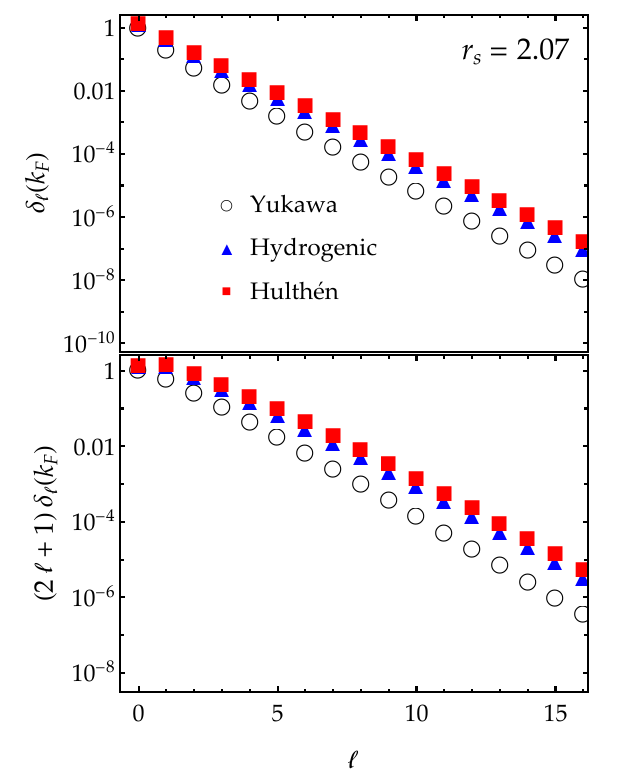}
    \caption{Partial-wave phase shifts for the three screened model potentials used in the static Friedel-sum analysis. The upper panel shows the physical phase shifts $\delta_{\ell}(\kf)$ as a function of angular momentum $\ell$, evaluated for $Z=1$, $\alpha=1$, and $\rs=2.07$, representative of Al. The lower panel shows the corresponding weighted Friedel-sum contributions $(2\ell+1)\delta_{\ell}(\kf)$. The rapid decay with $l$ demonstrates that the low partial waves dominate the phase-shift sum, while higher-$l$ channels contribute only exponentially small corrections over the range shown. For the same screening parameter, the Hulth\'en potential gives the largest phase shifts, the Yukawa potential the smallest, and the hydrogenic potential lies between them.}
\label{fig:delta_KF_Al_vs_l_AllPots}
\end{figure}

Now that we have examined how the running phase shift $\delta_{\ell}(k,r)$ evolves with radial distance $r$ and wave number $k$, we turn to its dependence on angular momentum $\ell$. Figure~\ref{fig:delta_KF_Al_vs_l_AllPots} shows the physical phase shifts $\delta_{\ell}(k_{\rm F})$ and the weighted Friedel contributions $(2\ell+1)\delta_{\ell}(k_{\rm F})$ for Al computed for the three screened model potentials considered in this work for $Z=1$ and screening parameter $\alpha=1$. Several important trends emerge from the figure. First, all three potentials exhibit an approximately exponential decay of $\delta_{\ell}(k_{\rm F})$ with increasing $l$, indicating that only the lowest partial waves contribute appreciably to the Friedel sum rule Eq.~\eqref{eq:static_friedel_sum_rule}. This rapid suppression originates from the centrifugal barrier $\ell(\ell+1)/r^2$, which increasingly prevents high-$l$ electronic partial waves from penetrating into the short-range region where the screened proton potential is strongest.  Consequently, the static screening problem is dominated by low-angular-momentum scattering channels, especially the $s$- and $p$-waves. Second, the hydrogenic and Hulth\'en potentials produce remarkably similar phase shifts over the entire $l$-range shown, whereas the Yukawa potential consistently yields smaller values of $\delta_{\ell}(k_{\rm F})$. This trend mirrors the behavior of the corresponding screening functions discussed earlier for Figure~\ref{fig:mod_pot_phi}. Although all three potentials share the same Coulombic short-range singularity, their intermediate- and long-range screening profiles differ. The Yukawa potential decays more rapidly with distance and therefore produces a weaker integrated scattering strength, leading to systematically smaller phase shifts. In contrast, the hydrogenic and Hulth\'en potentials retain a somewhat stronger tail and consequently generate larger partial-wave scattering phases. 

Another notable feature is that the lowest phase shifts are all of order unity near $\ell=0$, after which the logarithmic decay with $l$ becomes nearly linear on the semilogarithmic scale. This reflects the fact that the low-$l$ channels probe the strongly nonlinear screening region close to the proton, where the screened potential remains comparable to or larger than the electronic kinetic energy scale. Once the centrifugal barrier dominates, the scattering rapidly transitions into a weak-perturbation regime in which successive partial waves contribute only exponentially small corrections. The lower panel of Figure~\ref{fig:delta_KF_Al_vs_l_AllPots} further shows that the weighted quantities $(2\ell+1)\delta_{\ell}(k_{\rm F})$, which enter directly into the Friedel sum rule, also decay rapidly with $l$, despite the increasing degeneracy factor $(2\ell+1)$. This observation provides a direct numerical justification for truncating the partial-wave expansion at relatively modest values of $l_{\max}$ in practical Friedel calculations.

\section{Self-consistent DFT calculation of proton screening in a homogeneous electron gas}
\label{sec:almbladh_method}

We compute the electronic density induced by a static proton immersed in a homogeneous electron gas of density $\nz =
3/(4\pi \rs^{3})$ using the Kohn--Sham formulation of density
functional theory~\cite{1965KohnSelfConsistentEquationsIncluding}, following the prescription of Almbladh, von~Barth, Popovi\'c and Stott~\cite{1976AlmbladhScreeningProtonElectron}. The Hartree atomic units ($\hbar = m_{e} = e^{2}/4\pi\epsilon_{0} = 1$) are used throughout. Because the perturbing potential is spherically symmetric, the Kohn--Sham orbitals decouple into partial waves, and the problem reduces to a one-dimensional radial calculation.

\subsection{Radial grid and Kohn--Sham equation}

The radial equation for the reduced wavefunction
$u_{\ell}(r) = r R_{\ell}(r)$ is solved on a logarithmic grid
$r_{i} = e^{x_{i}}$, $x_{i}\in[\ln r_{\min},\ln r_{\max}]$, with
$N = 8000$ points, $r_{\min} = 10^{-4}\,a_{0}$ and
$r_{\max} = 50\,a_{0}$, corresponding to a uniform spacing
$\Delta x = 1.64\times10^{-3}$. Substituting $u_{\ell}(r) = r^{1/2}\phi_{\ell}(x)$
yields a Schr\"odinger-like equation in the uniform variable $x$,
\begin{align}
\begin{aligned}
  \phi_{\ell}''(x) &= Q_{\ell}(x)\phi_{\ell}(x), \\
  Q_{\ell}(x) &= \left(\ell+\frac{1}{2}\right)^{2}
  + 2 r^{2}\left[V(r)-E\right],
\end{aligned}
\label{eq:radial}
\end{align}
which is integrated with a Numerov scheme. To avoid floating-point
overflow in the classically forbidden region, the wavefunction is
rescaled adaptively whenever its magnitude leaves the range
$[10^{-50},10^{50}]$. All radial integrals are evaluated by the
trapezoidal rule; the resulting quadrature error in the observables
reported below is given in Table~\ref{tab:dft-convergence}.

\subsection{Phase shifts and the displaced density}

For each angular momentum $\ell$ and momentum $k\in[0,\kf]$, the
outward Numerov integration of Eq.~\eqref{eq:radial} is matched at
$R_{\mathrm{match}} = 40\,a_{0} = 0.8\,r_{\max}$ to the asymptotic form
\begin{align}
  u_{\ell}(r) \xrightarrow{r\to\infty}
  kr\left[\cos\delta_{\ell}(k)j_{\ell}(kr)
       - \sin\delta_{\ell}(k)n_{\ell}(kr)\right],
\end{align}
where $j_{\ell}$ and $n_{\ell}$ are spherical Bessel and Neumann
functions. The phase shift $\delta_{\ell}(k)$ and an overall normalization
constant are extracted from a $2\times2$ linear system using the values of
$u_{\ell}$ at two grid points near the matching radius. The separation of
those two points is chosen adaptively, at approximately $0.15$ wavelengths at
$k_{\rm F}$, so that it remains a fixed fraction of the local oscillation rather
than a fixed number of grid points; this avoids the near-degenerate
Wronskian that arises when the two points approach half-wavelength spacing.
Beyond $R_{\mathrm{match}}$, the numerical solution is replaced by its
analytical continuation. With the normalization used here, the continuum
contribution to the displaced density is
\begin{align}
\Delta n_{\mathrm{sc}}(r) = \frac{1}{\pi^{2}r^{2}} \sum_{\ell=0}^{\ell_{\max}}(2\ell+1) &\int_{0}^{k_{\rm F}} dk \times \nonumber \\
&\left[|u_{\ell}(r;k)|^{2} - |u_{\ell}^{(0)}(r;k)|^{2}\right],
\label{eq:dn_scattering}
\end{align}
where $u_{\ell}^{(0)}$ is the corresponding free-gas solution. Bound states, when present, are added as separate discrete spectral contributions. If $u_{\nu\ell}(r)$ denotes a normalized reduced radial bound-state wavefunction,
\begin{align}
  \int_{0}^{\infty}|u_{\nu\ell}(r)|^{2}dr = 1,
\end{align}
and $N_{\nu\ell}$ is the total occupation of that bound level, including spin and magnetic degeneracy, then the bound-state contribution to the spherically averaged density is
\begin{align}
  \Delta n_{\mathrm{b}}(r)
  =
  \sum_{\nu\ell}
  N_{\nu\ell}
  \frac{|u_{\nu\ell}(r)|^{2}}{4\pi r^{2}} .
\label{eq:dn_bound}
\end{align}
The total displaced density is therefore
\begin{align}
  \Delta n(r)
  =
  \Delta n_{\mathrm{sc}}(r)
  +
  \Delta n_{\mathrm{b}}(r).
\label{eq:drho}
\end{align}
For the proton-in-jellium problem in the metallic-density regime, the only bound level encountered is a shallow $s$-wave level. It is a doubly occupied $1s$ state, $N_{1s}=2$, so the total number of bound electrons is $N_{\rm b}=2$.

The momentum integral in Eq.~\eqref{eq:dn_scattering} is performed by Gauss--Legendre quadrature with $n_{k}=60$ nodes, and partial waves up to $\ell_{\max}=12$ are retained. Friedel oscillations in $\Delta n(r)$ become non-negligible only beyond $\sim 2.4\,a_{0}$; a smooth Fermi cutoff $\tfrac{1}{2}[1-\tanh((r-R_{\mathrm{match}})/a_{0})]$ is used only to damp residual numerical noise in the far tail.

The total Friedel sum
\begin{align}
  Z_{F}
  =
  \frac{2}{\pi}
  \sum_{\ell}(2\ell+1)\delta_{\ell}(k_{\rm F})
\label{eq:friedel}
\end{align}
serves as a diagnostic of self-consistency and total screening. At convergence, $Z_F$ should approach the impurity charge $Z=1$ for a proton.

When bound states are present, Levinson's theorem gives
\begin{align}
  \delta_{\ell}(0)=n_{\ell}^{\mathrm{b}}\pi ,
\label{eq:levinson_appendix}
\end{align}
where $n_{\ell}^{\mathrm{b}}$ is the number of radial bound states in the $\ell$ channel. Including spin and orbital degeneracy, the total number of bound electrons is
\begin{align}
  N_{\rm b}
  =
  2\sum_{\ell}(2\ell+1)n_{\ell}^{\mathrm{b}}
  =
  \frac{2}{\pi}
  \sum_{\ell}(2\ell+1)\delta_{\ell}(0).
\label{eq:Nb_appendix}
\end{align}
The continuum contribution to the induced charge is obtained from the Levinson-subtracted phase shifts,
\begin{align}
  Z_{\mathrm{sc}}
  =
  \frac{2}{\pi}
  \sum_{\ell}(2\ell+1)
  \left[
  \delta_{\ell}(k_{\rm F})-\delta_{\ell}(0)
  \right]
  =
  Z-N_{\rm b} .
\label{eq:Zsc_appendix}
\end{align}
For a single doubly occupied $1s$ bound level, $n_{0}^{\mathrm{b}}=1$ and $N_{\rm b}=2$. Thus the continuum contribution integrates to $Z_{\mathrm{sc}}=-1$, while the sum of the bound and continuum contributions gives the total induced screening charge $Z=1$, consistent with the charge-accounting convention used by Jena and Singwi~\cite{1978JenaElectronicStructureHydrogen}.

\subsection{Bound states}

For $\rs\gtrsim 1.9$, consistent with Refs.~\cite{1976AlmbladhScreeningProtonElectron,1978JenaElectronicStructureHydrogen}, the screened proton supports a shallow $1s$ bound state in the jellium description; in the present calculation the level becomes occupied between $\rs=1.9$ and $\rs=2.0$. The bound-state energy and wavefunction are located by outward/inward Numerov shooting, using bisection on the log-derivative mismatch at the classical turning point.

Formally, a true bound state is included as a discrete spectral contribution to the induced density, while the continuum contribution is obtained from the scattering states. With this separation there is no double counting.

The occupation is determined without any adjustable threshold. At every iteration of the self-consistency cycle the $\ell=0$ channel of the current effective potential is searched for a solution with $E<0$. If one exists it is occupied by two electrons; if none exists the bound contribution to $\Delta n$ vanishes. The occupation is therefore a property of the current potential alone, recomputed from scratch at each iteration, with no dependence on the iteration history and no binding-energy cutoff.

Very shallow levels require care because their spatial extent can exceed the finite radial box: as $E_{\rm b}\to 0^{-}$ the decay constant $\kappa=\sqrt{2|E_{\rm b}|}$ vanishes and $\kappa^{-1}$ diverges. The bound orbital is therefore normalized over all space rather than over the box, the contribution beyond $r_{\max}$ being added analytically,
\begin{align}
  \int_{0}^{\infty}|u_{\nu\ell}|^{2}dr
  =
  \int_{0}^{r_{\max}}|u_{\nu\ell}|^{2}dr
  +
  \frac{|u_{\nu\ell}(r_{\max})|^{2}}{2\kappa},
\label{eq:analytic_tail}
\end{align}
which follows from the asymptotic form $u_{\nu\ell}\propto e^{-\kappa r}$. This procedure includes the exterior tail in the normalization of the shallow bound orbital and its contribution to $\Delta n(r)$. Bound-orbital normalization alone does not establish convergence of the full bound-plus-continuum induced charge or of the density-derived contact Hartree potential $V_{\rm H}(0)$.

Across the binding threshold, the low-energy $\ell=0$ phase shift is checked against the bound-state count through Levinson's theorem, $\delta_0(0)=n_0^{\mathrm b}\pi$. The total Friedel sum $Z_F$ in Eq.~\eqref{eq:friedel} provides a separate spectral test of the required screening charge $Z=1$. The corresponding exact real-space screening condition is
\begin{align}
  Q = 4\pi\int_{0}^{\infty}\Delta n(r)r^{2}dr = Z .
\end{align}
A numerical charge integral evaluated over a finite radial domain, however, may differ from the infinite-domain quantity $Q$ because the induced-density tail is not fully included. This sensitivity can be especially pronounced near a shallow bound-state threshold. We therefore treat the finite-domain real-space charge integral as a supplementary diagnostic, distinct from the Friedel sum, Levinson's theorem, and the occupation count. In particular, it is not used to select the self-consistent branch or to estimate the uncertainty of $V_{\rm H}(0)$. No quantitative attribution of the finite-domain charge discrepancy to a specific tail contribution is made here.

\subsection{Density continuation}

The densities are solved as a sequence rather than independently. The calculation begins at the most dilute point, $\rs = 6.0$, converges it from a Thomas--Fermi initial potential, and uses each converged effective potential as the initial guess for the next density, proceeding in steps of $0.1$ to $\rs = 0.5$. Kohn and Majumdar~\cite{1965KohnContinuityBetweenBound} showed that observables vary continuously as a bound state emerges from the continuum, so following the solution continuously in $\rs$ is the numerical counterpart of that requirement. Density continuation is used to initialize the production sweep efficiently; the physical solution is not selected solely by this procedure, as confirmed by the independent-start comparison described below. 

The $1s$ level emerges with essentially zero binding ($E_{\rm b} = -3.3\times10^{-3}\,\mathrm{eV}$ at $\rs = 2.0$) and deepens continuously, the Levinson count switches in step with the occupation, and $V_{\rm H}(0)$ passes through the threshold with no step: its second difference over $\rs = 1.6$--$2.2$ is $4.1\times10^{-3}\,\mathrm{Ha}$, below the maximum of $4.1\times10^{-2}\,\mathrm{Ha}$ attained at $\rs = 0.6$, where the curve is steepest. To test whether branch selection depends on density continuation, we repeated the calculations at all 56 densities from independent Thomas--Fermi initial potentials. These calculations reproduce the continuation-based $V_H(0)$ values to within $3\times10^{-6}\,\mathrm{Ha}$, with identical bound-state occupations. No distinct converged branch was found under these initializations. 

\subsection{Self-consistency cycle}
\label{sec:sc_cycle}

At each iteration, the induced Hartree potential is obtained by direct integration,
\begin{align}
  V_{\rm H}(r)
  =
  \frac{4\pi}{r}\int_{0}^{r}\Delta n(r')r'^{2}dr'
  +
  4\pi\int_{r}^{\infty}\Delta n(r')r'dr',
\end{align}
and the exchange--correlation potential is evaluated from the chosen
functional. The implementation supports the
Perdew--Zunger~\cite{1981PerdewSelfinteractionCorrectionDensityfunctional} and
Hedin--Lundqvist~\cite{1971HedinExplicitLocalExchangecorrelationa}
parametrizations of the local-density approximation, as well as the
Perdew--Burke--Ernzerhof generalized-gradient
approximation~\cite{1996PerdewGeneralizedGradientApproximation}. The Hedin-Lundqvist parametrization is used for the LDA results reported here, enabling direct comparison with Almbladh \emph{et al.}~\cite{1976AlmbladhScreeningProtonElectron}; the PBE results use the generalized-gradient functional described below.

For the generalized-gradient calculations the exchange--correlation potential
is the full functional derivative of the PBE energy density $e[n]$ of
Ref.~\cite{1996PerdewGeneralizedGradientApproximation}, which depends on the
density and on $\sigma = |\nabla n|^{2}$,
\begin{align}
  V_{\rm xc}
  &= \frac{\partial e}{\partial n}
  - \nabla\cdot\!\left(2\frac{\partial e}{\partial\sigma}\nabla n\right)
  \nonumber\\
  &= \left.\frac{\partial e}{\partial n}\right|_{\sigma}
  - \frac{1}{r^{2}}\frac{\mathrm{d}}{\mathrm{d}r}
    \left[2r^{2}\left.\frac{\partial e}{\partial\sigma}\right|_{n}
      \frac{\mathrm{d}n}{\mathrm{d}r}\right],
\label{eq:vxc_pbe}
\end{align}
the second form holding for a spherically symmetric density with
$\sigma=(\mathrm{d}n/\mathrm{d}r)^{2}$. The two partial derivatives are
evaluated as central finite differences of $e(n,\sigma)$, each argument
perturbed with the other held fixed and with a relative increment
$h = 10^{-4}$, so that $n \to n(1 \pm h)$ at fixed $\sigma$ and
$\sigma \to \sigma(1 \pm h)$ at fixed $n$. Where $\sigma < 10^{-20}$ the step
is floored at that value so the central-difference denominator cannot vanish;
this is immaterial because $\partial e/\partial\sigma$ enters only multiplied
by $\mathrm{d}n/\mathrm{d}r$, which is zero wherever $\sigma$ is.
The derivative $\mathrm{d}n/\mathrm{d}r$ and the outer radial derivative are
second-order central differences on the logarithmic grid.

The new effective potential
\begin{align}
  V(r)
  =
  -\frac{1}{r}
  +
  V_{\rm H}(r)
  +
  \left[
  V_{\rm xc}(\nz+\Delta n)
  -
  V_{\rm xc}(\nz)
  \right]
\label{eq:effective_potential}
\end{align}
is shifted so that $V(r_{\max})=0$ and smoothly damped over a width of $1\,a_{0}$ beyond $0.8\,r_{\max}$. Linear mixing
$V \leftarrow \alpha V_{\mathrm{new}} + (1-\alpha)V_{\mathrm{old}}$ with $\alpha=0.01$ is iterated until two criteria are met simultaneously: the root-mean-square change of the potential falls below $10^{-5}\,\mathrm{Ha}$, and $V_{\rm H}(0)$ is stable to better than $10^{-7}\,\mathrm{Ha}$ over $50$ consecutive iterations. The second criterion measures self-consistency in the reported observable rather than in the potential residual that drives the stopping test. Convergence requires between $367$ and $1100$ iterations across the density range, with a median of $635$. Should the residual grow by more than a factor of $20$ relative to its initial value, $\alpha$ is halved and the iteration restarted, down to a floor of $5\times10^{-4}$; this safeguard was not triggered in any of the production runs reported here.

The cycle is initialized with a Thomas--Fermi screened Coulomb potential
$V_{0}(r) = -e^{-q_{\rm TF}r}/r$, $q_{\rm TF}=\sqrt{4k_{\rm F}/\pi}$, where the
negative sign reflects the attractive potential of the proton, at the first
(most dilute) density only; subsequent densities are initialized by
continuation as described above.
Spectral consistency is monitored through the total Friedel sum in Eq.~\eqref{eq:friedel} and the agreement between bound-state occupations and the Levinson count. The finite-domain real-space charge integral is retained as a supplementary diagnostic, not as a branch-selection criterion or as an error estimate for $V_{\rm H}(0)$. The principal observable reported for comparison with Almbladh \emph{et al.}\ is the induced Hartree potential at the origin, $V_{\rm H}(0)$, as a function of $\rs$, together with the radial profile of $\Delta n(r)$ shown in Figs.~1 and 2 of Ref.~\cite{1976AlmbladhScreeningProtonElectron}.
\begin{table*}[tp]
  \centering
  \caption{Convergence of the induced Hartree potential at the proton,
  $V_{\rm H}(0)$, with respect to every numerical parameter of the
  self-consistent Kohn--Sham calculation using the LDA exchange-correlation functional. Entries are the largest relative
  deviation $|\Delta V_{\rm H}(0)|/V_{\rm H}(0)$ found on varying the parameter over
  the stated range. Where the production value is itself the most accurate
  setting tested, the entry is the change over the final refinement step. The
  four outer-tail entries are prescription sensitivities quoted as the full
  spread over the range, these parameters having no more-accurate limit. The
  shallow-bound-state treatment carries no adjustable parameter and so admits
  no convergence entry.}
  \label{tab:dft-convergence}

  \small
  \renewcommand{\arraystretch}{1.12}

  \begin{tabularx}{\textwidth}{
    @{}
    >{\raggedright\arraybackslash}p{0.25\textwidth}
    >{\raggedright\arraybackslash}p{0.18\textwidth}
    >{\raggedright\arraybackslash}X
    >{\raggedleft\arraybackslash}p{0.18\textwidth}
    @{}
  }
    \hline
    \textbf{Parameter}
    & \textbf{Production}
    & \textbf{Varied over}
    & $\boldsymbol{|\Delta V_H(0)|/V_H(0)}$ \\
    \hline

    $r_{\min}$
    & $10^{-4}\,a_0$
    & $10^{-3}$--$10^{-6}\,a_0$
    & $2.6\times10^{-7}$ \\

    $r_{\max}$
    & $50\,a_0$
    & $15$--$50\,a_0$
    & $3.5\times10^{-4}$ \\

    Radial grid density $N$
    & $8000$
    & $2000$--$8000$
    & $2.8\times10^{-4}$ \\

    $\ell_{\max}$
    & $12$
    & $4$--$12$
    & $1.5\times10^{-4}$ \\

    $k$-quadrature nodes $n_k$
    & $60$
    & $15$--$60$
    & $5.7\times10^{-12}$ \\

    Self-consistency tolerance
    & $10^{-5}\,\mathrm{Ha}$
    & $10^{-3}$--$10^{-6}\,\mathrm{Ha}$
    & $4.9\times10^{-7}$ \\

    Mixing parameter $\alpha$
    & $0.01$
    & $0.01$ vs.\ $0.002$
    & $5.0\times10^{-7}$ \\[2pt]

    \multicolumn{4}{@{}l}{Outer tail} \\

    \quad $R_{\mathrm{match}}/r_{\max}$
    & $0.8$
    & $0.7$--$0.88$
    & $9.3\times10^{-4}$ \\

    \quad Density cutoff width
    & $1.0\,a_0$
    & $0.5$--$4\,a_0$
    & $6.4\times10^{-4}$ \\

    \quad $V$-damping radius$/r_{\max}$
    & $0.8$
    & $0.7$--$0.95$
    & $1.0\times10^{-4}$ \\

    \quad Phase-matching stencil
    & adaptive
    & $5$--$40$ points
    & $3.8\times10^{-4}$ \\[2pt]

    Shallow bound states
    & \multicolumn{2}{@{}l}{parameter-free (Sec.~\ref{sec:almbladh_method})}
    & --- \\[2pt]

    \hline
    \multicolumn{3}{@{}l}{Discretization and self-consistency, in quadrature}
    & $5\times10^{-4}$ \\

    \multicolumn{3}{@{}l}{Outer-tail treatment, in quadrature}
    & $1.2\times10^{-3}$ \\

    \multicolumn{3}{@{}l}{Total numerical uncertainty}
    & $1.3\times10^{-3}$ \\

    \hline
  \end{tabularx}
\end{table*}
%

\subsection{Convergence}

Convergence of $V_{\rm H}(0)$ with respect to every numerical parameter introduced above and using the LDA functional is demonstrated in Table~\ref{tab:dft-convergence}. Entries are relative deviations $|\Delta V_{\rm H}(0)|/V_{\rm H}(0)$; for scale, $V_{\rm H}(0)$ ranges from $1.159$ to $1.999\,\mathrm{Ha}$ over $\rs = 0.5$--$6.0$. 
\begin{table*}[tp]
  \centering
  \caption{Convergence of the induced Hartree potential at the proton,
  $V_{\rm H}(0)$, for the PBE calculations. Entries are the largest relative
  deviation $|\Delta V_{\rm H}(0)|/V_{\rm H}(0)$ found on varying the parameter
  over the stated range, at $\rs = 1.5$, $2$, $4$ and $6$. Where the
  production value is itself the most accurate setting that converges, the
  entry is the change over the final available refinement step; the
  finite-difference step $h$ and the four outer-tail entries have no
  more-accurate limit and are quoted as the full spread over the range.}
  \label{tab:pbe-convergence}

  \small
  \renewcommand{\arraystretch}{1.12}

  \begin{tabularx}{\textwidth}{
    @{}
    >{\raggedright\arraybackslash}p{0.25\textwidth}
    >{\raggedright\arraybackslash}p{0.18\textwidth}
    >{\raggedright\arraybackslash}X
    >{\raggedleft\arraybackslash}p{0.18\textwidth}
    @{}
  }
    \hline
    \textbf{Parameter}
    & \textbf{Production}
    & \textbf{Varied over}
    & $\boldsymbol{|\Delta V_H(0)|/V_H(0)}$ \\
    \hline

    $r_{\min}$
    & $10^{-4}\,a_0$
    & $10^{-3}$--$10^{-4}\,a_0$
    & $1.0\times10^{-5}$ \\

    $r_{\max}$
    & $50\,a_0$
    & $15$--$50\,a_0$
    & $2.5\times10^{-4}$ \\

    Radial grid density $N$
    & $8000$
    & $2000$--$8000$
    & $6.7\times10^{-7}$ \\

    $\ell_{\max}$
    & $12$
    & $4$--$12$
    & $3.8\times10^{-5}$ \\

    $k$-quadrature nodes $n_k$
    & $60$
    & $15$--$60$
    & $3.3\times10^{-6}$ \\

    Self-consistency tolerance
    & $10^{-4}\,\mathrm{Ha}$
    & $10^{-3}$--$10^{-4}\,\mathrm{Ha}$
    & $1.2\times10^{-7}$ \\

    Mixing parameter $\alpha$
    & $0.01$
    & $0.01$ vs.\ $0.002$
    & $4.4\times10^{-7}$ \\

    Finite-difference step $h$
    & $10^{-4}$
    & $10^{-3}$--$10^{-4}$
    & $1.8\times10^{-6}$ \\[2pt]

    \multicolumn{4}{@{}l}{Outer tail} \\

    \quad $R_{\mathrm{match}}/r_{\max}$
    & $0.8$
    & $0.7$--$0.88$
    & $8.7\times10^{-4}$ \\

    \quad Density cutoff width
    & $1.0\,a_0$
    & $0.5$--$4\,a_0$
    & $6.6\times10^{-5}$ \\

    \quad $V$-damping radius$/r_{\max}$
    & $0.8$
    & $0.7$--$0.95$
    & $2.9\times10^{-4}$ \\

    \quad Phase-matching stencil
    & adaptive
    & $5$--$40$ points
    & $1.9\times10^{-4}$ \\[2pt]

    \hline
    \multicolumn{3}{@{}l}{Discretization and self-consistency, in quadrature}
    & $2.5\times10^{-4}$ \\

    \multicolumn{3}{@{}l}{Outer-tail treatment, in quadrature}
    & $9.4\times10^{-4}$ \\

    \multicolumn{3}{@{}l}{Total numerical uncertainty}
    & $9.7\times10^{-4}$ \\

    \hline
  \end{tabularx}
\end{table*}

Most parameters were tested by varying one at a time about the production values at $\rs = 1.5$, $2$, $4$ and $6$, the range spanning the bound-state threshold. The mixing parameter was instead validated by the stricter test of repeating the entire sweep with $\alpha$ reduced by a factor of five, changing nothing else, and differencing the two $V_{\rm H}(0)(\rs)$ curves. Over all $56$ densities the two agree to a maximum relative difference of $5.0\times10^{-7}$ and an rms of $3.3\times10^{-7}$, with identical bound-state occupation at every density and binding energies agreeing to $7\times10^{-5}\,\mathrm{eV}$.

The total numerical uncertainty is $1.3\times10^{-3}$ in relative terms, equivalently $1.5$--$2.6\,\mathrm{mHa}$. The limiting error is not the radial grid, the partial-wave expansion or the self-consistency, each of which is converged to better than $5\times10^{-4}$, but the truncation of the Friedel tail at the matching radius. 

The following diagnostics are additionally evaluated at every density of the production run. $V_{\rm H}(0)$ is computed by three independent quadrature rules, which agree to within $1.2\times10^{-5}\,\mathrm{Ha}$. Its drift over the final fifty iterations, which measures self-consistency error in the reported observable rather than in the potential residual, is at most $1.0\times10^{-7}\,\mathrm{Ha}$. The total Friedel sum is monitored against $Z=1$, and the bound-state occupations are checked independently against the Levinson count. As discussed in the bound-state subsection, finite-domain real-space charge integration is sensitive to the long-range density tail and is not used as an uncertainty estimate for $V_{\rm H}(0)$. In Table~\ref{tab:pbe-convergence}, A null test with $Z = 0$, where $\Delta n \equiv 0$ and $V_{\rm H}(0) = 0$ identically, was performed as part of the convergence study. The residual is independent of $\ell_{\max}$ and $n_{k}$ and falls from $2.1\times10^{-7}$ to $9.3\times10^{-10}\,\mathrm{Ha}$ as $N$ increases from $2000$ to $8000$, indicating that the residual observed in this
null test is controlled by radial-grid discretization.

The PBE results were obtained with the same radial solver and the same partial-wave and $k$-quadrature construction, and with a self-consistency criterion adapted to that functional: because $V_{\rm xc}$ is built from finite differences the potential residual has a noise floor near $10^{-5}\,\mathrm{Ha}$, so a tolerance of $10^{-4}\,\mathrm{Ha}$ is used together with the same $V_{\rm H}(0)$ stability requirement. Production uses $h = 10^{-4}$: this floor grows as $1/h$, so $h = 10^{-5}$
fails to reach the tolerance at one of the four test densities and
$h = 10^{-6}$ at all of them.

The total numerical uncertainty of $V_{\rm H}(0)$ in the PBE calculations is
$9.7\times10^{-4}$ in relative terms, comparable to the $1.3\times10^{-3}$ of
the LDA calculation. The induced density at the proton was tracked alongside
and is converged to $6.4\times10^{-5}$ with respect to every parameter except
$r_{\min}$, where the contact density is sampled at radii differing by an
order of magnitude and the cusp in $\Delta n(r)$ accounts for most of the
resulting $4.1\times10^{-3}$.

\section{Convergence study for \texorpdfstring{$\Delta n^{\rm QM--VPA}(r)$ and $V_{\rm H}^{\rm QM--VPA}(0)$}{Delta n and V_H}}
\label{app:vpa_density_hartree_convergence}
The convergence of the FSR root and of the associated screening parameter
$\alpha_{\rm FSR}$ is documented separately in Table~\ref{tab:FSR_numerical_error} and
Appendix~\ref{sec:friedel_pipeline}. The present appendix concerns the
downstream QM--VPA reconstruction at exact $T=0$: once $\alpha_{\rm FSR}$ is
fixed, the phase--amplitude solution is used to construct the occupied bound
and continuum states, assemble the induced density, and evaluate the
density-derived contact Hartree potential.

The density decomposition is defined in
Eq.~\eqref{eq:vpa_density_decomposition}, with the bound contribution given by
Eq.~\eqref{eq:vpa_bound_density}.  The continuum kernel $\Phi(k;r)$ and the
corresponding scattering density are defined in
Eqs.~\eqref{eq:vpa_Phi_definition} and
\eqref{eq:vpa_scattering_density_Phi}, respectively, while the complete
zero-temperature density is given by Eq.~\eqref{eq:vpa_total_density_zeroT}.
The final observable is $V_{\rm H}^{\rm QM--VPA}(0)$ of
Eq.~\eqref{eq:vpa_VH0_definition}.  All convergence tests reported below are
therefore performed directly on quantities entering these equations.

\subsection{Numerical controls and zero-temperature tail treatment}

For the partial-wave sum entering $\Phi(k;r)$, define the weighted increment
\begin{equation}
c_\ell(k;r)
\equiv
\frac{2}{\pi}(2\ell+1)a_\ell^{\rm QM}(k;r),
\quad
\Phi_L(k;r)=\sum_{\ell=0}^{L}c_\ell(k;r).
\label{eq:vpa_phi_increment_appendix}
\end{equation}
The production angular controller increases $L$ until
$\max_r|c_\ell(k;r)|\le 10^{-7}$ for four consecutive channels, with
$L_{\min}=3$ and safety ceiling $L_{\rm cap}=266$; $L_{\max}(k)$ denotes the
final accepted cutoff at a retained momentum node.  This density-level
criterion is distinct from the partial-wave criterion used in determining
$\alpha_{\rm FSR}$ from the FSR.

The continuum integral in Eq.~\eqref{eq:vpa_scattering_density_Phi} is
evaluated over $0\le k\le k_{\rm F}$ with an adaptive retained-$k$ quadrature using
an embedded Gauss pair of orders $16/8$.  At exact $T=0$ the upper endpoint is
exactly $k_{\max}=k_{\rm F}$.  The radial density master uses a uniform spacing
$\Delta r=0.01\,a_0$.  We denote its finite radial extent by $R_{\max}$ to
distinguish it from the adaptive radial endpoint used in the phase-equation
integration.  The cumulative Hartree contact is
\begin{equation}
V_{\rm H}^{\rm QM--VPA}(0;R)
=
4\pi\int_0^R r\,\Delta n^{\rm QM--VPA}(r)\,dr,
\label{eq:vpa_VH0_cumulative_appendix}
\end{equation}
which approaches Eq.~\eqref{eq:vpa_VH0_definition} only after the
long-range Friedel contribution has been included.

At $T=0$ the late-$r$ Hartree integrand
\begin{equation}
g(r)\equiv4\pi r\,\Delta n^{\rm QM--VPA}(r)
\end{equation}
is completed with the form
\begin{equation}
\begin{split}
g_{\rm M23}(r)={}&
\frac{C_2\cos(2k_{\rm F}r)+D_2\sin(2k_{\rm F}r)}{r^2}
\\
&+
\frac{C_3\cos(2k_{\rm F}r)+D_3\sin(2k_{\rm F}r)}{r^3}.
\end{split}
\label{eq:vpa_M23_appendix}
\end{equation}
The label ``M23'' denotes the retained $r^{-2}$ and $r^{-3}$ terms; the
oscillation frequency $2k_{\rm F}$ is fixed rather than fitted.  The final production
domain is $R_{\max}=80\,a_0$.  Five fits over
$[45,80]$, $[50,80]$, $[55,80]$, $[60,80]$, and $[65,80]\,a_0$ are averaged,
and the remaining $80\,a_0\rightarrow\infty$ contribution is evaluated
analytically.  Thus
\begin{equation}
\begin{split}
V_{\rm H}^{\rm QM--VPA}(0)
={}&
V_{\rm H}^{\rm QM--VPA}(0;80\,a_0)
\\
&+
\left\langle
\Delta V_{\rm H}^{\rm M23}(80\rightarrow\infty)
\right\rangle_{\rm 5\,windows}.
\end{split}
\label{eq:vpa_VH0_M23_production_appendix}
\end{equation}
The tail model is validated by shifted withheld-shell tests entirely inside the
directly calculated domain: fits on $[40,60]$ and $[45,65]\,a_0$ predict the
$[60,70]$ and $[65,75]\,a_0$ shells, while fits on $[50,70]$ and
$[55,70]\,a_0$ predict the $[70,80]\,a_0$ shell.  The largest difference
between predicted and directly integrated withheld shells, together with the
spread among the five inferred contacts, is used as an empirical convergence
diagnostic rather than as a statistical error bar.

Near a shallow-bound-state threshold, the bound and scattering contributions
can be individually much less stable than their sum.  The bound-state radial
domain was therefore enlarged independently at $r_s=2.4$: the box-enlargement
factor $6.0$ failed the wavefunction-profile gate, whereas $6.5$, $7.0$, and
$7.5$ passed, and the production value was fixed at the coarsest passing
choice, $6.5$.  More importantly, applying M23 to the total
bound-plus-scattering density reduces the maximum withheld-shell error from
$7.71\times10^{-4}$ Ha for a scattering-only fit to
$2.44\times10^{-8}$ Ha.  The production tail treatment therefore acts on the
total density of Eq.~\eqref{eq:vpa_density_decomposition}.
\subsection{Observable-level convergence and validation}

Table~\ref{tab:vpa_density_hartree_convergence} summarizes the principal
zero-temperature sensitivity tests.  For direct refinement rows, the last
column is the largest observed relative change in the fully assembled
$V_{\rm H}^{\rm QM--VPA}(0)$.  The $\Phi(k;r)$ row instead reports the
relative omitted-tail diagnostic of the partial-wave sum, while the M23 row
uses a conservative relative equivalent of the largest withheld-shell
residual.

\begin{table*}[tp]
  \centering
  \caption{Convergence and validation of the exact-$T=0$ QM--VPA
  density/contact-Hartree calculation.  ``M23'' is defined by
  Eq.~\eqref{eq:vpa_M23_appendix}.  The last column gives the convergence
  measure defined in the text; the Popovi\'c--Stott row gives the largest
  relative external-benchmark discrepancy.}
  \label{tab:vpa_density_hartree_convergence}
  \small
  \renewcommand{\arraystretch}{1.12}
  \begin{tabularx}{\textwidth}{
    @{}
    >{\raggedright\arraybackslash}p{0.22\textwidth}
    >{\raggedright\arraybackslash}p{0.18\textwidth}
    >{\raggedright\arraybackslash}X
    >{\raggedleft\arraybackslash}p{0.17\textwidth}
    @{}
  }
    \hline
    \textbf{Parameter / validation}
    & \textbf{Production}
    & \textbf{Varied or tested over}
    & \textbf{Convergence measure} \\
    \hline

    $\Phi(k;r)$ angular stop
    & $10^{-7}$, 4 channels
    & overextended $L$; $r_s=1$
    & $4.47\times10^{-10}$ \\

    Radial spacing $\Delta r$
    & $0.01\,a_0$
    & $0.005$--$0.01\,a_0$; $r_s=4$
    & $2.3\times10^{-9}$ \\

    Density angular tolerance
    & $10^{-7}$
    & $10^{-8}$--$10^{-7}$; $r_s=4,6$
    & $3.1\times10^{-10}$ \\

    Retained-$k$ Gauss pair
    & $16/8$
    & $16/8$ vs.\ $32/16$; $r_s=4,6$
    & $1.7\times10^{-10}$ \\

    Radial domain $R_{\max}$
    & $80\,a_0$
    & $60$--$80\,a_0$; $r_s=6$
    & $2.3\times10^{-8}$ \\

    M23 tail completion
    & total density; 5 windows
    & shifted withheld shells; 51 states
    & $\leq1.4\times10^{-8}$ \\

    \multicolumn{4}{@{}l}{Independent external validation} \\

    Popovi\'c--Stott reconstruction
    & published potential parameters
    & Al and Mg
    & $4.4\times10^{-4}$ \\

    \hline
  \end{tabularx}
\end{table*}

The direct refinement tests show that the radial spacing, density-level
partial-wave truncation, and retained-$k$ quadrature change the final contact
only at the $10^{-9}$--$10^{-10}$ relative level.  The largest internal
sensitivity is associated with the long-range Friedel treatment.  At the
dilute-end stress state $r_s=6$, extending the directly calculated density
domain from $60$ to $80\,a_0$ changes the M23-completed contact by
$2.3\times10^{-8}$ relatively.  The $R_{\max}=60\,a_0$ withheld-shell test
failed the adopted empirical gate, whereas the shifted $R_{\max}=80\,a_0$
test passed; $R_{\max}=80\,a_0$ was therefore frozen for production.

Applied without retuning to the 51-state Yukawa sweep
$r_s=1.0,1.1,\ldots,6.0$, the same policy passed all state-level checks.  The
maximum shifted-$R=80$ withheld-shell residual is
$1.06\times10^{-8}$ Ha, and the largest realized density-level cutoff is
$L_{\max}=20\ll L_{\rm cap}=266$.  These production-wide checks complement
the individual refinements in Table~\ref{tab:vpa_density_hartree_convergence}.
\subsection{Independent external validation with the Popovi\'c--Stott potentials}
The internal sensitivity tests are supplemented by the density-informed
two-parameter potentials of Popovi\'c \textit{et al.}~
\cite{1976PopovicTheoryHeatSolution}, discussed in
Sec.~\ref{sec:Popovic_potential_vh0}.  Their published potential parameters
were propagated through the present zero-temperature QM--VPA density/Hartree
pipeline without refitting.  For the Hartree-only branch, the reconstructed
contacts are $1.015556$ Ha for Al and $0.937263$ Ha for Mg, compared with the
historical tabulated values $1.016$ and $0.937$ Ha, corresponding to relative
differences of $-0.0437\%$ and $+0.0281\%$.  For the
exchange--correlation-informed branch, the reconstructed values are
$1.185735$ and $1.156837$ Ha, compared with $1.186$ and $1.157$ Ha, giving
differences of $-0.0223\%$ and $-0.0140\%$.

This comparison exercises the complete numerical chain from a prescribed
screened potential through the bound and continuum wavefunctions, the radial
density, and the nonlocal Hartree moment of
Eq.~\eqref{eq:vpa_VH0_definition}.  Agreement at the few-$10^{-4}$ relative
level therefore provides an independent end-to-end validation of the
density/Hartree reconstruction.  It is not interpreted as an exact many-body
benchmark, since the Popovi\'c--Stott potentials themselves arise from an
approximate historical self-consistent construction and the published values
are tabulated to limited precision.

The tests in Table~\ref{tab:vpa_density_hartree_convergence} are not combined
into a single propagated uncertainty.  They are direct sensitivity and
certification tests rather than statistically independent random errors, and
the M23 window spreads and withheld-shell residuals are empirical diagnostics.
Together with the 51-state production replay and the independent
Popovi\'c--Stott reconstruction, they show that the reported
$\Delta n^{\rm QM--VPA}(r)$ and $V_{\rm H}^{\rm QM--VPA}(0)$ are numerically
stable with respect to the tested radial, angular, momentum-quadrature, and
long-range-tail controls.

\section{Numerical Verification of the Quantum-Density Cusp for FSR-Constrained Model Potentials}
\label{app:qm_vpa_cusp}

The exact Coulomb cusp condition constrains the total electronic density
at a point positive charge, rather than only the induced density
\cite{1957Kato_cusp,1982_Carlsson_cusp}.  For a proton ($Z=1$), in atomic
units, with $n(r)=n_0+\Delta n(r)$ and $n_0=3/(4\pi\rs^3)$,
\begin{align}
\left.\frac{dn(r)}{dr}\right|_{r=0}
&=-2Z\,n(0), \nonumber\\
\left.\frac{d\Delta n(r)}{dr}\right|_{r=0}
&=-2Z\,[n_0+\Delta n(0)].
\label{eq:qmvpa_cusp_exact_appendix}
\end{align}
As explained in Sec.~\ref{sec:cusp}, this is a local derivative relation;
it does not fix the value of $\Delta n(0)$ or the radial moment defining
$V_{\rm H}(0)$.
\subsection{Why the quantum reconstruction obeys the cusp}
The Yukawa, hydrogenic, and Hulth\'en model potentials considered in this
work share the bare Coulomb singularity,
\begin{equation}
V_{\nu}(r)=-\frac{Z}{r}+O(1),\qquad r\to0.
\label{eq:qmvpa_coulomb_short_range_appendix}
\end{equation}
For a regular $s$-wave bound or scattering eigenfunction of the corresponding
one-electron Schr\"odinger equation, the small-$r$ radial expansion has the
form $\psi_{\nu,\ell=0}(r)=\psi_{\nu,\ell=0}(0)
[1-Zr+O(r^2)]$.  Its density therefore obeys
$\bigl.\partial_r|\psi_{\nu,0}(r)|^2\bigr|_{0}
=-2Z|\psi_{\nu,0}(0)|^2$.  Regular higher-partial-wave densities
are $O(r^{2\ell})$ for $\ell\geq1$ and contribute neither to the
contact value nor to its first radial derivative.  Summing over the
occupied bound and continuum states therefore preserves the total-density
cusp.  Since the homogeneous background $n_0$ is constant, subtracting it
to form the induced density gives Eq.~\eqref{eq:qmvpa_cusp_exact_appendix}.

In the present one-parameter models, the screening parameters were
determined only from the zero-temperature Friedel sum rule (FSR).
No cusp condition was imposed in those parameter fits or in the subsequent
phase--amplitude reconstruction of the bound and scattering states.  Thus,
for the corresponding quantum density of
Eq.~\eqref{eq:vpa_total_density_zeroT}, satisfaction of the cusp is a
consequence of the Coulombic short-range wavefunctions, not an additional
potential-fitting constraint.  This observation concerns the
quantum-mechanically reconstructed density, not the distinct
Poisson-derived density associated electrostatically with each analytic
model potential [Sec.~\ref{sec:model_potentials}].
\subsection{Independent numerical test at \texorpdfstring{$\rs=2$}{rs=2}}

We test Eq.~\eqref{eq:qmvpa_cusp_exact_appendix} using the independently
computed $T=0$ QM--VPA master radial-density profiles for Yukawa,
hydrogenic, and Hulth\'en potentials at $\rs=2$.  Each profile includes the
bound and scattering contributions and was tabulated on a uniform radial
grid with spacing $\Delta r=0.01$ extending to $r=80$.  The contact density
$\Delta n^{\rm QM--VPA}(0)$ is taken directly from the separately recorded
$r=0$ value in each master profile; the radial derivative is evaluated
independently from strictly positive-radius points.  In particular, the
cusp slope is not prescribed in the derivative fit.

Throughout this appendix, lengths, densities, and their radial
derivatives are expressed in atomic units, so explicit powers of $a_0$
are suppressed.

\begin{figure}[htbp!]
\centering
\includegraphics[width=\columnwidth]{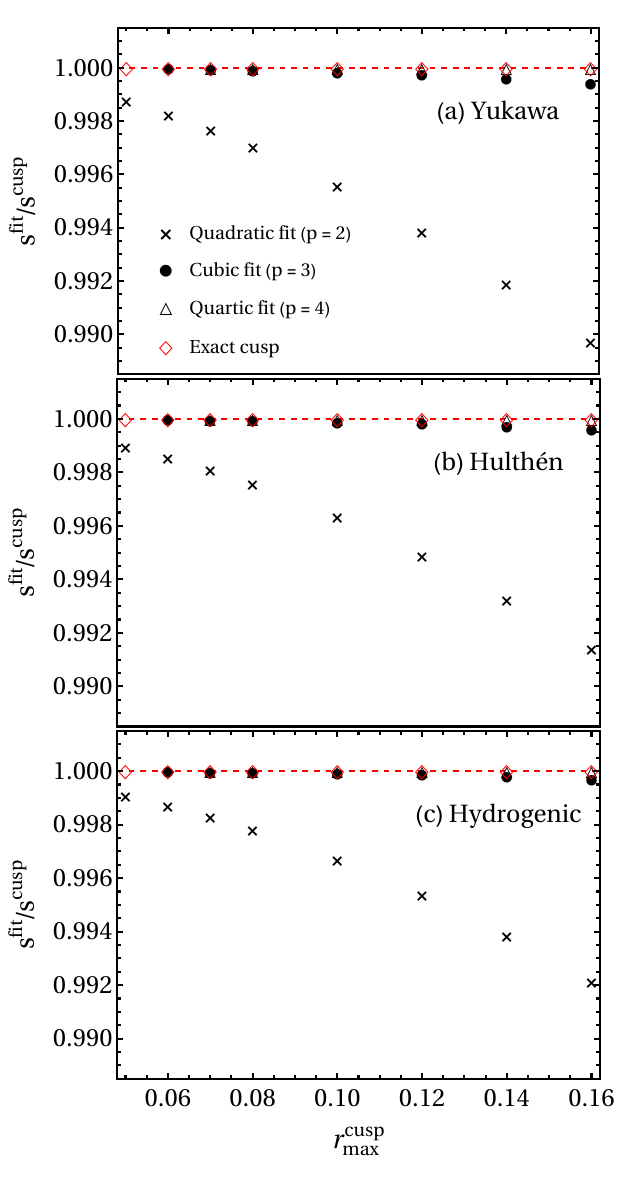}
\caption{Cusp-slope convergence at $\rs=2$ for the FSR-constrained
(a) Yukawa, (b) Hulth\'en, and (c) hydrogenic potentials. The ordinate
is $s_{\nu}^{\rm fit}(p,r_{\max}^{\rm cusp})/s_{\nu}^{\rm cusp}$,
shown in the panels with the abbreviated label
$s^{\rm fit}/s^{\rm cusp}$; the abscissa is the outermost included
radius $r_{\max}^{\rm cusp}$. Black crosses, filled circles, and open
triangles denote individual fixed-contact fits of degree $p=2$, $3$,
and $4$, respectively. Open red diamonds and the dashed red horizontal
line identify the exact total-density cusp ratio, unity. The underlying
radial grid has spacing $\Delta r=0.01$. Unlike the median of ten
primary cubic and quartic slopes reported in
Table~\ref{tab:qmvpa_cusp_test_rs2}, each fitted marker represents one
polynomial degree and one fitting window. The degree/window dependence
is an empirical sensitivity diagnostic, not a rigorous error bound.}
\label{fig:qmvpa_cusp_window_convergence}
\end{figure}

For each potential, we fit the near-origin data to
\begin{align}
\Delta n^{\rm QM--VPA}(r)
&-\Delta n^{\rm QM--VPA}(0) \nonumber\\
&=\sum_{j=1}^{p}c_j\left(\frac{r}{r_{\max}^{\rm cusp}}\right)^j,
\nonumber\\
\left.\frac{d\Delta n^{\rm QM--VPA}}{dr}\right|_{r=0}^{\!\rm fit}
&=\frac{c_1}{r_{\max}^{\rm cusp}}.
\label{eq:qmvpa_cusp_polyfit_appendix}
\end{align}
where $r_{\max}^{\rm cusp}$ denotes the outermost positive radial
grid point included in the cusp-slope fit. This dedicated notation
distinguishes the local fitting-window endpoint from the radial cutoffs
used in the VPA propagation and Hartree or charge integrals. The primary fitting windows are
$\mathcal W=\{0.08,0.10,0.12,0.14,0.16\}$, corresponding to 8, 10,
12, 14, and 16 positive-radius grid points, and the primary degrees
are $p=3,4$. The contact value is fixed to the $r=0$ master record. For each
potential $\nu$, polynomial degree $p$, and fitting radius $r_{\max}^{\rm cusp}$, we denote
the extrapolated contact slope by
\begin{align}
s_{\nu}^{\rm fit}(p,r_{\max}^{\rm cusp})
&\equiv \left.\frac{d\Delta n_{\nu}^{\rm QM--VPA}(r)}{dr}
\right|_{r=0}^{\!\rm fit}\nonumber\\
&=\frac{c_{1,\nu}(p,r_{\max}^{\rm cusp})}{r_{\max}^{\rm cusp}},
\nonumber\\
s_{\nu}^{\rm fit}
&\equiv \operatorname{median}_{\substack{p=3,4\\
r_{\max}^{\rm cusp}\in\mathcal W}}
\!s_{\nu}^{\rm fit}(p,r_{\max}^{\rm cusp}).
\label{eq:qmvpa_sfit_definition_appendix}
\end{align}
\begin{table*}[htbp!]
  \centering
  \caption{Numerical verification of the exact total-density cusp
  for the three FSR-constrained one-parameter potentials at $\rs=2$ and
  $Z=1$. All quantities are in atomic units. The fitted slope is the median
  of the ten primary cubic and quartic fits; the last column gives its
  relative deviation from the exact cusp slope, not an uncertainty in
  $\Delta n(0)$ or $V_{\rm H}(0)$.}
  \label{tab:qmvpa_cusp_test_rs2}
  \small
  \renewcommand{\arraystretch}{1.12}
  \begin{tabularx}{\textwidth}{
    @{}
    >{\raggedright\arraybackslash}p{0.17\textwidth}
    >{\raggedleft\arraybackslash}X
    >{\raggedleft\arraybackslash}p{0.19\textwidth}
    >{\raggedleft\arraybackslash}p{0.19\textwidth}
    >{\raggedleft\arraybackslash}p{0.12\textwidth}
    @{}
  }
    \hline
    \textbf{Potential}
    & $\boldsymbol{\Delta n^{\rm QM--VPA}(0)}$
    & $\boldsymbol{s_{\nu}^{\rm fit}}$
    & $\boldsymbol{s_{\nu}^{\rm cusp}}$
    & $\boldsymbol{\eta_{\nu}}$ (\%) \\
    \hline
    Yukawa     & 0.373435983 & $-0.806510851$ & $-0.806555069$ & 0.00548 \\
    Hulth\'en  & 0.445345857 & $-0.950340742$ & $-0.950374818$ & 0.00359 \\
    Hydrogenic & 0.450901787 & $-0.961463994$ & $-0.961486678$ & 0.00236 \\
    \hline
  \end{tabularx}
\end{table*}
Thus $s_{\nu}^{\rm fit}$ is the reported numerical estimate of
$\Delta n_{\nu}'(0)$, obtained as the median of ten fits, and is
not constrained to equal the theoretical cusp slope. The
wider fit-window and degree dependence, including quadratic fits, is
examined in Fig.~\ref{fig:qmvpa_cusp_window_convergence}.  Free-intercept
fits also recover the recorded contact densities to better than
$10^{-3}\%$ of the total contact density, providing an additional
consistency check.
At $\rs=2$, the uniform background density is
$n_0=3/(32\pi)=0.02984155$.  Table~\ref{tab:qmvpa_cusp_test_rs2}
compares the fitted slopes with the exact total-density prediction.
Here $s_{\nu}^{\rm cusp}$ is the slope predicted by the exact
total-density cusp condition, while $s_{\nu}^{\rm fit}$ is defined
independently by Eq.~\eqref{eq:qmvpa_sfit_definition_appendix}.
The tabulated relative deviation is
\begin{align}
s_{\nu}^{\rm cusp}
&=-2\,[n_0+\Delta n_{\nu}^{\rm QM--VPA}(0)],\nonumber\\
\eta_{\nu}
&=100\,\frac{\left|s_{\nu}^{\rm fit}-s_{\nu}^{\rm cusp}\right|}
{\left|s_{\nu}^{\rm cusp}\right|}.
\label{eq:qmvpa_cusp_percent_error_appendix}
\end{align}

Figure~\ref{fig:qmvpa_cusp_window_convergence} explicitly displays
the window-convergence test underlying Table~\ref{tab:qmvpa_cusp_test_rs2}.
Each panel plots $s_{\nu}^{\rm fit}(p,r_{\max}^{\rm cusp})/s_{\nu}^{\rm cusp}$
against the outermost cusp-fit radius $r_{\max}^{\rm cusp}$ for
quadratic, cubic, and quartic polynomials. Each symbol represents an
individual fit to a discrete radial window; the master data have
$\Delta r=0.01$. The cubic and quartic results remain close to unity
across the tested windows, whereas the quadratic estimates drift with
increasing $r_{\max}^{\rm cusp}$, consistent with truncation bias.  This visual comparison tests the
robustness of the numerical derivative without imposing the cusp.

All three extrapolated derivatives agree with the exact total-density
cusp to better than $0.006\%$.  By comparison, replacing the total density
by the induced density in the right-hand side would omit
$-2n_0=-0.05968310$ from the predicted slope.  This difference
corresponds to approximately $7.40\%$, $6.21\%$, and $6.28\%$ of the
exact slope magnitude for Yukawa, hydrogenic, and Hulth\'en, respectively,
well separated from the observed window-to-window fit scatter.  The
numerical results thus distinguish the exact total-density cusp from an
induced-density-only approximation and validate its recovery in the
QM--VPA density reconstruction at this representative state.  They do
not constitute a uniform numerical convergence proof for other $\rs$.

Finally, this validation clarifies the comparison with the two-range
Apagyi--Nagy construction.  In that construction, the exact cusp is
additionally imposed on an auxiliary Poisson-derived total-density
ansatz to determine its mixing coefficient
[Eq.~\eqref{eq:AN_weight}], before the FSR constrains the remaining
screening parameters.  The subsequent quantum density reconstructed from
the resulting Coulombic potential has its own automatic wavefunction cusp,
just as in the one-parameter models.  These two uses of the cusp condition
must not be conflated.  In particular, satisfying the total-density cusp
neither fixes the contact amplitude nor guarantees agreement of the
quantal contact Hartree moment with nonlinear DFT.
%
\clearpage
\bibliography{Refs}

\end{document}